\documentclass[twocolumn,aps,pra,showkeys,longbibliography,superscriptaddress,10pt]{revtex4-1}
\usepackage[breaklinks=true]{hyperref}
\hypersetup{
    colorlinks,
    linkcolor={red!50!black},
    citecolor={blue!50!black},
    urlcolor={blue!80!black}
}
\usepackage{blindtext}
\usepackage{amsmath,amssymb,amsthm,mathtools,mathrsfs}
\usepackage{tikz}
\usepackage{pgfplots}
\pgfplotsset{compat=1.18}
\usepackage{tikz-3dplot} \tdplotsetmaincoords{74}{130} %for 3d pictures
\usepgfplotslibrary{fillbetween}
\usetikzlibrary{arrows.meta}
\usetikzlibrary{patterns}
\usetikzlibrary{fpu}%floating point arithmetic??? used for more precision
\usepackage{pifont}% http://ctan.org/pkg/pifont %allows ding fonts
\usepackage{comment}
\usepackage{adjustbox}
\usepackage{dsfont}%this is for mathbb{1} but you write mathds{1}
\usepackage[normalem]{ulem}
\usepackage{enumerate}
\usepackage[all]{xy} %This is to make higher categorical stuff
\usetikzlibrary{circuits.ee.IEC}

\numberwithin{equation}{section}

\theoremstyle{plain}
\newtheorem{theorem}{Theorem}[section]
\newtheorem{lemma}[theorem]{Lemma}
\newtheorem{proposition}[theorem]{Proposition}
\newtheorem{corollary}[theorem]{Corollary}
\theoremstyle{definition}
\newtheorem{definition}[theorem]{Definition}
\newtheorem{remark}[theorem]{Remark}
\newtheorem{example}[theorem]{Example}

\newcommand\define[1]{\emph{\textbf{#1}}}

\newcommand{\bd}{\begin{definition}}
\newcommand{\ed}{\end{definition}}
\newcommand{\bt}{\begin{theorem}}
\newcommand{\et}{\end{theorem}}
\newcommand{\bc}{\begin{corollary}}
\newcommand{\ec}{\end{corollary}}
\newcommand{\blem}{\begin{lemma}}
\newcommand{\elem}{\end{lemma}}
\newcommand{\be}{\begin{equation}}
\newcommand{\ee}{\end{equation}}
\newcommand{\bx}{\begin{example}}
\newcommand{\ex}{\end{example}}
\newcommand{\bn}{\begin{proposition}}
\newcommand{\en}{\end{proposition}}
\newcommand{\bq}{\begin{question}}
\newcommand{\eq}{\end{question}}
\newcommand{\bprf}{\begin{proof}}
\newcommand{\eprf}{\end{proof}}
\newcommand{\br}{\begin{remark}}
\newcommand{\er}{\end{remark}}

\let\C=\Chi

\newcommand{\matr}{\mathbb{M}}%matrix algebra notation -- change as desired

\def\R{{{\mathbb R}}}
\def\C{{{\mathbb C}}}
\def\N{{{\mathbb N}}}

\newcommand{\id}{\mathrm{id}}

\newcommand{\Tr}{{\rm Tr} }

\newcommand{\<}{\langle}
\renewcommand{\>}{\rangle}

\newcommand{\Ad}{\mathrm{Ad}}

\newcommand{\Choi}{\mathscr{C}}
\newcommand{\Jamio}{\mathscr{J}}
\newcommand{\SWAP}{{\scriptstyle\mathrm{SWAP}}}

\DeclareFontFamily{OT1}{pzc}{}
\DeclareFontShape{OT1}{pzc}{m}{it}{ <-> s*[1.2] pzcmi7t }{}
\DeclareMathAlphabet{\mathpzc}{OT1}{pzc}{m}{it}
\newcommand{\Alg}[1]{\mathpzc{#1}}
\newlength\stateheight
\newlength\minimumstatewidth
\tikzset{width/.initial=\minimummorphismwidth}
\tikzset{colour/.initial=white}

\newif\ifblack\pgfkeys{/tikz/black/.is if=black}
\newif\ifwedge\pgfkeys{/tikz/wedge/.is if=wedge}
\newif\ifvflip\pgfkeys{/tikz/vflip/.is if=vflip}
\newif\ifhflip\pgfkeys{/tikz/hflip/.is if=hflip}
\newif\ifhvflip\pgfkeys{/tikz/hvflip/.is if=hvflip}

\pgfdeclarelayer{foreground}
\pgfdeclarelayer{background}
\pgfsetlayers{background,main,foreground}

\def\thickness{0.4pt}

\makeatletter
\pgfkeys{%
  /tikz/on layer/.code={
    \pgfonlayer{#1}\begingroup
    \aftergroup\endpgfonlayer
    \aftergroup\endgroup
  },
  /tikz/node on layer/.code={
    \gdef\node@@on@layer{%
      \setbox\tikz@tempbox=\hbox\bgroup\pgfonlayer{#1}\unhbox\tikz@tempbox\endpgfonlayer\pgfsetlinewidth{\thickness}\egroup}
    \aftergroup\node@on@layer
  },
  /tikz/end node on layer/.code={
    \endpgfonlayer\endgroup\endgroup
  }
}
\def\node@on@layer{\aftergroup\node@@on@layer}
\makeatother

\makeatletter
\pgfdeclareshape{state}
{
    \savedanchor\centerpoint
    {
        \pgf@x=0pt
        \pgf@y=0pt
    }
    \anchor{center}{\centerpoint}
    \anchorborder{\centerpoint}
    \saveddimen\overallwidth
    {
        \pgf@x=3\wd\pgfnodeparttextbox
        \ifdim\pgf@x<\minimumstatewidth
            \pgf@x=\minimumstatewidth
        \fi
    }
    \savedanchor{\upperrightcorner}
    {
        \pgf@x=.5\wd\pgfnodeparttextbox
        \pgf@y=.5\ht\pgfnodeparttextbox
        \advance\pgf@y by -.5\dp\pgfnodeparttextbox
    }
    \anchor{A}
    {
        \pgf@x=-\overallwidth
        \divide\pgf@x by 4
        \pgf@y=0pt
    }
    \anchor{B}
    {
        \pgf@x=\overallwidth
        \divide\pgf@x by 4
        \pgf@y=0pt
    }
    \anchor{text}
    {
        \upperrightcorner
        \pgf@x=-\pgf@x
        \ifhflip
            \pgf@y=-\pgf@y
            \advance\pgf@y by 0.4\stateheight
        \else
            \pgf@y=-\pgf@y
            \advance\pgf@y by -0.4\stateheight
        \fi
    }
    \backgroundpath
    {
       \begin{pgfonlayer}{foreground}
        \pgfsetstrokecolor{black}
        \pgfsetlinewidth{\thickness}
        \pgfpathmoveto{\pgfpoint{-0.5*\overallwidth}{0}}
        \pgfpathlineto{\pgfpoint{0.5*\overallwidth}{0}}
        \ifhflip
            \pgfpathlineto{\pgfpoint{0}{\stateheight}}
        \else
            \pgfpathlineto{\pgfpoint{0}{-\stateheight}}
        \fi
        \pgfpathclose
        \ifblack
            \pgfsetfillcolor{black!50}
            \pgfusepath{fill,stroke}
        \else
            \pgfusepath{stroke}
        \fi
       \end{pgfonlayer}
    }
}

\pgfdeclareshape{my ground}
{
  \inheritsavedanchors[from=rectangle ee]
  \inheritanchor[from=rectangle ee]{center}
  \inheritanchor[from=rectangle ee]{north}
  \inheritanchor[from=rectangle ee]{south}
  \inheritanchor[from=rectangle ee]{east}
  \inheritanchor[from=rectangle ee]{west}
  \inheritanchor[from=rectangle ee]{north east}
  \inheritanchor[from=rectangle ee]{north west}
  \inheritanchor[from=rectangle ee]{south east}
  \inheritanchor[from=rectangle ee]{south west}
  \inheritanchor[from=rectangle ee]{input}
  \inheritanchor[from=rectangle ee]{output}
  \anchorborder{\csname pgf@anchor@my ground@input\endcsname}

  \backgroundpath{
    \pgf@process{\pgfpointadd{\southwest}{\pgfpoint{\pgfkeysvalueof{/pgf/outer xsep}}{\pgfkeysvalueof{/pgf/outer ysep}}}}
    \pgf@xa=\pgf@x \pgf@ya=\pgf@y
    \pgf@process{\pgfpointadd{\northeast}{\pgfpointscale{-1}{\pgfpoint{\pgfkeysvalueof{/pgf/outer xsep}}{\pgfkeysvalueof{/pgf/outer ysep}}}}}
    \pgf@xb=\pgf@x \pgf@yb=\pgf@y
    \pgfpathmoveto{\pgfqpoint{\pgf@xa}{\pgf@ya}}
    \pgfpathlineto{\pgfqpoint{\pgf@xa}{\pgf@yb}}
    \pgfmathsetlength\pgf@xa{.5\pgf@xa+.5\pgf@xb}
    \pgfmathsetlength\pgf@yc{.16666\pgf@yb-.16666\pgf@ya}
    \advance\pgf@ya by\pgf@yc
    \advance\pgf@yb by-\pgf@yc
    \pgfpathmoveto{\pgfqpoint{\pgf@xa}{\pgf@ya}}
    \pgfpathlineto{\pgfqpoint{\pgf@xa}{\pgf@yb}}
    \advance\pgf@ya by\pgf@yc
    \advance\pgf@yb by-\pgf@yc
    \pgfpathmoveto{\pgfqpoint{\pgf@xb}{\pgf@ya}}
    \pgfpathlineto{\pgfqpoint{\pgf@xb}{\pgf@yb}}
  }
}
\makeatother

\tikzset{inline text/.style =
  {text height=1.2ex,text depth=0.25ex,yshift=0.5mm}}
\tikzset{arrow box/.style =
  {rectangle,inline text,fill=white,draw,
    minimum height=5mm,yshift=-0.5mm,minimum width=5mm}}
\tikzset{bubble/.style =
  {inner sep=0mm,minimum width=3mm,minimum height=3mm,
    draw,shape=circle,fill=white}}
\tikzset{dot/.style =
  {inner sep=0mm,minimum width=1mm,minimum height=1mm,
    draw,shape=circle}}
\tikzset{white dot/.style = {dot,fill=white,text depth=-0.2mm}}
\tikzset{scalar/.style = {diamond,draw,inner sep=1pt}}
\tikzset{square/.style =
  {inner sep=0mm,minimum width=2mm,minimum height=2mm,
    draw,shape=rectangle}}%%me trying to make a fakecopier

\tikzset{star/.style = {dot,fill=white,text depth=-0.2mm}}
\tikzset{copier/.style = {dot,fill,text depth=-0.2mm}}
\tikzset{fakecopier/.style = {square,fill,text depth=-0.2mm}}
\tikzset{discarder/.style = {my ground,draw,inner sep=0pt,
    minimum width=4.2pt,minimum height=11.2pt,anchor=input,rotate=90}}

\tikzset{xshiftu/.style = {shift = {(#1, 0)}}}
\tikzset{yshiftu/.style = {shift = {(0, #1)}}}
\tikzset{scriptstyle/.style={font=\everymath\expandafter{\the\everymath\scriptstyle}}}

\makeatletter
\renewcommand*\env@matrix[1][\arraystretch]{%
  \edef\arraystretch{#1}%
  \hskip -\arraycolsep
  \let\@ifnextchar\new@ifnextchar
  \array{*\c@MaxMatrixCols c}}
\makeatother

\def\meter{\raisebox{-2pt}{\begin{tikzpicture}[scale=0.75]\draw[-{Latex[length=1mm]}] (0,0) -- ({0.5*cos(60)},{0.5*sin(60)}); \draw[smooth,samples=40,domain=30:150] plot ({0.35*cos(\x)},({0.35*sin(\x)-0.1}); \end{tikzpicture}}}

\begin{document}
\title{Causal confusion in quantum systems: \texorpdfstring{\\}{ } distinguishing direct cause and common cause}
\author{Krisztina Tolotti}
\affiliation{Experimental Study Group, Massachusetts Institute of Technology, Cambridge, Massachusetts 02139, USA}
\affiliation{Department of Physics, Massachusetts Institute of Technology, Cambridge, Massachusetts 02139, USA}
\author{Riley Blank-Matthews}
\affiliation{Experimental Study Group, Massachusetts Institute of Technology, Cambridge, Massachusetts 02139, USA}
\affiliation{Department of Chemical Engineering, Massachusetts Institute of Technology, Cambridge, Massachusetts 02139, USA}
\author{Arthur J.~Parzygnat}
\email[Corresponding author: ]{arthurjp@mit.edu}
\affiliation{Experimental Study Group, Massachusetts Institute of Technology, Cambridge, Massachusetts 02139, USA}

\begin{abstract}
The joint expectation values of quantum observables measured by two parties typically depend on the causal structure of the measurement protocol. The two causal structures we investigate here are the direct-cause scenario, in which a single quantum state is measured twice in series with a general quantum channel between measurements, and the common-cause scenario, in which a shared bipartite quantum state is measured in parallel. Through a variety of examples of qubit channels with tunable parameters, we identify the full set of input states where the direct-cause joint expectation values of Pauli observables coincide with those obtained from a common cause, a situation that we call \emph{causal confusion}. More generally, for arbitrary finite-dimensional quantum systems and light-touch observables, we prove that a channel exhibits causal confusion for some input state if and only if it is a positive-partial-transpose (PPT) channel, thereby endowing PPT channels with a new operational meaning in terms of causal confusion. Using this result, we characterize the set of qubit channels for which there exists an input state leading to causal confusion as the set of entanglement-breaking channels. 
\end{abstract}


\maketitle

\section{Introduction}

\subsection{Causal confusion from joint expectation values}

When do quantum correlations reveal causal structure, such as common cause or direct cause? 
Here, we investigate the spatiotemporal quantum correlations compatible with two types of measurement protocols involving two parties called Alice (labeled by $\Alg{A}$) and Bob (labeled by $\Alg{B}$), with each performing local measurements on a quantum system~\cite{RAVJSR15,BMKG13,SongParzygnat2025,CRGWF18,MengEtAl2024,ZPTGVF18,KublerBraun2018}. We define these protocols more precisely in Section~\ref{sec:STdefns}, while here we provide a brief sketch with a diagrammatic visualization in order to summarize the achievements of this paper. In each experiment, Alice and Bob perform measurements of a particular collection of observables $L^{(\Alg{A})}_{1},\dots,L^{(\Alg{A})}_{M}$ for Alice and $L^{(\Alg{B})}_{1},\dots,L^{(\Alg{B})}_{N}$ for Bob. They repeat the experiments until they gather enough statistics to determine the probabilities of outcome occurrences. 

In the \emph{spatial scenario} (also sometimes called a \emph{common-cause} scenario~\cite{RAVJSR15}), Alice and Bob share a bipartite quantum state $\tau$ and perform these measurements locally, multiplying the outcomes and weighing them by the joint probability of their occurrence. The diagram 
\[
\begin{tikzpicture}[font=\small]
	\node[state] (omega) at (0,0) {\;\;\;$\tau$\;\;\;};
	\coordinate (B) at (0.25,2.00) {};
	\node[discarder] (X) at (0.85,1.65) {};
	\coordinate (A) at (-0.85,2.00) {};
	\node[discarder] (Y) at (-0.25,1.65) {};
	\node[arrow box] (e) at (-0.55,0.85) {\;$\meter$\;};
	\node[arrow box] (g) at (0.55,0.85) {\;$\meter$\;};
	\draw (omega) ++(0.55, 0) to (g);
	\draw[double] (g) ++(-0.30,0.25) to (B);
	\draw (g) ++(0.30,0.25) to (X);
	\draw (omega) ++(-0.55, 0) to (e);
	\draw[double] (e) ++(-0.30,0.25) to (A);
	\draw (e) ++(0.30,0.25) to (Y);
	\path[scriptstyle]
	node at (-0.75,0.35) {$\Alg{A}$}
	node at (0.70,0.35) {$\Alg{B}$};
\end{tikzpicture}
\]
depicts this spatial scenario visually, and it is to be read from bottom to top~\cite{ChJa18,Fr20,FGGPS22,Pa24,SongParzygnat2025,SSGC22}. At the bottom, a bipartite state $\tau$ is prepared. The two wires correspond to the two halves of the quantum system that Alice and Bob measure. They perform their local measurements and get certain outcomes, with the classical outcomes drawn as double wires. The grounding symbol indicates that Alice and Bob no longer need the quantum system after measurements (i.e., they discard the output quantum state after measurement). Each observable $L_{\gamma}$ has outcomes $\lambda_{\gamma z}\in\R$ labeled by a discrete parameter $z$ enumerating the number of such outcomes (this number may depend on $\gamma$). Let $\mathbb{P}(x,y|\alpha,\beta;\tau)$ denote the probability of Alice measuring outcome $\lambda^{(\Alg{A})}_{\alpha x}$ and Bob measuring outcome $\lambda^{(\Alg{B})}_{\beta y}$ given that Alice measures observable $L^{(\Alg{A})}_{\alpha}$ and Bob measures observable $L^{(\Alg{B})}_{\beta}$ on the bipartite state $\tau$. Their joint expectation value for this pair of observables is given by 
\be
\mathbb{E}\left[L^{(\Alg{A})}_{\alpha},L^{(\Alg{B})}_{\beta}\right]=\sum_{x,y}\lambda^{(\Alg{A})}_{\alpha x}\lambda^{(\Alg{B})}_{\beta y}\mathbb{P}(x,y|\alpha,\beta;\tau).
\ee
The collection of these numbers, indexed by $\alpha$ and $\beta$, forms an $M\times N$ matrix (a table) of joint expectation values among the observables Alice and Bob measure. 

In the \emph{temporal scenario} (also sometimes called a \emph{direct-cause} scenario~\cite{RAVJSR15}), Alice performs a measurement on a single quantum state $\rho$, then she sends the quantum state to Bob via some quantum channel $\mathcal{E}$, and finally Bob measures the received quantum state (unlike earlier works~\cite{RAVJSR15,KublerBraun2018}, we do not assume that $\rho$ is maximally mixed nor do we assume that $\mathcal{E}$ is a unital quantum channel). Alice and Bob also multiply their resulting outcomes and weigh them by the joint probability of their occurrence. The diagram 
\[
\begin{tikzpicture}[font=\small]
\node[state] (omega) at (0,-0.8) {\;$\rho$\;};
\node[arrow box] (q) at (0,-0.1) {\;\;\;$\meter$\;\;\;};
\node[arrow box] (g) at (0.40,0.8) {$\mathcal{E}$};
\node[arrow box] (N) at (0.40,1.7) {\;$\meter$\;};
\node[discarder] (d) at (0.65,2.45) {};
\draw (omega) to (q);
\draw (q) ++(0.40,0.25) to (g);
\draw[double] (q) ++(-0.40,0.25) to (-0.40,2.85);
\draw (g) to (N);
\draw (N) ++(0.25,0.25) to (d);
\draw[double] (N) ++(-0.25,0.25) to (0.15,2.85);
\path[scriptstyle]
node at (0.20,-0.55) {$\Alg{A}$}
node at (0.60,0.35) {$\Alg{A}$}
node at (0.60,1.25) {$\Alg{B}$};
\end{tikzpicture}
\] 
depicts this temporal scenario visually. Alice and Bob measure the same observables as they did in the spatial scenario, so that their possible outcomes are the same as in the spatial scenario, but the joint probabilities are calculated differently. This is because Alice's first measurement and the quantum channel $\mathcal{E}$ typically have an effect on the outcome of Bob's measurement~\cite{BCL90,BLM1996}. Let $\mathbb{P}(x|\alpha,\rho)$ denote the probability of Alice measuring outcome $\lambda^{(\Alg{A})}_{\alpha x}$ given that Alice measures observable $L^{(\Alg{A})}_{\alpha}$ on the state $\rho$. Note that this probability does not depend on Bob's measurement. Let $\mathbb{P}(y|\alpha,x,\beta;\rho,\mathcal{E})$ be the probability of Bob measuring outcome $\lambda^{(\Alg{B})}_{\beta y}$ given that Bob measures observable $L^{(\Alg{B})}_{\beta}$ \emph{and} given that Alice first measured outcome $x$ for the observable $L^{(\Alg{A})}_{\alpha}$ from the initial state $\rho$ and then Alice transferred the resulting state to Bob via the quantum channel $\mathcal{E}$. The joint probability is then the product of these two probabilities, and so the joint expectation value for this pair of observables is given by 
\be
\mathbb{E}\left[L^{(\Alg{A})}_{\alpha},L^{(\Alg{B})}_{\beta}\right]=\sum_{x,y}\lambda^{(\Alg{A})}_{\alpha x}\lambda^{(\Alg{B})}_{\beta y}\mathbb{P}(y|\alpha,x,\beta;\rho,\mathcal{E})\mathbb{P}(x|\alpha,\rho).
\ee
The collection of these numbers, indexed by $\alpha$ and $\beta$, also forms an $M\times N$ matrix (a table) of joint expectation values (a special case was considered in Ref.~\cite{BMKG13}). 

The numbers that can appear in these tables are not arbitrary. Moreover, the set of possible tables for the spatial and temporal scenarios may differ in general. Given only such a table of values, a natural question to ask is whether that table could have been generated by either the spatial or temporal scenario~\cite{RAVJSR15,SongParzygnat2025,MengEtAl2024,ZPTGVF18} (or any other causal structure not mentioned here~\cite{liu2023quantum,Hayashi2026,PearlCausality2009}). If both scenarios are possible for a given table, we would say that there is \emph{causal confusion}, since one would not be able to determine which of the two causal structures led to that specific table. Conversely, if one causal structure is compatible with a given table, while the other causal structure is not, then one could rule out causal structures from the data. These types of problems fall under the general subject of causal inference, causal discovery, and causal compatibility, and are heavily investigated in both classical and quantum settings~\cite{RAVJSR15,PearlCausality2009,FrKiSpWo2025,FrKi2024,BaLoOr19,ABHLS17,navascues2020inflation,wolfe2021qinflation,WolfeSpekkensFritz2019,CostaShrapnel16,Pienaar2020,JacobsKissingerZanasi2019,CaAnWoEv2025,WoodSpekkens2015,Evans2016,MengEtAl2024,OuWaSe2026,SmWoSp2026,KublerBraun2018}. 

Ref.~\cite{SongParzygnat2025} answered the question: what properties of a bipartite quantum state $\tau$ guarantee that the spatial joint expectation values of light-touch observables can also be obtained as the temporal joint expectation values of those same observables for some input state $\rho$ and quantum channel $\mathcal{E}$? Light-touch observables are observables that square to a nonnegative multiple of the identity (with all Pauli observables as examples)~\cite{liu2023quantum,FuPa24a}. One result of Ref.~\cite{SongParzygnat2025} says that the joint expectation values of all \emph{separable} states can also be obtained temporally via some process $(\mathcal{E},\rho)$, where $\mathcal{E}$ is entanglement breaking~\cite{horodecki2003ebc}. In fact, such \emph{causal confusion}, where it becomes uncertain which spatiotemporal scenario gave rise to a set of joint expectation values, is still present for positive-partial-transpose (PPT) states, except that the associated channel $\mathcal{E}$ is not necessarily entanglement breaking. However, there exist negative-partial-transpose states that also admit causal confusion, and a general necessary and sufficient condition for testing such causal confusion was provided in Ref.~\cite{SongParzygnat2025} (see also Ref.~\cite{LorenzBrunnerBenedetti2026} for more recent related developments). 

In this paper, we investigate the converse direction, which splits into two separate questions. First, what properties of the quantum channel $\mathcal{E}$ guarantee that there exists a state $\rho$ such that the temporal joint expectation values associated with $(\mathcal{E},\rho)$ can also be obtained as the spatial joint expectation values for some bipartite quantum state $\tau$? Second, for such a channel $\mathcal{E}$, what is the set of all input states $\rho$ such that the process $(\mathcal{E},\rho)$ leads to causal confusion? To answer this, we first explore qubit systems, Pauli observables, and the von~Neumann projection postulate after a measurement~\cite{vN18,Lu06} (as opposed to the general theory of quantum instruments~\cite{DaLe70}). More specifically, we investigate several quantum channels $\mathcal{E}$ that depend on at most one parameter (unitary, discard-and-prepare, depolarizing, amplitude-damping, dephasing, and mixed Pauli channels~\cite{NiCh11}) and we identify the precise subset of states $\rho$ such that $(\mathcal{E},\rho)$ leads to causal confusion. This provides a visual representation for causal confusion in terms of subsets of the Bloch ball. 

\begin{table*}
\begin{tabular}{l | l | l}
Earlier results & This paper & Result \\
\hline 
    \begin{tabular}{l}
    Bipartite state $\implies$ causal confusion for some\\ pseudoprocess $(\mathcal{E},\rho)$ with $\mathcal{E}$ PTP~\cite[Remark III.10]{SongParzygnat2025} 
    \end{tabular}
& \begin{tabular}{l}
    Bipartite state $\implies$ causal confusion for some\\ pseudoprocess $(\mathcal{E},\rho)$ with $\mathcal{E}$ CCPTP
    \end{tabular}
&Theorem~\ref{thm:Eisalwayscopositive} \\
\hline 
    \begin{tabular}{l} 
    Bipartite PPT state $\implies$ causal confusion for some\\ process $(\mathcal{E},\rho)$ with $\mathcal{E}$ CPTP~\cite[Corollary III.12]{SongParzygnat2025} 
    \end{tabular}
    &     
    \begin{tabular}{l} 
    Process $(\mathcal{E},\rho)$ leads to causal confusion\\
    $\implies$ $\mathcal{E}$ is CCPTP and hence a PPT channel
    \end{tabular}
    & Theorem~\ref{thm:PPTEQSOTstate} \\
\hline 
&     $\mathcal{E}$ admits causal confusion $\iff$ $\mathcal{E}$ is PPT & Corollary~\ref{cor:causalconfusioniffPPT} \\
\hline
    \begin{tabular}{l} 
    Bipartite separable state $\implies$ causal confusion for \\
    some process $(\mathcal{E},\rho)$ with $\mathcal{E}$ EB~\cite[Theorem III.7]{SongParzygnat2025}; \\
    Bipartite separable state with maximally-mixed\\
    marginals is equivalent to $(\mathcal{E},\frac{\mathds{1}_{m}}{m})$ with $\mathcal{E}$ EB~\cite{RAVJSR15} 
    \end{tabular}
& 
    \begin{tabular}{l} $\mathcal{E}$ is EB $\iff$ $\mathcal{E}\star\frac{\mathds{1}}{m}$ is separable \\
    but $\mathcal{E}\star\rho$ need not be separable for general $\rho$ 
    \end{tabular}
& 
    \begin{tabular}{l} 
    Proposition~\ref{prop:EBandQSOT} \\ 
    Example~\ref{ex:EBdniState} 
    \end{tabular}
    \\
\hline 
\begin{tabular}{l} Partial results on causal confusion for unital qubit \\ channels and maximally-mixed marginals~\cite{RAVJSR15,KublerBraun2018} \end{tabular}
& \begin{tabular}{l} Classification of states leading to causal confusion\\ for unitary, amplitude-damping, dephasing, \\ depolarizing, and mixed two-Pauli channels \end{tabular}  
& Section~\ref{sec:examples} \\
\hline 
& \begin{tabular}{l} Causal confusion exists for a generic qubit \\ quantum channel $\mathcal{E}$ $\iff$ $\mathcal{E}$ is EB\end{tabular} 
& Theorem~\ref{thm:generalqubitchannelcausalconfusion} \\
\hline 
 \begin{tabular}{l} Characterization for when a qubit process has\\
  a Bayesian inverse~\cite[Theorem 2]{TiFuWu2026} \end{tabular}
& 
    \begin{tabular}{l} $\mathcal{E}$ is a PPT channel \\
    $\implies$ $(\mathcal{E},\frac{\mathds{1}_{m}}{m})$ has a PPT Bayesian inverse \end{tabular} 
& Theorem~\ref{thm:PPTBayesianinverse} 
\end{tabular}
\caption{A summary of some of the key results in this paper. The quantity $\mathcal{E}\star\rho$ is the quantum state over time associated with a process $(\mathcal{E},\rho)$ and is defined in Section~\ref{sec:twocausalscenarios} (pseudoprocess means that $\mathcal{E}$ is not necessarily a quantum channel). The acronyms PTP, CPTP, CCPTP, EB, and PPT stand for positive trace preserving, completely positive trace preserving, completely copositive trace preserving, entanglement breaking, and positive-partial transpose, respectively. Bayesian inverses~\cite{FuPa22a}, forms of operational time-reversal symmetry, are defined in Section~\ref{sec:Bayesianinverses}.}
\label{table:summary}
\end{table*}

Motivated by the examples, we identify general structural results for arbitrary finite-dimensional systems, light-touch observables, and the L\"uders instrument. For example, we prove that a channel $\mathcal{E}$ admits causal confusion for some state if and only if $\mathcal{E}$ is a positive-partial-transpose (PPT) channel~\cite{ChMuWo2019}. As a consequence, a qubit channel $\mathcal{E}$ admits causal confusion for some state if and only if $\mathcal{E}$ is entanglement breaking. Surprisingly, we show that a channel being entanglement breaking is not sufficient for causal confusion with arbitrary input states, despite the fact that the converse is true by Ref.~\cite{SongParzygnat2025} and also despite the fact that causal confusion does hold if the initial state is maximally mixed and the channel is entanglement breaking. We provide a concrete counter-example of an entanglement-breaking quantum channel and a family of input qubit states (that are necessarily not maximally mixed) for which the temporal correlations in time \emph{cannot} be realized as spatial correlations of a bipartite qubit state. Our results therefore extend those of Refs.~\cite{RAVJSR15,KublerBraun2018}, since we avoid the restriction to maximally-mixed input states and unital channels, and our results also illustrate the importance of the initial state in determining causal confusion. Table~\ref{table:summary} summarizes our main contributions together with connections to earlier works. 

The motivation for investigating these topics stems from the fact that understanding the differences between classical and quantum causal inference highlights a key distinction between classical and quantum versions of Bayes' theorem~\cite{FuPa22a} (see also Refs.~\cite{Le06,Le07,LeSp13,Ts22,Ts22b,SASDS23,GPRR21,PaRuBayes,PaBu22,Fuchs01,Fuchs2003,FuchsSchack2013}). Indeed, if the states were replaced with classical probability distributions and the channel with a classical noisy channel (a collection of conditional probabilities), then classical Bayes' rule would imply an equivalence between the spatial and temporal scenarios in the following sense. Given a probability distribution $\rho$ and classical noisy channel $\mathcal{E}$, there exists a bipartite probability distribution $\tau$ that equals the joint probability distribution associated with $\mathcal{E}$ and $\rho$. Conversely, given a bipartite probability distribution $\tau$, there exists a probability distribution $\rho$ and a channel $\mathcal{E}$ for which the joint probability distribution associated with $\mathcal{E}$ and $\rho$ equals $\tau$. Thus, causal confusion between the spatial and temporal settings is guaranteed in the classical case. The fact that this is no longer the case for quantum systems illustrates an essential difference between spatial and temporal correlations in quantum systems that demands further investigation. 

Our paper is structured as follows. Section~\ref{sec:twocausalscenarios} sets up the precise definitions for the spatial and temporal scenarios that we study in this paper. Section~\ref{sec:FPtheorem} rephrases a result from Ref.~\cite{FuPa24a} that provides an algebraic test for causal confusion that we will use throughout the rest of the paper. Section~\ref{sec:examples} completely identifies which states lead to causal confusion for certain channels in finite-dimensional systems, mainly focusing on qubit systems. The two non-trivial examples that we work out in full include the depolarizing channel and a non-uniform mixed Pauli channel. Section~\ref{sec:generalresults} proves a variety of structural results for finite-dimensional systems motivated by the preceding examples, the main result being that a channel admits causal confusion if and only if it is a PPT channel. Section~\ref{sec:generalqubit} revisits the case of qubits and provides necessary and sufficient conditions for a general qubit channel to admit causal confusion, showing that such channels must be entanglement breaking. Appendix~\ref{sec:background} contains basic definitions of quantum information theory and establishes our conventions. Appendix~\ref{sec:exampletables} contains examples of tables of joint expectation values to help further clarify the approach we use throughout. Appendix~\ref{sec:proofs} contains proofs of standard results scattered in the literature. Appendix~\ref{sec:Ruskai} contains an argument justifying the characterization for causal confusion in qubit systems.

\section{Joint expectation values based on two causal scenarios}
\label{sec:twocausalscenarios}

\subsection{Notation}

Physical systems will be represented by algebras of operators, where \emph{operators} are complex linear combinations of observables and where \emph{algebra} refers to the fact that one can take linear combinations and products of operators in a way that satisfies associativity, distributivity, etc.\ (see Ref.~\cite{Fa01} for more details). Such algebras will be denoted by $\Alg{A},\Alg{B},$ 
etc. If the systems are purely quantum-mechanical, the Hilbert spaces on which these algebras act on will be exclusively finite-dimensional.  
Such Hilbert spaces are most often taken to be of the form $\C^{m}$, complex-valued vectors of length $m$, for some $m\in\N$, where $\N$ denotes the set of natural numbers. In this case, the algebra of all operators coincides with $\matr_{m}$, the matrix algebra of all complex-valued $m\times m$ matrices. 
The calligraphic letters $\mathcal{E}$ and $\mathcal{F}$ will be used for linear maps on algebras, which are often channels. 
We assume familiarity with linear algebra~\cite{Strang2022,Fa01}, and we recall concepts from quantum theory while establishing notation~\cite{HallQuantum13,Watrous2025,BCRS19,NiCh11} in Appendix~\ref{sec:background}.

\subsection{The spatial and temporal scenarios}
\label{sec:STdefns}

Throughout, we assume that Alice and Bob perform local measurements under one of the following two protocols, along with their associated definitions of joint expectation values. Both scenarios involve algebras $\Alg{A}$ and $\Alg{B}$ but different assumptions about what data are given to compute joint expectation values. 
A \define{bipartite state} on system $\Alg{A}\otimes\Alg{B}$ is a density matrix $\tau$ in $\Alg{A}\otimes\Alg{B}$.
A \define{process} from $\Alg{A}$ to $\Alg{B}$ is a pair $(\mathcal{E},\rho)$, where $\rho$ is a density matrix in $\Alg{A}$ and $\mathcal{E}:\Alg{A}\to\Alg{B}$ is a quantum channel.

\begin{enumerate}
\item[(S)] (Spatial scenario) Alice and Bob perform measurements of observables of the form $A\otimes B$ on a bipartite quantum state $\tau\in\Alg{A}\otimes\Alg{B}$. The \define{joint expectation value} of the observable $A\otimes B$ in the state $\tau$ is given by 
\be
\mathbb{E}_{\tau}[A\otimes B]:=\Tr\big[\tau(A\otimes B)\big].
\ee
If $A=\sum_{j}\lambda_jP_{j}$ and $B=\sum_{k}\mu_{k}Q_{k}$ are the spectral decompositions of $A$ and $B$, respectively, then 
\be
\mathbb{E}_{\tau}[A\otimes B]=\sum_{j,k}\mathbb{P}(j,k)\lambda_{j}\mu_{k},
\ee
where the joint probability $\mathbb{P}(j,k)$ is given by 
\be
\mathbb{P}(j,k):=\Tr\big[\tau(P_{j}\otimes Q_{k})\big].
\ee
Thus, the joint expectation value as defined in the spatial scenario is just the expectation value specialized to observables of the form $A\otimes B$. 

\item[(T)] (Temporal scenario) Alice first performs a measurement of an observable $A=\sum_{j}\lambda_{j}P_{j}$ on a quantum state $\rho\in\Alg{A}$, which upon measuring outcome $\lambda_{j}$ gets updated to the quantum state 
\be
\rho\mapsto
\frac{P_{j}\rho P_{j}}{\Tr[\rho P_{j}]} 
\ee
via \define{L\"uders' update rule}. Afterwards, the updated quantum state gets transmitted to Bob under a quantum channel $\mathcal{E}$. Finally, Bob measures observable $B=\sum_{k}\mu_{k}Q_{k}$. The \define{joint expectation value} of measuring $A$ followed by $B$ associated with the process $(\mathcal{E},\rho)$ is given by 
\be
\label{eqn:temporalJEV}
\mathbb{E}_{(\mathcal{E},\rho)}[A,B]:=\sum_{j,k}\mathbb{P}(k|j)\mathbb{P}(j)\lambda_{j}\mu_{k},
\ee
where 
$
\mathbb{P}(j):=\Tr[\rho P_{j}]
$
is the probability of Alice measuring outcome $\lambda_j$ for observable $A$ and 
\be
\mathbb{P}(k|j):=\Tr\left[\mathcal{E}\left(\frac{P_{j}\rho P_{j}}{\Tr[\rho P_{j}]}\right)Q_{k}\right]
\ee
is the conditional probability of Bob measuring outcome $\mu_{k}$ for observable $B$ given that Alice measured outcome $\lambda_{j}$ for observable $A$. The joint expectation value can therefore also be written as 
\be
\label{eqn:Texpval}
\mathbb{E}_{(\mathcal{E},\rho)}[A,B]=\sum_{j,k}\lambda_{j}\mu_{k}\Tr\big[\mathcal{E}(P_j \rho P_j) Q_k \big].
\ee
\end{enumerate}

In the temporal scenario, we assume a particular state-update rule for the quantum state after Alice measures. This is the \emph{L\"uders' update rule}~\cite{BuLa2009,khrennikov2009luders,Lu06,Jacobs2014}, which is a generalization of von~Neumann's projection postulate to allow for projection operators with rank larger than 1. For the qubit examples studied in this paper, we work mostly with qubits and Pauli operators so that von~Neumann's projection postulate suffices. Although this is not the most general setting possible, it leads to a fruitful middle ground where we can go beyond maximally-mixed input states and unital quantum channels, while still remaining computationally tractable. 

In both spatial and temporal scenarios, Alice and Bob record the outcomes of their experiment (the eigenvalues of the observables they measure). In the expressions for the joint expectation values, the eigenvalues are always multiplied together as in $\lambda_{j}\mu_{k}$ and then they are multiplied by their joint probability of occurrence, which can be obtained in practice by repeating the experiment multiple times (cf.\ Appendix~\ref{sec:background}). The sum is then taken in both cases to arrive at their expectation values. Importantly, we note that we are not considering other spatiotemporal scenarios in this work~\cite{liu2023quantum,Hayashi2026}. One example of such as a scenario includes a third party, Charlie, who produces a bipartite state that is shared between Alice and Bob. Alice measures one part of that state, the full state undergoes some evolution, and then Bob measures the other part of that state. The reason for us focusing on the spatial (S) and temporal (T) scenarios is to isolate an important distinction between classical and quantum systems already in this simple situation. 

\subsection{Expectation value tables for qubits}

Since the Pauli matrices $(\sigma_0,\sigma_1,\sigma_2,\sigma_3)$ form an orthogonal basis for all $2\times 2$ observables, their $d$-fold tensor products also form an orthogonal basis for observables of $d$ qubits, which corresponds to the tensor product $\matr_{2}\otimes\cdots\otimes\matr_{2}\cong\matr_{2^{d}}$. Namely, 
\be
\Tr[(\sigma_{j_1}\otimes\cdots\otimes\sigma_{j_d})(\sigma_{k_1}\otimes\cdots\otimes\sigma_{k_d})]=2^{d}\delta_{j_1 k_1}\cdots\delta_{j_d k_d}
\ee
for all $j_1,k_1,\dots,j_d,k_d\in\{0,1,2,3\}$. We will primarily be concerned with the $d=2$ case, which will cover the spatial and temporal scenarios introduced earlier in the case that Alice and Bob perform single-qubit measurements. For the rest of this section, let $\Alg{A}=\Alg{B}=\matr_{2}$.

By the above orthogonality of tensor products of Pauli matrices, every bipartite density matrix $\tau\in\Alg{A}\otimes\Alg{B}$ can be expressed as the linear combination 
\be
\label{eqn:tauasPDM}
\tau=\frac{1}{4}\sum_{j,k=0}^{3}\Tr\big[\tau(\sigma_j\otimes\sigma_k)\big] (\sigma_j\otimes\sigma_k).
\ee
In other words, the bipartite state $\tau$ is uniquely characterized by the expectation values of measurements of local Pauli observables, i.e., the $16$ observables $\{\sigma_{j}\otimes\sigma_{k}\}$ with $j,k\in\{0,1,2,3\}$ (technically, only 15 are needed because the expected value for $\sigma_0\otimes\sigma_0$ is guaranteed to be $1$ by the assumption that $\tau$ is a density matrix). These 16 expectation values can be written as the table 
\begin{center}
\noindent
\begin{tabular}{c|cccc}
 & $\sigma_0$ & $\sigma_1$ & $\sigma_2$ & $\sigma_3$ \\
\hline 
$\sigma_0$ & \,$\mathbb{E}_{\tau}[\sigma_0\otimes\sigma_0]$\, & \,$\mathbb{E}_{\tau}[\sigma_0\otimes\sigma_1]$\, & \,$\mathbb{E}_{\tau}[\sigma_0\otimes\sigma_2]$\, & \,$\mathbb{E}_{\tau}[\sigma_0\otimes\sigma_3]$ \\ 
$\sigma_1$ & $\mathbb{E}_{\tau}[\sigma_1\otimes\sigma_0]$ & $\mathbb{E}_{\tau}[\sigma_1\otimes\sigma_1]$ & $\mathbb{E}_{\tau}[\sigma_1\otimes\sigma_2]$ & $\mathbb{E}_{\tau}[\sigma_1\otimes\sigma_3]$ \\
$\sigma_2$ & $\mathbb{E}_{\tau}[\sigma_2\otimes\sigma_0]$ & $\mathbb{E}_{\tau}[\sigma_2\otimes\sigma_1]$ & $\mathbb{E}_{\tau}[\sigma_2\otimes\sigma_2]$ & $\mathbb{E}_{\tau}[\sigma_2\otimes\sigma_3]$ \\
$\sigma_3$ & $\mathbb{E}_{\tau}[\sigma_3\otimes\sigma_0]$ & $\mathbb{E}_{\tau}[\sigma_3\otimes\sigma_1]$ & $\mathbb{E}_{\tau}[\sigma_3\otimes\sigma_2]$ & $\mathbb{E}_{\tau}[\sigma_3\otimes\sigma_3]$
\end{tabular}
\end{center}
whose entries are based on Alice's and Bob's possible measurement outcomes. Here, Alice's measured observables are shown on the leftmost column and Bob's measured observables are shown on the topmost row. Several examples involving factorized states, separable states, Werner states, etc.\ are provided in Appendix~\ref{sec:exampletables}. 

In the case that Alice and Bob measure in the temporal scenario, so that the datum associated with the computation of the expectation values is a process $(\mathcal{E},\rho)$, a similar table of expectation values can be written. However, the $jk$ entry of such a table is instead given by $\mathbb{E}_{(\mathcal{E},\rho)}[\sigma_j,\sigma_k]$ as defined in~\eqref{eqn:Texpval}. Since we investigate many examples of processes in Appendix~\ref{sec:exampletables} and later in this paper, we only illustrate one motivating example for now. 

\begin{example}
Given a table of joint expectation values such as 
\begin{center}
\noindent
\begin{tabular}{c|cccc}
 & $\sigma_0$ & $\sigma_1$ & $\sigma_2$ & $\sigma_3$ \\
\hline 
$\sigma_0$ & \,\;\;$1$\,\;\; & \,\;\;$0$\,\;\; & \,\;\;$0$\,\;\; & \,\;\;$0$\;\; \\ 
$\sigma_1$ & $0$ & $\frac{1}{4}$ & $0$ & $0$ \\
$\sigma_2$ & $0$ & $0$ & $\frac{1}{4}$ & $0$ \\
$\sigma_3$ & $0$ & $0$ & $0$ & $\frac{1}{4}$
\end{tabular}
\end{center}
which of the spatial and/or temporal scenarios is this table compatible with? The answer is both, so that it exhibits causal confusion. On the one hand, this table can be obtained from the spatial scenario by using the bipartite Werner state~\eqref{eqn:wernerstates} with $s=\frac{15}{16}$. On the other hand, this table can also be obtained from the temporal scenario of Example~\ref{ex:temporaldepolarizing} by using the depolarizing channel with $\lambda=\frac{3}{4}$ and the input state being the maximally-mixed state.
\end{example}

One might wonder if all temporal joint expectation value tables can equivalently be obtained as spatial joint expectation value tables and vice-versa. This is not the case, and we will see why later in this paper. For example, the spatial table in Example~\ref{ex:Wernerentangled} cannot be obtained as a temporal one, and the temporal table in Example~\ref{ex:temporaldepolarizing} cannot be obtained as a spatial one. Appendix~\ref{sec:exampletables} contains more examples for motivation.

\subsection{Pseudo-density matrices}

Given a table of joint expectation values $\mathbb{E}[\sigma_{j},\sigma_{k}]$, but for which the causal structure leading to these values is not known, one can define the associated \define{pseudodensity matrix}~\cite{FJV15,FuPa24a}
\be
\label{eqn:PDMfromExpVals}
R=\frac{1}{4}\sum_{j,k=0}^{3}\mathbb{E}[\sigma_{j},\sigma_{k}] (\sigma_{j}\otimes\sigma_{k}).
\ee
The pseudodensity matrix then satisfies the condition that the joint expectation values $\mathbb{E}[\sigma_{j},\sigma_{k}]$ can be obtained from it by the formula 
\be
\mathbb{E}[\sigma_{j},\sigma_{k}]=\Tr\big[R(\sigma_{j}\otimes\sigma_{k})\big]
\ee
for all $j,k$. For example, suppose that the expectation values are obtained from a bipartite state $\tau$ as in the formula $\mathbb{E}[\sigma_{j},\sigma_{k}]=\Tr\big[\tau(\sigma_j\otimes\sigma_k)\big]$. Then the pseudodensity matrix $R$ associated with these values is exactly equal to $\tau$, which we already saw in~\eqref{eqn:tauasPDM}. In the case that the table of joint expectation values is obtained from an initial qubit state $\rho$ and a qubit quantum channel $\mathcal{E}$, the pseudodensity matrix need not be positive, though it is Hermitian and has trace $1$. Moreover, its partial traces are density matrices. The pseudodensity matrix for qubit systems was introduced in Ref.~\cite{FJV15} precisely in order to distinguish spatial versus temporal correlations in quantum systems of qubits~\cite{BMKG13}, and to therefore identify causality from correlations. Ref.~\cite{HHPBS17} provided a simple formula for $R$ in terms of a Jordan product involving $\rho$ and $\mathcal{E}$ (with earlier closely-related observations established in Refs.~\cite{Fritz10,BMKG13}), and they formalized the pseudodensity matrix as an example of a quantum state over time, which we turn to next. 

\subsection{Quantum states over time}

The \define{Jordan product quantum state over time} (QSOT for short) 
associated with the process $(\mathcal{E},\rho)$ from $\Alg{A}=\matr_{m}$ to $\Alg{B}=\matr_{n}$ is the operator in $\Alg{A}\otimes\Alg{B}$ given by 
\be
\label{eqn:JPQSOT}
\mathcal{E}\star\rho=\frac{1}{2}\big\{\rho\otimes\mathds{1}_{n},\Jamio[\mathcal{E}]\big\}, 
\ee
where $\{C,D\}=CD+DC$ denotes the anti-commutator of $C$ and $D$, and where 
\be
\Jamio[\mathcal{E}]=\sum_{i,j}|i\>\<j|\otimes\mathcal{E}\big(|j\>\<i|\big)
\ee
is the \define{Jamio{\l}kowski matrix} of $\mathcal{E}$~\cite{Ja72,LeSp13,HHPBS17,FuPa22}. We will describe the importance of the QSOT shortly. 

Notice that the Jamio{\l}kowski matrix is related to the Choi matrix $\Choi[\mathcal{E}]$~\eqref{eqn:Choimatrix} by a partial transpose operation. In more detail, let $\mathcal{T}:\matr_{m}\to\matr_{m}$ denote the \define{transpose} operation, which sends an $m\times m$ matrix $A$ to its transpose $\mathcal{T}(A):=A^{T}$. The \define{partial transpose} is the tensor product of the transpose with the identity channel. Thus, the partial transpose of the Jamio{\l}kowski matrix is 
\begin{align}
(\mathcal{T}\otimes\id_{\Alg{B}})\Big(\Jamio[\mathcal{E}]\Big)
&=\sum_{i,j}\mathcal{T}\big(|i\>\<j|\big)\otimes\mathcal{E}\big(|j\>\<i|\big) \nonumber \\
&=\sum_{i,j}|j\>\<i|\otimes\mathcal{E}\big(|j\>\<i|\big) \nonumber \\
&=\Choi[\mathcal{E}].
\label{eqn:PTofJamiol}
\end{align}
And since the transpose operation is the inverse of itself, we also have that the partial transpose of the Choi matrix is the Jamio{\l}kowski matrix, namely, 
\be
\label{eqn:TonChoi}
(\mathcal{T}\otimes\id_{\Alg{B}})\Big(\Choi[\mathcal{E}]\Big)=\Jamio[\mathcal{E}].
\ee

There is a close relationship between the Jordan product state over time and pseudodensity matrices associated with joint expectation values obtained from a temporal scenario~\cite{HHPBS17,liu2023quantum,FuPa24a}. This connection holds beyond qubit systems, and the proof is simpler in the more general case, so we review the key definition and theorem here. A \define{light-touch observable} in $\matr_{m}$ is an observable $A$ such that $A^2=\lambda\mathds{1}_{m}$ for some $\lambda\in[0,\infty)$. Pauli matrices and their tensor products are all examples ($\lambda=1$ in this case). The main result of Ref.~\cite{FuPa24a} connects the QSOT to temporal joint expectation values. 

\bt
Let $(\mathcal{E},\rho)$ be a process. Then 
\be
\mathbb{E}_{(\mathcal{E},\rho)}[A,B]=\Tr\big[(\mathcal{E}\star\rho)(A\otimes B)\big]
\ee
for all light-touch observables $A$ and $B$, where $\mathbb{E}_{(\mathcal{E},\rho)}[A,B]$ is the temporal joint expectation value defined in~\eqref{eqn:temporalJEV}. 
\et

In particular, $\mathcal{E}\star\rho$ is the pseudodensity matrix associated with the temporal joint expectation values for a qubit density matrix $\rho$ undergoing evolution dictated by $\mathcal{E}$. This result for \emph{Pauli observables} and a \emph{single qubit} was first proved in Ref.~\cite{HHPBS17}. The mathematical consistency of the Jordan product state over time was proved for arbitrary quantum systems in Ref.~\cite{FuPa22}, while three different characterization theorems for it appeared in Refs.~\cite{LiNg23,PFBC23,FuPa24a}. Ref.~\cite{liu2023quantum} proved that the Jordan product state over time provides a formula for the temporal pseudodensity matrix for \emph{multiqubit} systems. Ref.~\cite{FuPa24a} extended this result to arbitrary finite-dimensional quantum systems by introducing light-touch observables mimicking some properties of Pauli observables, following terminology of Ref.~\cite{liu2023quantum}. Besides its usage in distinguishing spatial versus temporal correlations, which we turn to next, the QSOT has also been used in optimal transport~\cite{HoCaWi2026,Hoogsteder2026SOT}, time-symmetric temporal correlations~\cite{PaFu24TSC}, quasiprobabilistic approaches to temporal correlations~\cite{FuMaWu2026}, information recovery~\cite{BHKK2024}, interferometry~\cite{LiKw25}, the hermitian part of the operator representing two-point correlation functions~\cite{BDOV13,BDOV14}, quantum metrology~\cite{HanKwonLie2026}, and quantum chaos~\cite{Das2026JHEP}, to name a few. 

\subsection{Distinguishing causation from correlation}
\label{sec:FPtheorem}

Using pseudodensity matrices and the QSOT, it now becomes possible to provide a simple algebraic method to distinguish spatial (S) versus temporal (T) joint expectation values~\cite{FuPa24a,liu2023quantum,SongParzygnat2025,SNREG23,LCD24,JSK23,RAVJSR15}. 

\bn
Let $(\mathcal{E},\rho)$ be a process from $\Alg{A}$ to $\Alg{B}$. Then a table of temporal joint expectation values $\mathbb{E}_{(\mathcal{E},\rho)}[A,B]$ for a basis of light-touch observables $A\in\Alg{A}$ and $B\in\Alg{B}$ is equal to a table of spatial joint expectation values $\mathbb{E}_{\tau}[A,B]$ for some bipartite state $\tau\in\Alg{A}\otimes\Alg{B}$ if and only if $\mathcal{E}\star\rho$ is positive, in which case $\tau=\mathcal{E}\star\rho$. 
\en

Thus, if $\mathcal{E}\star\rho$ is a state, we say that $(\mathcal{E},\rho)$ exhibits \define{causal confusion}. We also say that $\mathcal{E}$ \define{admits causal confusion} iff there exists a state $\rho$ such that $(\mathcal{E},\rho)$ exhibits causal confusion. In Section~\ref{sec:examples}, we will look at specific parameterized families of quantum channels $\mathcal{E}$. For each value of the parameter, we will identify the subset of states $\rho$ for which $(\mathcal{E},\rho)$ exhibits causal confusion. Since we mostly deal with qubit systems, this will correspond to a region in the Bloch ball, and so we will visualize these regions for such quantum channels.  
Later, in Section~\ref{sec:generalresults}, we will use the examples to make some general observations about the properties of quantum channels that admit causal confusion for some nonempty subset of states.

\section{Examples distinguishing correlation and causation}
\label{sec:examples}

To analyze causal confusion, we will look at several examples in this section. We first begin with the discard-and-prepare channel and unitary channels between arbitrary quantum systems. The rest of the examples focus on the case of qubits and will follow a similar outline. First, the channel $\mathcal{E}$ will be defined, oftentimes parameterized by a single parameter. A visualization of what this channel does to the Bloch sphere might be shown for certain values of the parameters. Second, the matrix $\mathcal{E}\star\rho$ will be shown with its dependence on the parameters from the channel $\mathcal{E}$ and the dependence on the vector $\mathbf{r}=(x,y,z)$ used to parameterize the density matrix following~\eqref{eqn:qubitrho}. Third, given a fixed value of the parameter for the channel $\mathcal{E}$, it will then be determined what values of $\mathbf{r}$ satisfy the condition that $\mathcal{E}\star\rho$ is positive semidefinite. The corresponding set of density matrices $\rho$ such that $\mathcal{E}\star\rho\ge0$ may then be drawn on the Bloch sphere to indicate which input states $\rho$ induce causal confusion. Section~\ref{sec:generalresults} will contain more general observations.

\subsection{Discard-and-prepare channels}

A \define{discard-and-prepare} channel $\mathcal{E}$ from $\Alg{A}$ to $\Alg{B}$ is one of the form
\be
\mathcal{E}(A)=\Tr[A]\nu
\ee
for all $A\in\Alg{A}$, for some density matrix $\nu$ in $\Alg{B}$. In this case, $\mathcal{E}\star\rho=\rho\otimes\nu$ for every density matrix $\rho$ in $\Alg{A}$. Thus, every process $(\mathcal{E},\rho)$, with $\mathcal{E}$ a discard-and-prepare channel, exhibits causal confusion.

\subsection{Unitary channels}

\bn
\label{prop:identitychannel}
The identity channel $\id_{\Alg{A}}:\Alg{A}\to\Alg{A}$ \emph{never} admits causal confusion, i.e., $\id_{\Alg{A}}\star\rho$ is \emph{not} positive semidefinite for \emph{any} density matrix $\rho\in\Alg{A}$ as long as $\Alg{A}=\matr_{m}$ with $m\ge2$. 
\en

This result follows from~\cite[Lemma A.5]{FuPa23}, and we provide some details in Appendix~\ref{sec:idcausalconf} to be self-contained and to simplify the proof for this particular case. In this proof and some later proofs, we utilize the \define{swap operator} for $\Alg{A}=\matr_{m}$, which is defined by 
\be
\label{eqn:SWAP}
\SWAP_{\Alg{A}}:=\sum_{i,j}|i\>\<j|\otimes|j\>\<i|, 
\ee
where $\{|i\>\}$ is the standard basis of $\C^{m}$. 
The swap operator $\SWAP_{\Alg{A}}$ is an element of $\Alg{A}\otimes\Alg{A}$ that satisfies many properties, the most relevant for us being 
\be
\Jamio[\mathcal{E}]=(\id\otimes\mathcal{E})(\SWAP_{\Alg{A}})
\ee
for all linear maps $\mathcal{E}:\Alg{A}\to\Alg{B}$. Other standard properties are provided at the beginning of Appendix~\ref{sec:idcausalconf}. 

Proposition~\ref{prop:identitychannel} extends to all unitary channels, as the following proposition shows. 

\bn
Let $\Alg{A}=\matr_{m}$ with $m\ge2$, and let $U\in\Alg{A}$ be a unitary matrix. Then the channel $\mathcal{E}:\Alg{A}\to\Alg{A}$ given by $\mathcal{E}(A)=U A U^{\dag}$ for all $A\in\Alg{A}$ \emph{never} admits causal confusion, i.e., $\mathcal{E}\star\rho$ is \emph{not} positive semidefinite for \emph{any} density matrix $\rho\in\Alg{A}$ and \emph{any} unitary $U\in\Alg{A}$. 
\en

\bprf
Since
\begin{align}
\mathcal{E}\star\rho
&=\frac{1}{2}\big\{\rho\otimes\mathds{1}_{\Alg{A}},(\mathds{1}_{\Alg{A}}\otimes U)\SWAP_{\Alg{A}}(\mathds{1}_{\Alg{A}}\otimes U^{\dag})\big\} \nonumber \\
&=(\mathds{1}_{\Alg{A}}\otimes U)(\id_{\Alg{A}}\star\rho)(\mathds{1}_{\Alg{A}}\otimes U)^{\dag},
\end{align}
we have that $\mathcal{E}\star\rho$ and $\id_{\Alg{A}}\star\rho$ are similar matrices and therefore have the same eigenvalues. Since Proposition~\ref{prop:identitychannel} showed that $\id_{\Alg{A}}\star\rho$ always has at least one negative eigenvalue, this proves that $\mathcal{E}\star\rho$ also has at least one negative eigenvalue and therefore cannot be positive semidefinite. 
\eprf

\subsection{Amplitude-damping channel}
\label{sec:ADC}

\begin{figure}
\includegraphics{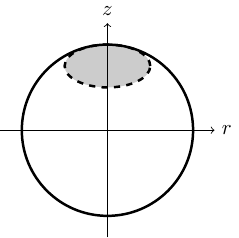}
\caption{A 2d cross section visualization of how the amplitude-damping channel~\eqref{eqn:ADC} acts on the Bloch sphere for $\gamma=\frac{3}{4}$. The image depicts the cross section of an ellipsoid (shaded) whose center is at $(0,0,\gamma)$. The horizontal axis is the radial distance $r=\sqrt{x^2+y^2}$ from the $z$ axis and the vertical axis is the $z$ direction, so that cylindrical coordinates are used. The negative radial direction is included for easier visualization since the image is rotationally symmetric about the $z$ axis. 
The dashed curve indicates the boundary of the image---it does not indicate an open set~\cite{Munkres2000Topology}.}
\label{fig:ADCBlochball}
\end{figure}

The qubit \define{amplitude-damping channel} is a qubit quantum channel $\mathcal{E}:\matr_{2}\to\matr_{2}$ of the form 
\be
\label{eqn:ADC}
\mathcal{E}\left( \left[ \setlength{\arraycolsep}{3pt} \begin{array}{cc} \\[-15pt] a &b \\ c & d  \end{array}\right]\right) = \begin{bmatrix} a+\gamma d & b\sqrt{1-\gamma} \\ c\sqrt{1-\gamma} & (1-\gamma)d \end{bmatrix},
\ee 
where $\gamma$ is called the \define{damping parameter} and lies in the range $0\leq\gamma\leq1$. When inputting a qubit density matrix of the form $\rho=\frac{1}{2}(\mathds{1}_{2}+\mathbf{r}\cdot\boldsymbol{\sigma})$~\eqref{eqn:qubitrho} into $\mathcal{E}$, we obtain 
$
\mathcal{E}\big(\frac{1}{2}(\mathds{1}_{2}+\mathbf{r}\cdot\boldsymbol{\sigma})\big)=\frac{1}{2}(\mathds{1}_{2}+\mathbf{s}\cdot\boldsymbol{\sigma}),
$
where 
\begin{align}
\mathbf{s}&:=(x',y',z') \nonumber \\
&=\Big(\Tr\big[\mathcal{E}(\rho)\sigma_1],\Tr\big[\mathcal{E}(\rho)\sigma_2],\Tr\big[\mathcal{E}(\rho)\sigma_3]\Big) \nonumber \\
&=\left(\sqrt{1-\gamma}\,x,\sqrt{1-\gamma} \, y, (1-\gamma)z+\gamma\right). 
\end{align}
This shows that the effect of the amplitude-damping channel on the Bloch ball is to reduce it in size and shift towards the north pole as $\gamma$ increases (cf.\ Figure~\ref{fig:ADCBlochball}). 
More specifically, since $\lVert\mathbf{r}\rVert\le 1$ holds because $\rho$ is an arbitrary state, we get
\be
\left(\frac{r'}{\sqrt{1-\gamma}}\right)^2+\left(\frac{z'-\gamma}{1-\gamma}\right)^2\le 1
\ee
where $r':=\sqrt{x'^2+y'^2}$ is the distance from the $z$ axis using cylindrical coordinates. This shows that the Bloch ball becomes an ellipsoid centered at $(0,0,\gamma)$.  
When $\gamma=1$, the image is just the north pole of the Bloch sphere, which corresponds to the state $|0\>\<0|$.

\bn
Let $\mathcal{E}$ be the amplitude-damping channel with damping parameter $\gamma$ as in~\eqref{eqn:ADC}. Then $\mathcal{E}$ admits causal confusion only when $\gamma=1$. When $\gamma=1$, $(\mathcal{E},\rho)$ exhibits causal confusion for all density matrices $\rho$. 
\en

\bprf
Writing $\gamma^{\perp}:=1-\gamma$ and $\rho=\frac{1}{2}(\mathds{1}_{2}+\mathbf{r}\cdot\boldsymbol{\sigma})$ with $\mathbf{r}=(x,y,z)$, we have 
\be
\mathcal{E}\star\rho
=\frac{1}{2}\begin{bmatrix}1+z & \frac{r_{+}}{2}\sqrt{\gamma^{\perp}} & \frac{r_{+}}{2}(1+\gamma ) & 0 \\ \frac{r_{-}}{2}\sqrt{\gamma^{\perp}} & 0&\sqrt{\gamma^{\perp}}&\frac{r_{+}}{2}\gamma^{\perp} \\ \frac{r_{-}}{2}(1+\gamma ) &\sqrt{\gamma^{\perp}} &(1-z)\gamma & \frac{r_{+}}{2}\sqrt{\gamma^{\perp}}\\ 0 & \frac{r_{-}}{2}\gamma^{\perp} & \frac{r_{-}}{2}\sqrt{\gamma^{\perp}} & (1-z)\gamma^{\perp} \end{bmatrix}
\ee
where we introduced the shorthand notation 
$r_{\pm}:=x\pm i y.$ 
It must be the case that, if a matrix is positive, all sub $2\times 2$ matrices must also be positive~\cite{Gantmacher1959,Strang2022}.
We begin by looking at the middle $2\times 2$ submatrix. By taking its determinant, we are left with the inequality $-(1-\gamma)= \gamma-1\ge 0$. Since our bounds for the parameter are $0\le\gamma\le1$, the inequality can only be true when $\gamma =1$. Thus, $\mathcal{E}\star\rho$ is not positive semidefinite for $\gamma\in[0,1)$. But when $\gamma=1$, the matrix $\mathcal{E}\star\rho$ is reduced to 
\be
\mathcal{E}\star\rho=\frac{1}{2}\begin{bmatrix} 1+z & 0 & x+iy & 0 \\ 0 & 0 & 0 & 0 \\ x-iy & 0 & 1-z & 0 \\ 0 & 0 & 0 & 0 \end{bmatrix}
=\rho^{T}\otimes|0\>\<0|, 
\ee
which is positive for all density matrices $\rho$. This is because $\rho^{T}$, where ${}^{T}$ denotes the transpose, is positive. 
\eprf

\subsection{Dephasing channel}
\label{sec:dephasing}

\begin{figure}
\includegraphics{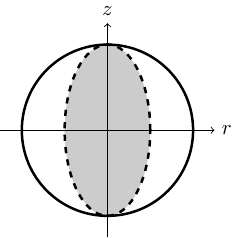}
\caption{A 2d cross section visualization of how the dephasing channel~\eqref{eqn:QDC} acts on the Bloch sphere for $p=\frac{3}{8}$. The image is the cross section of an ellipsoid (shaded) whose center is at $(0,0,0)$. As in Figure~\ref{fig:ADCBlochball}, cylindrical coordinates are used.}
\label{fig:QDCBlochball}
\end{figure}

The qubit \define{dephasing channel} is a qubit quantum channel $\mathcal{E}:\matr_{2}\to\matr_{2}$ of the form 
\be
\label{eqn:QDC}
\mathcal{E}\left( \left[ \setlength{\arraycolsep}{3pt} \begin{array}{cc} \\[-15pt] a &b \\ c & d  \end{array}\right]\right) = \begin{bmatrix} a & (1-2p)b \\ (1-2p)c & d\end{bmatrix}
\ee 
where $p$ is a parameter in the range $0 \le p \le 1$. 
It describes a channel in which a state loses a portion of its relative phase information depending on a parameter $p$, and it is an example of a mixed unitary channel since it can be expressed as 
\be
\mathcal{E}(A)=(1-p)A+p\sigma_{3}A\sigma_{3}
\ee
for all $A\in\matr_{2}$~\cite{NiCh11}. It is an example of a Pauli noise model, in the sense that with probability $1-p$, an initial state is left unchanged, while with probability $p$, it is acted on by the Pauli $Z$ operator $\sigma_{3}$. 
When inputting a qubit density matrix of the form $\rho=\frac{1}{2}(\mathds{1}_{2}+\mathbf{r}\cdot\boldsymbol{\sigma})$~\eqref{eqn:qubitrho} into $\mathcal{E}$, we obtain $\mathcal{E}(\rho)=\frac{1}{2}(\mathds{1}_{2}+\mathbf{s}\cdot\boldsymbol{\sigma})$, where 
\be
\mathbf{s}=(x',y',z')=\big((1-2p)x,(1-2p)y,z\big). 
\ee
Assuming $\lVert\mathbf{r}\rVert\le 1$, this yields the (filled-in) ellipsoid
\be
\left(\frac{r'}{\sqrt{1-2p}}\right)^2+(z')^2\le 1
\ee
described in cylindrical coordinates, which is shown in Figure~\ref{fig:QDCBlochball} (compare with Section~\ref{sec:ADC}).

\bn
Let $\mathcal{E}$ be the dephasing channel with parameter $p\in[0,1]$. Then $\mathcal{E}$ admits causal confusion if and only if $p=\frac{1}{2}$. When $p=\frac{1}{2}$, $(\mathcal{E},\rho)$ exhibits causal confusion only for density matrices $\rho$ of the form $\rho=\frac{1}{2}(\mathds{1}_{2}+z\sigma_3)$ with $z\in[-1,1]$. 
\en

\bprf
The matrix $\mathcal{E}\star\rho$ is given by 
\be
\mathcal{E}\star\rho=\frac{1}{2}\begin{bmatrix}1+z&\frac{r_{+}}{2}q&\frac{r_{+}}{2}&0\\\frac{r_{-}}{2}q&0&q&\frac{r_{+}}{2} \\ \frac{r_{-}}{2}&q&0&\frac{r_{+}}{2}q\\0&\frac{r_{-}}{2}&\frac{r_{-}}{2}q&1-z\end{bmatrix},
\ee
where we used the shorthand notation $r_{\pm}:=x\pm i y$ and $q=1-2p$. By looking at the middle $2\times 2$ submatrix, demanding nonnegativity of its determinant gives us the inequality $-(1-2p)^2 \ge 0$. The only way for this to be true is if $p = \frac{1}{2}$. This shows us that the only parameter for which $\mathcal{E}\star\rho$ could be positive is $p = \frac{1}{2}$. Now, when $p=\frac{1}{2}$, the matrix $\mathcal{E}\star\rho$ becomes
\be
\mathcal{E}\star\rho=
\frac{1}{2}\begin{bmatrix}1+z &0&\frac{x+iy}{2}&0\\ 0&0&0&\frac{x+iy}{2}\\ \frac{x-iy}{2}&0&0&0 \\ 0& \frac{x-iy}{2}&0&1-z\end{bmatrix}.
\ee
Removing the 2nd and 4th rows and columns gives a principal $2\times2$ determinant whose nonnegativity leads to the constraint that $x=y=0$. There are no further constraints for positivity of $\mathcal{E}\star\rho$ so that $\mathcal{E}\star\rho$ is positive when $p=\frac{1}{2}$ if and only if $\rho$ of the form $\rho=\frac{1}{2}(\mathds{1}_{2}+z\sigma_3)$ with $z\in[-1,1]$. 
\eprf

\subsection{Depolarizing channel}
\label{sec:depolarizing}

\begin{figure}
\includegraphics{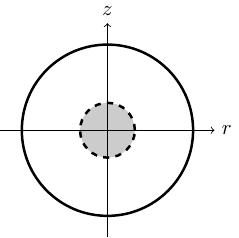}
\caption{A 2d cross section visualization of how the partially depolarizing channel~\eqref{eqn:PDC} acts on the Bloch ball for $\lambda=0.68$. The image is the cross section of a ball (shaded) whose center is at $0$. As in Figure~\ref{fig:ADCBlochball}, cylindrical coordinates are used.}
\label{fig:PDCBlochball}
\end{figure}

The preceding examples had relatively simple solutions to the problem for when $(\mathcal{E},\rho)$ exhibits causal confusion. In this and the next example, the situation is significantly more delicate. Let $\mathcal{E}$ be the qubit partially depolarizing channel from Example~\ref{ex:temporaldepolarizing}, with parameter $\lambda\in[0,\frac{4}{3}]$, so that 
\be
\label{eqn:PDC}
\mathcal{E}\left( \left[ \setlength{\arraycolsep}{3pt} \begin{array}{cc} \\[-15pt] a &b \\ c & d  \end{array}\right]\right) = \begin{bmatrix} \left(1-\frac{\lambda}{2}\right)a+\frac{\lambda}{2}d & (1-\lambda)b \\ (1-\lambda)c & \frac{\lambda}{2} a+\left(1-\frac{\lambda}{2}\right) d\end{bmatrix}
\ee
on a general input matrix. 
When plugging in a qubit density matrix of the form $\rho=\frac{1}{2}(\mathds{1}_{2}+\mathbf{r}\cdot\boldsymbol{\sigma})$~\eqref{eqn:qubitrho} into $\mathcal{E}$, we obtain $\mathcal{E}\left(\frac{1}{2}(\mathds{1}_{2}+\mathbf{r}\cdot\boldsymbol{\sigma})\right)=\frac{1}{2}(\mathds{1}_{2}+\mathbf{s}\cdot\boldsymbol{\sigma})$, where
\be
\mathbf{s}:=(x',y',z')=\big((1-\lambda)x,(1-\lambda)y,(1-\lambda)z\big).
\ee
This shows that the effect of the partially depolarizing channel on the Bloch ball is to reduce it to a ball of radius $|1-\lambda|$ still centered at the origin (cf.\ Figure~\ref{fig:PDCBlochball}). 

\bn
\label{prop:DepChanEstarRho}
If $\mathcal{E}$ is the qubit depolarizing channel with parameter $\lambda\in[0,\frac{4}{3}]$ as in~\eqref{eqn:PDC}, then a necessary condition for $\mathcal{E}$ to admit causal confusion is that $\lambda\in[\frac{2}{3},\frac{4}{3}]$. Moreover, when $\lambda$ is in this range, let $S_{\lambda}$ denote the subset of states $\rho$ for which $\mathcal{E}\star\rho$ is also a state, i.e., $(\mathcal{E},\rho)$ exhibits causal confusion. Then $S_{\lambda}$ corresponds to the closed ball $B_{\lambda}\subseteq\R^{3}$ of radius 
\be
\label{eqn:Rlambda}
R(\lambda) 
:=\frac{\sqrt{(2-\lambda)(3\lambda-2)}}{\lambda}
\ee
centered at $0$ and inside the Bloch ball. 
In other words, $\rho=\frac{1}{2}(\mathds{1}_{2}+\mathbf{r}\cdot\boldsymbol{\sigma})$ satisfies $\mathcal{E}\star\rho\ge0$ if and only if $\lVert \mathbf{r}\rVert\le R(\lambda)$. 
\en 

\bprf
The state over time $\mathcal{E}\star\rho$ is given by 
\be
\mathcal{E}\star\rho = 
\begin{bmatrix}
a(1-\frac{\lambda}{2}) & \frac{b}{2}\lambda^{\perp} & \frac{b}{2} & 0 \\
\frac{\overline{b}}{2}\lambda^{\perp} & a\frac{\lambda}{2} & \frac{1}{2}\lambda^{\perp} & \frac{b}{2} \\
\frac{\overline{b}}{2} & \frac{1}{2}\lambda^{\perp} & c\frac{\lambda}{2} & \frac{b}{2}\lambda^{\perp} \\
0 & \frac{\overline{b}}{2} & \frac{\overline{b}}{2}\lambda^{\perp} & c(1-\frac{\lambda}{2}) \\
\end{bmatrix}, 
\ee
where we have used the shorthand notation $\lambda^{\perp}:=1-\lambda$, $a=\frac{1+z}{2}$, $b=\frac{x - i y}{2}$, $c=\frac{1-z}{2}$, and $\overline{b}=\frac{x+iy}{2}$ denotes the complex conjugate of $b$. 
We aim to determine when $\mathcal{E}\star\rho $ is positive semidefinite. 

\begin{figure}
\includegraphics{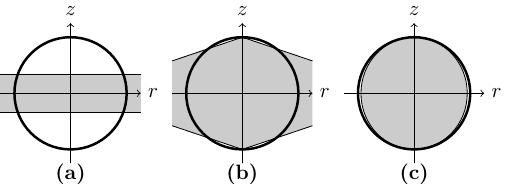}
    \caption{2d cross sections of the regions inside the Bloch ball from all $2\times2$ principal minors of $\mathcal{E}\star\rho$ with nonegative determinant, where $\mathcal{E}$ is the partially depolarizing channel (the actual shape in the 3d Bloch ball is obtained by rotating the region about the $z$-axis). In all figures, $\lambda=0.68$. (a) The strip given by $|z| \le \frac{\sqrt{(2-\lambda)(3\lambda-2)}}{\lambda}$. (b) The region between two cones $|z| \le 1-\sqrt{\frac{(1-\lambda)^2}{\lambda(2-\lambda)}} \; |r|$. (c) The region inside the ellipse $\frac{r^2}{2(2-\lambda)}+z^2\le1$.}
\label{fig:PDC2by2regions}
\end{figure}

We begin by demanding that the central $2\times 2$ determinant must be nonnegative, which leads to the equation 
\be
\label{eq:2by2detPDCfirst}
0\le \left(\frac{\lambda}{2}\right)^2\frac{1-z^2}{4}-\left(\frac{1-\lambda}{2}\right)^2 ,
\ee
which yields
\be
\label{eqn:PDCmiddle2by2det}
|z| \le \frac{\sqrt{(2-\lambda)(3\lambda-2)}}{\lambda}.
\ee
This region is shown in Figure~\ref{fig:PDC2by2regions}~(a).
Since $|z|\ge0$, this restricts $\lambda$ to lie in the range $\frac{2}{3} \leq \lambda \leq \frac{4}{3}$. We therefore get information on what $z$ must be at some specific values. For example, when $\lambda = \frac{2}{3}$, then $z = 0$. When $\lambda = \frac{4}{3}$, then $ |z| \le \frac{\sqrt{3}}{2}$. When $\lambda = 1$, then $|z| \le 1$. 
Requiring that the lower-right $2\times 2$ principal minor has nonnegative determinant says
\be
\label{eqn:PDCcone1}
\frac{\lambda}{2}\left(1-\frac{\lambda}{2}\right)(1-z)^{2} - (1-\lambda)^2 \left(\frac{r^2}{4}\right)\ge 0.
\ee
This gives a portion of the Bloch ball carved out by a cone that opens down along the $z$ axis with apex at $(0,0,1)$ and whose angle with the $z$ axis varies depending on the value of $\lambda$. 
A similar result 
\be
\label{eqn:PDCcone2}
\frac{\lambda}{2} \left(1-\frac{\lambda}{2}\right) (1+z)^2 - (1-\lambda)^2 \left(\frac{r^2}{4} \right) \ge 0
\ee
is obtained from the upper-left $2\times2$ principal minor, except that the cone has apex at $(0,0,-1)$ and opens up along the $z$ axis. The cones are described more explicitly by the inequalities
\be
\label{eqn:PDCtopleftbottomright2by2det}
|z| \le 1-\sqrt{\frac{(1-\lambda)^2}{\lambda(2-\lambda)}} \; |r|
\ee 
in the $(r,z)$ plane, and their intersection is shown in Figure~\ref{fig:PDC2by2regions}~(b).  
The only other $2\times2$ principal minor determinant that leads to further constraints is the one associated with the 2nd and 4th rows and columns. It yields 
\be
\frac{\lambda}{2}\left(1-\frac{\lambda}{2}\right)(1-z^2)-\frac{r^2}{4}\ge0,
\ee
which describes an ellipse, since it can be re-expressed as 
\be
\label{eqn:PDC2ndand4th2by2det}
\left(\frac{r}{\sqrt{2(2-\lambda)}}\right)^2+z^2\le1. 
\ee
This ellipse is shown in Figure~\ref{fig:PDC2by2regions}~(c).

Next, we will examine the entire $4\times4$ determinant of $\mathcal{E}\star\rho$ (before examining the principal $3\times3$ determinants). Simplifying this determinant and demanding it to be nonnegative yields 
\be
(-1 + r^2+z^2)\big(4 - 8 \lambda + (3 + r^2+z^2) \lambda^2\big) \ge 0. 
\ee
Since $r^2+z^2$ cannot be greater than $1$, this yields two regions, one given by $r^2+z^2=1$ (the Bloch sphere, i.e., all pure states) and the ball given by the equation
\be
r^2+z^2\le \frac{-3\lambda^2+8\lambda-4}{\lambda^2}=\frac{(2-\lambda)(3\lambda-2)}{\lambda^2}.
\ee
The latter region is the proposed ball $B_{\lambda}$ in the statement of the proposition. For values of $\lambda\in[\frac{2}{3},1)\cup(1,\frac{4}{3}]$, the Bloch sphere $r^2+z^2=1$ is excluded (this follows from nonnegativity of the determinants of the $2\times 2$ principal minors, but we will show this more simply using one of the $3\times3$ principal minors later).

Therefore, our goal moving forward will be to prove that all of the regions obtained from the nonnegativity of all other principal minor determinants contain $B_{\lambda}$, i.e., all principal minors do not further constrain this region. 

The strip~\eqref{eqn:PDCmiddle2by2det} clearly contains $B_{\lambda}$ since the height of the strip equals the radius of $B_{\lambda}$.

We next verify that the cones~\eqref{eqn:PDCtopleftbottomright2by2det} contain $B_{\lambda}$ and only intersect $B_{\lambda}$ when $\lambda=1$, in which case the lines are horizontal and tangent to the full Bloch sphere. More specifically, we will show that the distance between the center of the Bloch sphere $(0,0,0)$ and the cone is always greater than or equal to the radius of $B_{\lambda}$. By symmetry, we can focus on the $(r,z)$ plane. Let $\mathbf{p}:=(0,1)$ and $\mathbf{w}=(-m,1)$, where $m=\sqrt{\frac{(1-\lambda)^2}{\lambda(2-\lambda)}}$, as depicted here
\begin{center}
\begin{tikzpicture}[scale=1.25] 
\pgfmathsetmacro{\m}{1/(2*sqrt(2))} 
\draw[->] (-1.25,0) -- (1.25,0) node[right] {$r$}; \draw[->] (0,-1.25) -- (0,1.25) node[above] {$z$};  
\draw[thick,domain=-1.25:1.25] plot (\x,{1+\m*\x}); \draw[thick,domain=-1.25:1.25] plot (\x,{1-\m*\x});  
\draw[thick,domain=-1.25:1.25] plot (\x,{-1+\m*\x}); \draw[thick,domain=-1.25:1.25] plot (\x,{-1-\m*\x}); 
\draw[->,very thick,black] (0,0) -- node[left]{$\mathbf{w}$} ({-\m},1); 
\draw[->,very thick,black] (0,0) -- node[right]{$\mathbf{p}$} (0,1); 
\end{tikzpicture}
\end{center}
so that $\mathbf{w}$ is perpendicular to two of the cone boundaries intersected in the $(r,z)$ plane. 
The (minimum perpendicular) squared distance from the center of the Bloch ball to the cones is then given by 
\be
\left(\frac{|\mathbf{p}\cdot\mathbf{w}|}{\lVert\mathbf{w}\rVert}\right)^2
=\frac{\lambda(2-\lambda)}{(1-\lambda)^2+\lambda(2-\lambda)}
=\lambda(2-\lambda).
\ee
Therefore, 
\be
\left(\frac{|\mathbf{p}\cdot\mathbf{w}|}{\lVert\mathbf{w}\rVert}\right)^2 \ge R(\lambda)^2
\ee
is equivalent to 
\be
\lambda\ge\frac{3\lambda-2}{\lambda^2}
\iff
\lambda^3 - 3\lambda+2\ge0,
\ee
which is true for all $\lambda \in [\frac{2}{3},\frac{4}{3}]$ with equality holding if and only if $\lambda=1$ in this range. By symmetry, we therefore conclude that the cones only intersect $B_{\lambda}$ when $\lambda = 1$.
 
The ellipse~\eqref{eqn:PDC2ndand4th2by2det} has semiminor axis $\sqrt{2(2-\lambda)}$ (and semimajor axis $1$), and because $2\ge\frac{3\lambda-2}{\lambda^2}$ for all $\lambda\in[\frac{2}{3},\frac{4}{3}]$, the ellipse~\eqref{eqn:PDC2ndand4th2by2det} contains $B_{\lambda}$. 

\begin{figure}
\includegraphics{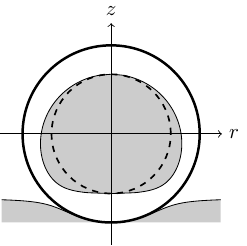}
    \caption{2d cross section of the region $\kappa_{4}(r,z,\lambda)\ge0$ (shaded), with $\kappa_{4}$ defined in~\eqref{egn:kappa3by3}, for the value $\lambda=0.73$. Only a part of the region is contained in the Bloch ball. The region contained within the dashed circle depicts the ball $B_{\lambda}$.}
    \label{fig:depolchan3by3region}
\end{figure}

We next examine the top-left $3\times 3$ principal minor of $\mathcal{E}\star\rho$, which yields 
\be
\label{eqn:PDCtopleft3by3det}
\kappa_{4}(r,z,\lambda)\ge 0,
\ee
where
\begin{align}
\kappa_{4}(r,z,\lambda)&:=
r^{2}\big(2(1-\lambda)^2-(1+z)\lambda^2\big) \nonumber \\ 
&+\lambda^2 (1+z)\big(R(\lambda)^2-z^2\big). \label{egn:kappa3by3}
\end{align}
This region is depicted in Figure~\ref{fig:depolchan3by3region}. 
We will prove that $B_{\lambda}$ is contained in this region by taking an arbitrary point $(r,z)\in B_{\lambda}$ and showing it satisfies~\eqref{eqn:PDCtopleft3by3det}. To do this, first note that such a point $(r,z)\in B_{\lambda}$ satisfies $r^2\le R(\lambda)^2-z^2$ by definition. Furthermore, since 
\be
2(1-\lambda)^2-(1+z)\lambda^2 
\le 2(1-\lambda)^2-(1-R(\lambda))\lambda^2 
\le 0 
\ee
for all $\lambda\in\left[\frac{2}{3},\frac{4}{3}\right]$ and $(r,z)\in B_{\lambda}$, this implies 
\begin{align}
r^{2}&\big(2(1-\lambda)^2-(1+z)\lambda^2 \big) \nonumber \\ 
&\ge \big(R(\lambda)^2-z^2\big)\big(2(1-\lambda)^2-(1+z)\lambda^2 \big) \label{eqn:simplifrlambda}
\end{align}
for all $z\in\big[-R(\lambda),R(\lambda)\big]$ and $\lambda\in\left[\frac{2}{3},\frac{4}{3}\right]$. Therefore, plugging in a point $(r,z)\in B_{\lambda}$ into $\kappa_{4}$ from~\eqref{egn:kappa3by3} yields
\be
\kappa_{4}\ge 2 (1-\lambda)^2 \big(R(\lambda)^2-z^2\big)\ge0,
\ee
where the first inequality follows from~\eqref{eqn:simplifrlambda}, and the second inequality follows from the fact that $z^2\le R(\lambda)^2$. The inequality $\kappa_{4}\ge0$ can also be used to exclude the Bloch sphere $r^2+z^2=1$ from the solution set whenever $\lambda\ne1$. To see this, let's assume this region intersects with $r^2+z^2 = 1$. Then plugging $r^2=z^2-1$ into $\kappa_{4}\ge0$ leads to 
\be
-2(1-\lambda)^2(1+z)^2\ge0,
\ee
which is possible only when $\lambda=1$ or $z=-1$. However, the point $z=-1$ is excluded when $\lambda\ne0$ due to the constraint~\eqref{eqn:PDCmiddle2by2det}. 

It remains to show that regions obtained from demanding nonnegativity of the determinants of the remaining three $3\times3$ principal minors all contain $B_{\lambda}$. We do this with fewer details since the calculations are similar to those in the previous paragraph. By removing the third column and row from $\mathcal{E}\star\rho$, nonnegativity of the resulting determinant yields 
\be
\label{eqn:DPC3by3remove3}
\kappa_{3}(r,z,\lambda) \ge0, 
\ee
where
\be
\kappa_{3}(r,z,\lambda):=
(1-z)(1+z)^2 (2-\lambda)\lambda+r^2\big((1-z)(2-\lambda)\lambda-2\big).
\ee
Plugging in $(r,z)\in B_{\lambda}$ into $\kappa_{3}$ and using the fact that \be
(1-z)(2-\lambda)\lambda-2
\le -2 (1-\lambda)^2
\le0
\ee 
for all $z\in\big[-1,1\big]$ and $\lambda\in\left[\frac{2}{3},\frac{4}{3}\right]$ yields 
\be
\kappa_{3}\ge2\left(\frac{(1-\lambda) (2-(1-z)\lambda)}{\lambda}\right)^2\ge0.
\ee
Hence, this principal $3\times3$ determinant does not further restrict the solution set $B_{\lambda}$. Removing the second row and column from $\mathcal{E}\star\rho$ leads to the inequality $\kappa_{2}(r,z,\lambda)\ge0$, where $\kappa_{2}(r,z,\lambda):=\kappa_{3}(r,-z,\lambda)$. Plugging in $(r,z)\in B_{\lambda}$ into $\kappa_{2}$ yields 
\be
\kappa_{2}\ge 2 \left(\frac{(1-\lambda) (2-(1+z)\lambda)}{\lambda}\right)^2\ge0.
\ee
Hence, this principal $3\times3$ determinant does not further restrict the solution set $B_{\lambda}$. Finally, removing the first row and column yields $\kappa_{1}(r,z,\lambda)\ge0$, where $\kappa_{1}(r,z,\lambda)=\kappa_{4}(r,-z,\lambda)$. Plugging in $(r,z)\in B_{\lambda}$ into $\kappa_{1}$ yields 
\be
\kappa_{1}\ge 2 (1-\lambda)^2 \big(R(\lambda)^2-z^2\big)\ge0. 
\ee
By the Sylvester test for positive-semidefiniteness~\cite{Gantmacher1959,Strang2022}, this proves that $\mathcal{E}\star\rho$ is positive if and only if $\rho=\frac{1}{2}(\mathds{1}_{2}+\mathbf{r}\cdot\boldsymbol{\sigma})$ satisfies $\lVert\mathbf{r}\rVert\le R(\lambda)$ whenever $\lambda\in\left[\frac{2}{3},\frac{4}{3}\right]$. 
\eprf

\begin{figure}
\includegraphics{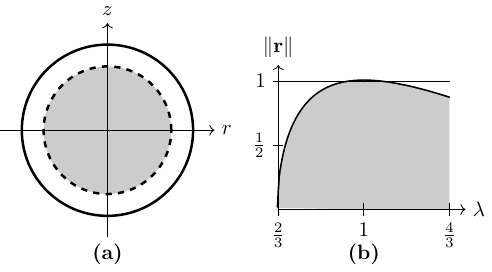}
\caption{(a) A 2d cross section of the region inside the Bloch ball for which the partially depolarizing channel admits causal confusion (the actual shape in the 3d Bloch ball is obtained by rotating the region about the $z$-axis). This is a ball of radius $R(\lambda)=\frac{\sqrt{(2-\lambda)(3\lambda-2)}}{\lambda}$ (dashed). The specific choice of $\lambda$ for the region shown is $\lambda=\frac{3}{4}$. (b) A plot of the radius $R(\lambda)$ used to describe the region in the Bloch ball admitting causal confusion for the partially depolarizing channel. Only the states within this radius admit causal confusion.}
\label{fig:PDCcausalconfusionregion}
\end{figure}

Figure~\ref{fig:PDCcausalconfusionregion} shows the region $B_{\lambda}$ for $\lambda=\frac{3}{4}$ and the function $R(\lambda)$, which describes the radius of $B_{\lambda}$.

\subsection{Mixed two-Pauli channel}

The fully mixed Pauli channel is defined by 
\be
\mathcal{E}(A)= \sum_{j=0}^3 p_{j} \sigma_{j}A \sigma_{j}
\ee
for any probability distribution $(p_0,p_1,p_2,p_3)$. We have already seen two examples in this class. 
In the special case that $p_1=p_2=p_3=\frac{p}{3}$ and $p_0=1-p$, this reproduces the partially depolarizing channel from Section~\ref{sec:depolarizing} with the relationship
$
3\lambda = 4p
$
between the parameters $\lambda$ and $p$. 
In the special case that $p_1=p_2=0$, $p_3=p$, and $p_0=1-p$, this reproduces the dephasing channel from Section~\ref{sec:dephasing}. 
Therefore, we analyze the next natural example in this class, the mixed two-Pauli channel defined by $p_1=p_2=\frac{p}{2}$, $p_3=0$, $p_0=1-p$.  The values of $p$ and the corresponding set of states admitting causal confusion is already quite intricate in this class. Before formulating the result, we first set some notation for this section. Let $\mathcal{E}:\matr_{2}\to\matr_{2}$ be the \define{mixed two-Pauli channel} defined by 
\be
\label{eqn:MPCKrausForm}
\mathcal{E}(A):=qA+\frac{p}{2}(\sigma_1A\sigma_1+\sigma_2A\sigma_2)
\ee
for all $A\in\matr_{2}$ and for $p\in[0,1]$, where
$
q:=1-p.
$
\begin{figure}
\includegraphics{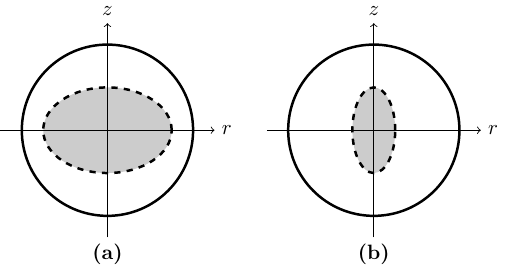}
\caption{A 2d cross section visualization of how the mixed Pauli channel~\eqref{eqn:MPC} acts on the Bloch sphere for $p=\frac{1}{4}$ (a) and $p=\frac{3}{4}$ (b). The image is the cross section of an ellipsoid (shaded) whose center is at $(0,0,0)$. As in Figure~\ref{fig:ADCBlochball}, cylindrical coordinates are used.}
\label{fig:MPCBlochball}
\end{figure}
%%%%%%%%%%%%%%%%%%%%%%%%%%%%%%%%%%%%%
\begin{figure*}[t]
\includegraphics{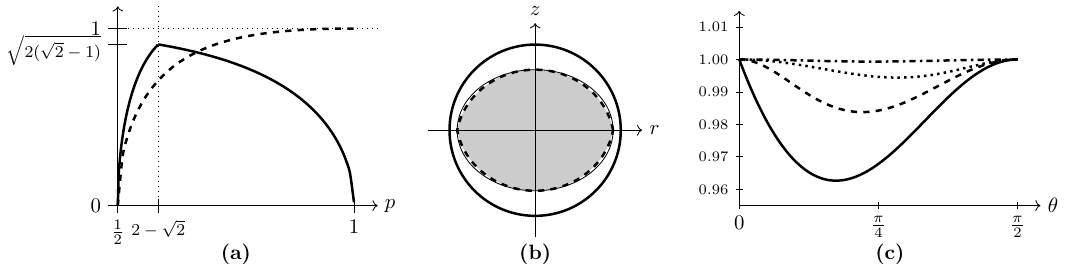}
\caption{(a) A plot of $r_{\min}$ (solid) from~\eqref{eq:rmin} and $z_{\min}$ (dashed) from~\eqref{eq:zmin}, both expressed as functions of $p\in[\frac{1}{2},1]$. 
(b) A 2d cross section of the region $D_{p}$ inside the Bloch ball for which the mixed two-Pauli channel~\eqref{eqn:MPC} exhibits causal confusion.  
The region $D_{p}$ (shaded inside the dashed curve) is approximated by an ellipsoid $E_{p}$ (drawn in a thin solid curve). The value $p=2-\sqrt{2}$ is chosen because it illustrates the greatest difference between $D_{p}$ and $E_{p}$ (upon observation, the two look almost indistinguishable). As in Figure~\ref{fig:ADCBlochball}, cylindrical coordinates are used.
(c) The deviation of $D_{p}$ from the ellipse $E_{p}$ is measured by the function $\ell$ from~\eqref{eqn:elltheta}, which equals $1$ precisely at the values of $p$ when $D_{p}=E_{p}$. The fact that $\ell(\theta)\le1$ means that $D_{p}\subseteq E_{p}$. The function $\ell(\theta)$ depends on $p$ as well, so four values of $p$ were chosen to show how close $\ell(\theta)$ is to $1$:  $p=0.55$ (dashed), $p=0.585786\approx 2-\sqrt{2}$ (solid), $p=0.65$ (dash-dot), $p=0.75$ (dotted). Only the interval $0\le\theta\le\frac{\pi}{2}$ is shown because the function $\ell(\theta)$ is even in $\theta$ and periodic with period $\pi$.}
\label{fig:MPCDpEp}
\end{figure*}
In matrix form, 
\be
\label{eqn:MPC}
\mathcal{E}\left( \left[ \setlength{\arraycolsep}{3pt} \begin{array}{cc} \\[-15pt] a &b \\ c & d  \end{array}\right]\right) = \begin{bmatrix} qa+p d & qb \\ qc & q d+ p a\end{bmatrix}.
\ee
When plugging in a qubit density matrix of the form $\rho=\frac{1}{2}(\mathds{1}_{2}+\mathbf{r}\cdot\boldsymbol{\sigma})$~\eqref{eqn:qubitrho} into $\mathcal{E}$, we obtain $\mathcal{E}\left(\frac{1}{2}(\mathds{1}_{2}+\mathbf{r}\cdot\boldsymbol{\sigma})\right)=\frac{1}{2}(\mathds{1}_{2}+\mathbf{s}\cdot\boldsymbol{\sigma})$, where
\be
\mathbf{s}:=(x',y',z')=\big((1-p)x,(1-p)y,(1-2p)z\big).
\ee
This shows that the effect of the mixed two-Pauli channel on the Bloch ball is to deform it to an ellipsoid still centered at the origin (cf.\ Figure~\ref{fig:MPCBlochball}). 

\bn
\label{prop:mixedPaulichannel}
If $\mathcal{E}$ is the mixed Pauli channel with parameter $p\in[0,1]$ as in~\eqref{eqn:MPCKrausForm}, then a necessary condition for $\mathcal{E}$ to admit causal confusion is that $p\in[\frac{1}{2},1]$. Moreover, when $p$ is in this range, let $S_{p}$ denotes the subset of states $\rho$ for which $(\mathcal{E},\rho)$ exhibits causal confusion. 
Then $S_{p}$ corresponds to the region $D_{p}\subseteq\R^{3}$ whose elements $(x,y,z)$ are described in parametric form by 
\begin{align}
x(s,\theta,\phi)&:=s \ell(\theta)\cos(\theta)\cos(\phi) \nonumber \\
y(s,\theta,\phi)&:=s \ell(\theta)\cos(\theta)\sin(\phi) \nonumber \\
z(\zeta,\theta,\phi)&:=\zeta \ell(\theta)\sin(\theta) \label{eqn:MPCparametric}
\end{align} 
with $\theta\in[-\frac{\pi}{2},\frac{\pi}{2})$, $\phi\in[0,2\pi)$, 
$s\in[0,r_{\min}]$, $\zeta\in[0,z_{\min}]$, where 
\begin{align}
r_{\min} &:= \frac{2\sqrt{q(p^3-4p^2+6p-2-q|p^2-4p+2|)}}{p(2-p)} \nonumber \\
&=\begin{cases} 
\frac{2\sqrt{q(2p-1)}}{p} & \mbox{ for }\frac{1}{2}\le p \le 2-\sqrt{2} \\
\frac{2\sqrt{q}}{2-p} & \mbox{ for } 2-\sqrt{2}\le p \le 1, 
\end{cases}
\label{eq:rmin}
\end{align}
\be
\label{eq:zmin}
z_{\min}:=\frac{\sqrt{2p-1}}{p}, 
\ee
and 
\be
\label{eqn:elltheta}
\ell(\theta):=\sqrt{\frac{-B-\sqrt{B^2-4AC}}{2A}}, 
\ee
where 
\be
\label{eqn:MPCA}
A := \left(\frac{p\big(4q z_{\min}^2\sin^2(\theta)+(2-p)r_{\min}^2\cos^2(\theta)\big)}{4}\right)^2, 
\ee
\be
\label{eqn:MPCB}
B := q^2(q^2-2p^2)z_{\min}^2\sin^2(\theta)+\frac{q}{2}\big((2-p)q^2-p\big)r_{\min}^2\cos^2(\theta),
\ee
and 
\be
\label{eqn:MPCC}
C:=q^2 (2p-1). 
\ee
Moreover, $D_{p}$ is contained in the ellipsoid $E_{p}$ defined in cylindrical coordinates $(r,z)$ by 
\be
\left(\frac{r}{r_{\min}}\right)^2+\left(\frac{z}{z_{\min}}\right)^2 \le 1, 
\ee
so that states outside of $D_{p}$ are guaranteed to not lead to causal confusion for Pauli observables and their joint expectation values. 
\en

The values $r_{\min}$ and $z_{\min}$ are plotted in Figure~\ref{fig:MPCDpEp} (a) and are used to describe the semi-major and semi-minor axes of the ellipsoid $E_{p}$. The set $D_{p}$ and approximating ellipsoid $E_{p}$ are shown in Figure~\ref{fig:MPCDpEp} (b) for a specific value of $p$ chosen to illustrate the largest deviation between $D_{p}$ and $E_{p}$. The two regions $D_{p}$ and $E_{p}$ are nearly identical. In fact, the deviation of $D_{p}$ from this ellipsoid $E_{p}$ is described by the function $\ell$, which satisfies the bound $0.96\le\ell(\theta)\le 1$ for all $\theta\in[-\frac{\pi}{2},\frac{\pi}{2})$ and is plotted in Figure~\ref{fig:MPCDpEp} (c).

\bprf
[Proof of Proposition~\ref{prop:mixedPaulichannel}]
The state over time $\mathcal{E}\star\rho$ is given by 
\be
\mathcal{E}\star\rho = \frac{1}{2}\begin{bmatrix}2aq & bq & b & 0 \\ cq & 2ap & q & b \\ c & q & 2dp & bq \\ 0 &c & cq & 2dq \end{bmatrix},
\ee
where $q=1-p$, $a = \frac{1+z}{2}$, $b = \frac{x-iy}{2}$, $c = \bar{b}$, and $d = \frac{1-z}{2}$. Our goal is to determine the conditions on $(x,y,z)$ such that $\mathcal{E}\star\rho$ is positive. The proof will be similar to the proof of Proposition~\ref{prop:DepChanEstarRho} with the exception that we will show fewer details regarding checking all principal determinants, and we will instead focus on the $4\times 4$ determinant and how positivity of it leads to the parametric form~\eqref{eqn:MPCparametric} and the approximating ellipse. 

\begin{figure}
\includegraphics{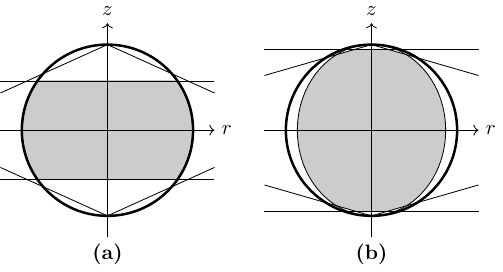}
\caption{2d cross section of the intersection of the regions inside the Bloch ball obtained by demanding nonnegativity of the principal $2\times2$ matrices of $\mathcal{E}\star\rho$ for the mixed two-Pauli channel $\mathcal{E}$ defined in~\eqref{eqn:MPCKrausForm} for the values $p=0.55$ (a) and $p=0.75$ (b). The thick solid curve depicts the Bloch sphere. The shaded region is obtained from the intersection of the regions~\eqref{eqn:MPCzheight2by2}, \eqref{eqn:MPCellipse2by2}, and~\eqref{eqn:MPCcone2by2} (the ellipse from~\eqref{eqn:MPCellipse2by2} is not visible when $p=0.55$ because the semiminor axis is $0.995\approx1$).}
\label{fig:MPC2by2regions}
\end{figure}

We begin by demanding that the central $2\times 2$ determinant must be nonnegative, which leads to the inequality 
\be
\label{eqn:MPCcentral2by2}
0\le (1-z^2)p^2-(1-p)^2\le p^2-(1-p)^2,
\ee
which itself implies that $p\ge\frac{1}{2}$ (the second inequality follows from demanding $-1\le z \le 1$). Thus, $p\in[\frac{1}{2},1]$. We also obtain 
\be
\label{eqn:MPCzheight2by2}
|z|\le \frac{\sqrt{2p-1}}{p}
\ee
from~\eqref{eqn:MPCcentral2by2}. Positivity of the other $2\times 2$ principal determinants yield the inequalities
\be
\label{eqn:MPCellipse2by2}
\left(\frac{r}{2\sqrt{p(1-p)}}\right)^2+z^2\le 1
\ee
and 
\be
\label{eqn:MPCcone2by2}
\frac{1}{2}\sqrt{\frac{1-p}{p}}\,|r|+|z|\le 1.
\ee
The intersection of these regions is shown in Figure~\ref{fig:MPC2by2regions} for two values of $p$. 

\begin{figure}
\includegraphics{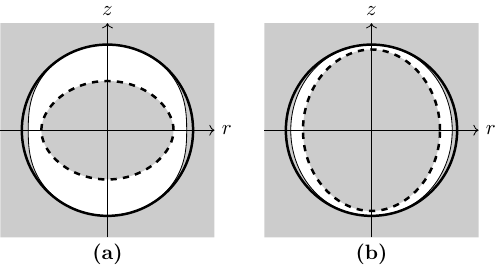}
\caption{2d cross section of the region $\det(\mathcal{E}\star\rho)\ge0$ (shaded), for the mixed two-Pauli channel $\mathcal{E}$ defined in~\eqref{eqn:MPCKrausForm} for the values $p=0.55$ (a) and $p=0.75$ (b). The thick solid curve depicts the Bloch sphere. Only the region contained within the dashed curve depicts the region $D_{p}$. The regions outside these inner pseudo-ellipses that are still contained in the shaded region inside the Bloch ball are excluded by the nonnegativity of the $3\times3$ principal determinants of $\mathcal{E}\star\rho$. The boundary of these outer regions is obtained in an analogous way as for the inner dashed curve except that we use $\ell(\theta)=\sqrt{\frac{-B+\sqrt{B^2-4AC}}{2A}}$, where $A,B,C$ are defined as in~\eqref{eqn:MPCA}, \eqref{eqn:MPCB}, and~\eqref{eqn:MPCC}, but $r_{\mathrm{max}}$ and $z_{\mathrm{max}}$ are used instead of $r_{\mathrm{min}}$ and $z_{\mathrm{min}}$. }
\label{fig:MPCinnerouterregions}
\end{figure}

Before looking at the $3\times3$ principal determinants, we analyze the determinant of $\mathcal{E}\star\rho$, which yields a degree-$2$ polynomial in $r^2$ and $z^2=x^2+y^2$. Rather than writing this polynomial out in a single line, we write down a table of the coefficients (after multiplying the determinant by $256$ to avoid fractions)
\[
\begin{tabular}{c|ccc}
& $1$ & $r^2$ & $r^4$ \\
\hline
$1$ & $16(2p-1)q^2$ & $8q(2-6p+4p^2-p^3)$ & $(2-p)^2p^2$ \\
$z^2$ & $16q^2(1-2p-p^2)$ & $8(2-p)qp^2$ & $0$ \\
$z^4$ & $16q^2p^2$ & $0$ & $0$ \\
\end{tabular}
\]
The region obtained from the inequality $\det(\mathcal{E}\star\rho)\ge0$ is shown in Figure~\ref{fig:MPCinnerouterregions} for two values of $p$. 
This region is \emph{approximately} given by an inequality of the form 
\be
\label{eqn:approximatedtwoellipses}
\left(\frac{r^2}{r_{\min}^2}+\frac{z^2}{z_{\min}^2} -1\right)
\left(\frac{r^2}{r_{\max}^2}+\frac{z^2}{z_{\max}^2} -1\right)
\ge0,
\ee
where $r_{\min},r_{\max}$ and $z_{\min},z_{\max}$ are determined by setting $z=0$ and $r=0$, respectively, in the expression ${\det(\mathcal{E}\star\rho)=0}$. We will first obtain $r_{\min},r_{\max},z_{\min},z_{\max}$, and save the explanation in what sense the region is approximated by the inequality~\eqref{eqn:approximatedtwoellipses}, which describes the region inside the smaller ellipse and outside the larger ellipse, for later. 
Setting $r=0$ in the expression ${\det(\mathcal{E}\star\rho)=0}$ yields
\be
q^2 (1-z^2) \big(p^2 (1-z^2) -q^2\big)=0
\ee
If $p\ne1$, then $q^2$ can be cancelled from all terms and solving for nonnegative $z$ gives two solutions 
\be
z_{\min}=\frac{\sqrt{2p-1}}{p}
\quad\text{ and }\quad
z_{\max}=1
.
\ee
%%%%%%%%%%%%%%%%%%%%%%%%%%%%%
Meanwhile, following a similar procedure by setting $z=0$ in the expression ${\det(\mathcal{E}\star\rho)=0}$ gives
\be
16(2p-1)q^2+8q(2-6p+4p^2-p^3) r^2+(2-p)^2p^2 r^4=0
\ee
and solving for nonnegative $r$ gives two solutions
$r_{\min}$ exactly as in~\eqref{eq:rmin}
and
\be
r_{\max}
=\begin{cases}
\frac{2\sqrt{(1-p)(2p-1)}}{p}  & \mbox{ if $\frac{1}{2}\le p\le 2-\sqrt{2}$} \\
\frac{2\sqrt{1-p}}{2-p} & \mbox{ if $2-\sqrt{2}\le p \le 1$.} 
\end{cases}
\ee

Next, we introduce a polar coordinate angle $\theta$ for the $(r,z)$ plane and we parameterize $r$ and $z$ via the ans\"atz $r= r_{\min}\ell(\theta)\cos(\theta)$ and $z = z_{\min}\ell(\theta)\sin(\theta)$, with $\theta\in[0,2\pi)$ and $\ell(\theta)$ a function that will be determined momentarily with the purpose of obtaining the boundary of the inner region of Figure~\ref{fig:MPCinnerouterregions}, i.e., the inner region described by the inequality $\det(\mathcal{E}\star\rho)\ge0$. We can interpret $\ell(\theta)$ as the relative distortion between the ellipse $\frac{r^2}{r_{\min}^2}+\frac{z^2}{z_{\min}^2}=1$ and the boundary of the inner region in Figure~\ref{fig:MPCinnerouterregions} ($\ell(\theta)=1$ would correspond to a true ellipse). By plugging this ans\"atz into our determinant equation $\det(\mathcal{E}\star\rho)=0$ and factoring out powers of $\ell$, we get 
\be
A\ell^4 + B\ell^2 + C=0,
\ee
where $A,B,$ and $C$ are functions of $\theta$ and $p$ given by~\eqref{eqn:MPCA}, \eqref{eqn:MPCB}, and \eqref{eqn:MPCC}, respectively. Using the quadratic formula, we can get a closed form for $\ell^2$ and, in turn, $\ell$, which is given by~\eqref{eqn:elltheta}, and which is plotted in Figure~\ref{fig:MPCDpEp} (c). 

%%%%%%%%%%%%%%%%%%%%%%%%%%%%%

\begin{figure}
\includegraphics{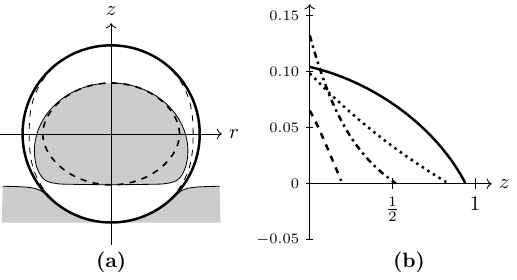}
    \caption{(a) 2d cross section of the region $\kappa_{4}(r,z,p)\ge0$ (shaded), with $\kappa_{4}$ defined in~\eqref{egn:TPCkappa3by3}, for the value $p=0.55$. Only a part of the region is contained in the Bloch ball. The region contained within the dashed circle depicts $D_{p}$. (b) Solving $\kappa_{4}(r,z,p)=0$ and $\det(\mathcal{E}\star\rho)=0$ for $r^2$ as functions of $z$ (and $p$) and subtracting the latter from the former shows that the inner region obtained from $\det(\mathcal{E}\star\rho)\ge0$ is completely contained in the inner region obtained from $\kappa_{4}(r,z,p)\ge0$. These differences are shown for the values $p=0.505$ (dashed), $p=0.54$ (dash-dot), $p=0.65$ (dotted), and $p=0.75$ (solid). All such differences are thus nonnegative on the interval $[0,\frac{\sqrt{2p-1}}{p}]$.}
    \label{fig:TPMchan3by3region}
\end{figure}

We now return to the $3\times3$ determinants. Removing the fourth column and row from $\mathcal{E}\star\rho$, nonnegativity of that determinant says $\kappa_{4}(r,z,p)\ge0$, where 
\begin{align}
\kappa_{4}(r,z,p)&:= r^2 \big(2-6p+2p^2(2-z)-p^3(1-z)\big) \nonumber \\
&-4(1-p)(1+z)\big(1-2p+p^2z^2\big).
\label{egn:TPCkappa3by3}
\end{align}
This region is depicted in Figure~\ref{fig:TPMchan3by3region}. 
Removing the first column and row yields $\kappa_{1}(r,z,p)=\kappa_{4}(r,-z,p)$, which is a reflection of the previous region across the plane $z=0$. One can show that $D_{p}$ is contained inside the intersections of the regions determined by $\kappa_{1}(r,z,p)\ge0$ and $\kappa_{4}(r,z,p)\ge0$, which are contained within the Bloch ball and moreover exclude the outer region obtained from $\det(\mathcal{E}\star\rho)\ge0$. 
Finally, removing the second, resp.\ third, column and row from $\mathcal{E}\star\rho$ yields $\kappa_{2}(r,z,p)$, resp.\ $\kappa_{3}(r,z,p)$, where 
\begin{align}
\kappa_{2}(r,z,p)&
:=r^2(1-p)\big(-2+p(2-p)(1+z)\big) \nonumber \\
	&+4(1-p)^2p(1-z)^2(1+z)
\end{align}
and $\kappa_{3}(r,z,p)=\kappa_{2}(r,-z,p)$. 
The region $D_{p}$ is also contained in the regions $\kappa_{2}(r,z,p)\ge0$ and $\kappa_{3}(r,z,p)\ge0$. Since this exhausts all subdeterminants, this shows that the subset of states corresponding to causal confusion for the mixed Pauli channel is indeed $D_{p}$. 
\eprf

The examples we have analyzed indicate that determining the set of states leading to causal confusion is a nontrivial task already in the case of two local measurements on qubits even for the simplest quantum channels. These examples have also shown that channels that are sufficiently noisy tend to lead to larger regions of state space admitting causal confusion. But how much noise is necessary for the \emph{existence} of causal confusion? These general questions will be answered in the next section.

\section{General results on causal confusion}
\label{sec:generalresults}

Before examining qubit causal confusion further, we draw some general conclusions about arbitrary finite-dimensional systems. The first observations we make concern entanglement-breaking (EB) channels, positive partial transpose (PPT) channels, and causal confusion. These results are summarized in the following implication flow chart, all under the assumption that $\mathcal{E}:\matr_{m}\to\matr_{n}$ is a quantum channel and $\rho\in\matr_{m}$ is a density matrix (more inferred implications are not shown to avoid clutter). 
\[
\xy0;/r.25pc/:
(-28,15)*+{\begin{tabular}{c}$\mathcal{E}\star\frac{\mathds{1}_{m}}{m}$ is a \\ separable state\end{tabular}}="sep";
(0,15)*+{\begin{tabular}{c}$\mathcal{E}\star\frac{\mathds{1}_{m}}{m}$ is a \\PPT state\end{tabular}}="PPTstate";
(28,15)*+{\begin{tabular}{c}$\mathcal{E}\star\frac{\mathds{1}_{m}}{m}$ is a \\ state\end{tabular}}="state";
(-15,0)*+{\mathcal{E}\text{ is EB}}="EB";
(15,0)*+{\mathcal{E}\text{ is PPT}}="PPT";
(-28,-15)*+{\begin{tabular}{c}$\mathcal{E}\star\rho$ is a \\ separable state\end{tabular}}="Erhosep";
(0,-15)*+{\begin{tabular}{c}$\mathcal{E}\star\rho$ is a \\ PPT state\end{tabular}}="ErhoPPT";
(28,-15)*+{\begin{tabular}{c}$\mathcal{E}\star\rho$ is a \\ state\end{tabular}}="Erhostate";
{\ar@{=>}"sep";"PPTstate"};
{\ar@{<=>}"PPTstate";"state"};
{\ar@{=>}"EB";"PPT"};
{\ar@{<=>}"PPT";"state"_{\text{Prop~\ref{prop:PPTQSOT}}}};
{\ar@{=>}"Erhosep";"ErhoPPT"};
{\ar@{=>}"ErhoPPT";"Erhostate"};
{\ar@{<=>}"sep";"EB"_{\text{Prop~\ref{prop:EBandQSOT}}}};
{\ar@{=>}"Erhosep";"EB"^{\text{Thm~\ref{thm:separableimpliesEB}}}}; 
{\ar@{=>}"Erhostate";"PPT"_{\text{Thm~\ref{thm:PPTEQSOTstate}}}}; 
\endxy
\]
The converse of any unidirectional implication is not true in general, and we provide counter-examples to illustrate each case (the cases we do not provide examples for are already known in the literature). 

\subsection{Entanglement-breaking channels and separable states over time}

Let $\Alg{A}=\matr_{m}$ and $\Alg{B}=\matr_{n}$. A channel $\mathcal{E}:\Alg{A}\to\Alg{B}$ is called \define{entanglement breaking} iff $(\id_{\Alg{A}}\otimes\mathcal{E})\big(|\Omega\>\<\Omega|\big)$ is a separable state~\cite{horodecki2003ebc,Ruskai2003,PaPaPrRa2020}, where  
\be
|\Omega\>=\frac{1}{\sqrt{m}}\sum_{j=1}^{m}|j\>\otimes|j\>
\ee
and $|1\>, \dots, |m\>$ are the standard unit vectors (this convention differs from the one we used for qubits, where $|0\>$ was the first unit vector). 

In this section, we will prove some facts relating entanglement-breaking channels and spatiotemporal correlations. In particular, we prove that $\mathcal{E}$ is entanglement breaking if and only if $\mathcal{E}\star\frac{\mathds{1}_{m}}{m}$ is a separable state (Proposition~\ref{prop:EBandQSOT}). We recall one of the main results of Ref.~\cite{SongParzygnat2025}, which states that if $\mathcal{E}\star\rho$ is a separable state, then $\mathcal{E}$ is entanglement breaking, and we provide an alternative proof that will be useful for our later results. Additionally, we provide an example of an entanglement-breaking channel $\mathcal{E}$ and a state $\rho$ such that $\mathcal{E}\star\rho$ is an entangled state, thereby showing that the converse of the previous statement is not true in general. We also provide an example where $\mathcal{E}$ is entanglement breaking and $\rho$ is a state such that $(\mathcal{E},\rho)$ does not exhibit causal confusion. 

In what follows, it will be useful to note that 
\be
\label{eqn:ChoiEBidentity}
\frac{1}{m}\Choi[\mathcal{E}]=(\id_{\Alg{A}}\otimes\mathcal{E})\big(|\Omega\>\<\Omega|\big),
\ee
which follows from the definition of the Choi martrix~\eqref{eqn:Choimatrix}. 
In Ref.~\cite{horodecki2003ebc}, the authors provided a list of several equivalent conditions for a channel $\mathcal{E}$ to be entanglement breaking. We now add to this list of equivalent conditions and provide further properties relating entanglement-breaking channels to quantum states over time and hence causal confusion. 

\bn
\label{prop:EBandQSOT}
Let $\Alg{A}=\matr_{m}$,  $\Alg{B}=\matr_{n}$, and let $\mathcal{E}:\Alg{A}\to\Alg{B}$ be a quantum channel. Then $\mathcal{E}$ is entanglement breaking if and only if $\mathcal{E}\star\frac{\mathds{1}_{m}}{m}$ is a separable state. 
\en

Before proving Proposition~\ref{prop:EBandQSOT}, we recall the fact that if $\tau$ is a separable state in $\Alg{A}\otimes\Alg{B}$, then its partial transpose is also a separable state~\cite{peres1996separability,horodecki1996necessary}. 

\bprf[Proof of Proposition~\ref{prop:EBandQSOT}]
First note that 
\be
\mathcal{E}\star\frac{\mathds{1}_{m}}{m}
=\frac{1}{2}\left\{\frac{\mathds{1}_{m}}{m}\otimes\mathds{1}_{n},\Jamio[\mathcal{E}]\right\}
=\frac{1}{m}\Jamio[\mathcal{E}].
\ee
Taking the partial transpose $\mathcal{T}\otimes\id_{\Alg{B}}$ of both sides yields
\be
(\mathcal{T}\otimes\id_{\Alg{B}})\left(\mathcal{E}\star\frac{\mathds{1}_{m}}{m}\right)=\frac{1}{m}\Choi[\mathcal{E}]
\ee
due to~\eqref{eqn:PTofJamiol}.
Therefore, if $\mathcal{E}\star\frac{\mathds{1}_{m}}{m}$ is a separable state, then its partial transpose $\frac{1}{m}\Choi[\mathcal{E}]$ is a separable state. By the identity~\eqref{eqn:ChoiEBidentity}, $\mathcal{E}$ is entanglement breaking. Conversely, if $\mathcal{E}$ is entanglement breaking, then $\frac{1}{m}\Choi[\mathcal{E}]$ is separable. Applying $\mathcal{T}\otimes\id_{\Alg{B}}$ to it therefore yields another separable state. But since $\mathcal{T}^2=\id_{\Alg{A}}$, this means $\mathcal{E}\star\frac{\mathds{1}_{m}}{m}$ is a separable state. 
\eprf

Thus, Proposition~\ref{prop:EBandQSOT} relates entanglement-breaking channels to causal confusion by providing a correspondence between spatial correlations associated with separable states that have a maximally mixed marginal and temporal correlations involving maximally mixed states and entanglement-breaking channels. The precise statement is given in the following corollary. 

\bc
\label{cor:EBandseptau}
Let $\Alg{A}=\matr_{m}$,  $\Alg{B}=\matr_{n}$, and let $\tau\in\Alg{A}\otimes\Alg{B}$ be a matrix satisfying $\Tr_{\Alg{B}}[\tau]=\frac{\mathds{1}_{m}}{m}$. 
Then $\tau$ is a separable state if and only if there exists an entanglement-breaking channel $\mathcal{E}:\Alg{A}\to\Alg{B}$ such that $\tau=\mathcal{E}\star\frac{\mathds{1}_{m}}{m}$. 
\ec

Proposition~\ref{prop:EBandQSOT} and Corollary~\ref{cor:EBandseptau} are consistent with the claims of Ref.~\cite{RAVJSR15}, where it was stated (in the Supplementary Information Section III.E) that ``when confronted with a channel that is entanglement-breaking, or a bipartite state that is separable, an observational scheme does not provide \emph{any} information about the causal relations.'' However, it is important to note that Ref.~\cite{RAVJSR15} assumed in this statement that the input state $\rho$ and the marginal $\Tr_{\Alg{B}}[\tau]$ are maximally mixed. We will soon use the QSOT~\eqref{eqn:JPQSOT} to show that these claims are no longer necessarily true when $\rho$ and $\Tr_{\Alg{B}}[\tau]$ are not necessarily maximally mixed, thereby illustrating that causal confusion is a state-dependent phenomenon. 

Before proving Corollary~\ref{cor:EBandseptau}, we recall two definitions and a useful fact about entanglement-breaking channels. A (finite) \define{positive operator-valued measure} (POVM) in $\Alg{A}=\matr_{m}$ is a collection $\{M_{\alpha}\}$ of positive semidefinite matrices $M_{\alpha}\in\Alg{A}$ indexed by $\alpha\in\{1,\dots,s\}$, such that $\sum_{\alpha=1}^{s}M_{\alpha}=\mathds{1}_{m}$~\cite{Kr83,NiCh11}. Second, a \define{measure-and-prepare channel} from $\Alg{A}$ to $\Alg{B}=\matr_{n}$ is a channel $\mathcal{E}:\Alg{A}\to\Alg{B}$ of the form 
\be
\label{eq:temporalchannelMPform}
    \mathcal{E}(A) = \sum_{\alpha=1}^{s} \Tr[M_{\alpha} A] \nu_{\alpha},
\ee
where $\{M_{\alpha}\}$ is a POVM, and $\{\nu_{\alpha}\}$ is a collection of density matrices in $\Alg{B}$~\cite{horodecki2003ebc}. Equivalently, a quantum channel $\mathcal{E}:\Alg{A}\to\Alg{B}$ is entanglement breaking if and only if there exists a classical system $\C^{s}$ (viewed as a commutative $*$-algebra) and positive (necessarily completely positive~\cite{St55}) trace-preserving maps $\mathcal{M}:\Alg{A}\to\C^{s}$ and $\mathcal{P}:\C^{s}\to\Alg{B}$ such that $\mathcal{E}=\mathcal{P}\circ\mathcal{M}$, i.e., the diagram 
\be
\xy0;/r.25pc/:
(-12.5,-5)*+{\Alg{A}}="A";
(0,5)*+{\C^{s}}="S";
(12.5,-5)*+{\Alg{B}}="B";
{\ar"A";"B"_{\mathcal{E}}};
{\ar"A";"S"^{\mathcal{M}}};
{\ar"S";"B"^{\mathcal{P}}};
\endxy
\ee
commutes, i.e., $\mathcal{E}$ \emph{factors through} a classical system~\cite{Ma98,FuPa22a}. 
Ref.~\cite{horodecki2003ebc} proved that $\mathcal{E}$ is a measure-and-prepare channel if and only if $\mathcal{E}$ is entanglement breaking. 

\bprf[Proof of Corollary~\ref{cor:EBandseptau}]
Suppose $\tau\in\Alg{A}\otimes\Alg{B}$ is a separable state and $\Tr_{\Alg{B}}[\tau]=\frac{\mathds{1}_{m}}{m}$. Since $\tau$ is separable, decompose it as  
\be
\tau=\sum_{\alpha=1}^{r}t_{\alpha}\rho_{\alpha}\otimes\nu_{\alpha}
\ee
with $\{t_{\alpha}\}$ a probability distribution and $\rho_{\alpha}\in\Alg{A}$ and $\nu_{\alpha}\in\Alg{B}$ states. Let $\mathcal{E}:\Alg{A}\to\Alg{B}$ be the map defined by 
\be
\Alg{A}\ni A\mapsto\mathcal{E}(A):=m\sum_{\alpha=1}^{r}t_{\alpha}\Tr[\rho_{\alpha}A] \nu_{\alpha}. 
\ee
Set $M_{\alpha}:=mt_{\alpha}\rho_{\alpha}$. Notice that each $M_{\alpha}$ is positive semidefinite and 
\be
\sum_{\alpha=1}^{r}M_{\alpha}=m\sum_{\alpha=1}^{r}t_{\alpha}\rho_{\alpha}=m\Tr_{\Alg{B}}[\tau]=m\frac{\mathds{1}_{m}}{m}=\mathds{1}_{m}
\ee
so that $\{M_{\alpha}\}$ defines a POVM. This shows that $\mathcal{E}$ is a measure-and-prepare channel and is therefore entanglement breaking. 
Moreover, 
\begin{align}
\mathcal{E}\star\frac{\mathds{1}_{m}}{m}
&=\frac{1}{m}\Jamio[\mathcal{E}]
=\sum_{j,k=1}^{m} |j\>\<k| \otimes \sum_{\alpha=1}^{r} t_{\alpha} \Tr\big[ \rho_{\alpha} |k\>\<j| \big] \nu_{\alpha} \nonumber \\
&=\sum_{\alpha=1}^{r} t_{\alpha} \sum_{j,k=1}^{m}  \<j|\rho_{\alpha} |k\> |j\>\<k| \otimes \nu_{\alpha} \nonumber \\ 
&=\sum_{\alpha=1}^{r} t_{\alpha}  \rho_{\alpha}\otimes\nu_{\alpha} = \tau.
\end{align}
This proves the forward direction. 
The converse direction follows from Proposition~\ref{prop:EBandQSOT}.
\eprf

Proposition~\ref{prop:EBandQSOT} and Corollary~\ref{cor:EBandseptau} are closely related to, but distinct from, the following result proved in Ref.~\cite{SongParzygnat2025}, though we make one assumption here for simplicity. 

\bt
\label{thm:separableimpliesEB}
Let $\Alg{A}=\matr_{m}$,  $\Alg{B}=\matr_{n}$, and let $\mathcal{E}:\Alg{A}\to\Alg{B}$ be a quantum channel. Suppose that there exists a full rank (no zero eigenvalues) state $\rho\in\Alg{A}$ such that $\mathcal{E}\star\rho$ is a separable state. Then $\mathcal{E}$ is entanglement breaking. 
\et

We remark that Ref.~\cite{SongParzygnat2025} proved a slightly more general claim, namely that if $\tau\in\Alg{A}\otimes\Alg{B}$ is a separable state, then there exists an entanglement-breaking channel $\mathcal{E}:\Alg{A}\to\Alg{B}$ such that $\tau=\mathcal{E}\star\rho$, where $\rho=\Tr_{\Alg{B}}[\tau]$. For completeness, and also to provide an alternative proof using the definition of separability (as opposed to the proof of constructing an explicit measure-and-prepare channel as in Ref.~\cite{SongParzygnat2025}), we prove Theorem~\ref{thm:separableimpliesEB} here. The approach presented here will also be applicable in Section~\ref{sec:PPT} to positive-partial transpose (PPT) channels. Therefore, we will single out an important lemma, since it will be used in several proofs moving forward. 

\blem
\label{lem:BhatiaRosenberg}
Let $A$ be a positive-definite $m\times m$ matrix, and let $B$ and $X$ be $m\times m$ matrices. The equation 
\be
\label{eqn:SylvesterEquation}
\{A,X\}=B
\ee
has unique solution given by 
\be
\label{eqn:SylvesterSolution}
X=\int_{0}^{\infty}e^{-tA}Be^{-tA}\,dt.
\ee
Moreover, 
if $B$ is positive semidefinite, then $X$ is positive semidefinite. 
\elem

The last fact say that the linear function 
\be
\label{eqn:integralcombo}
\matr_{m}\ni B\mapsto \int_{0}^{\infty}e^{-tA}Be^{-tA}\,dt
\ee
is a positive map; in fact, it is completely positive but it is not necessarily a quantum channel. Notice also that it is important that $A$ be positive definite, and not just positive semidefinite (otherwise, the integral in~\eqref{eqn:integralcombo} might not converge). 
Although Lemma~\ref{lem:BhatiaRosenberg} is well-known~\cite{BhRo97,LCD24}, we provide a proof in Appendix~\ref{sec:Sylvester} since the proof techniques are used throughout.

\bprf
[Proof of Theorem~\ref{thm:separableimpliesEB}]
Suppose that $\mathcal{E}\star\rho$ is a separable state. Then there exists a natural number $s$, a probability distribution $\{c_{\alpha}\}$ and density matrices $\rho_{\alpha}\in\Alg{A}$ and $\nu_{\alpha}\in\Alg{B}$ with $\alpha\in\{1,\dots,s\}$, such that 
\be
\mathcal{E}\star\rho=\sum_{\alpha=1}^{s}c_{\alpha}\rho_{\alpha}\otimes\nu_{\alpha}.
\ee
To show $\mathcal{E}$ is entanglement breaking, we will use Proposition~\ref{prop:EBandQSOT} to show $\mathcal{E}\star\frac{\mathds{1}_{m}}{m}=\frac{1}{m}\Jamio[\mathcal{E}]$ is a separable state. 
Since $\mathcal{E}\star\rho=\frac{1}{2}\big\{\rho\otimes\mathds{1}_{n},\Jamio[\mathcal{E}]\big\}$, setting $X=\Jamio[\mathcal{E}]$, $A=\frac{1}{2}\rho\otimes\mathds{1}_{n}$, $B=\mathcal{E}\star\rho$ in Lemma~\ref{lem:BhatiaRosenberg} gives us
\begin{align}
\frac{1}{m}\Jamio[\mathcal{E}]&=\frac{1}{m}\int_{0}^{\infty}\sum_{\alpha=1}^{s} c_{\alpha} e^{-\frac{t}{2}\rho}\rho_{\alpha}e^{-\frac{t}{2}\rho} \otimes \nu_{\alpha}\, dt \nonumber \\
&=\sum_{\alpha=1}^{s} \frac{c_{\alpha}}{m} \left(\int_{0}^{\infty}e^{-\frac{t}{2}\rho}\rho_{\alpha}e^{-\frac{t}{2}\rho} \,dt\right)\otimes \nu_{\alpha}, \label{eqn:separableJE}
\end{align}
where we have used the fact that 
$
e^{-\frac{t}{2}\rho\otimes\mathds{1}_{n}}=e^{-\frac{t}{2}\rho}\otimes\mathds{1}_{n}. 
$
The term inside the parentheses in~\eqref{eqn:separableJE} is positive semidefinite by Lemma~\ref{lem:BhatiaRosenberg}. 
Let $\rho=\sum_{k=1}^{m}p_{k}|\mathbf{u}_{k}\>\<\mathbf{u}_{k}|$ be an eigendecomposition of $\rho$ so that $\{p_{k}\}$ defines a probability distribution and so that $\{\mathbf{u}_{k}\}$ is an orthonormal basis of $\C^{m}$. 
Now, let 
\be
b_{\alpha}:=\Tr\left[\int_{0}^{\infty}e^{-\frac{t}{2}\rho} \rho_{\alpha} e^{-\frac{t}{2}\rho} \,dt\right]
=\sum_{k=1}^{m}\frac{\<\mathbf{u}_{k}|\rho_{\alpha}|\mathbf{u}_{k}\>}{p_{k}},
\ee
where the second expression follows from a similar calculation to that in~\ref{eqn:uSylvestoru}. 
Also, let 
\be
\varrho_{\alpha}:=\frac{1}{b_{\alpha}}\int_{0}^{\infty}e^{-\frac{t}{2}\rho} \rho_{\alpha} e^{-\frac{t}{2}\rho} \,dt
\ee
so that $\varrho_{\alpha}$ is a state and so that 
\be
\frac{1}{m}\Jamio[\mathcal{E}]=\sum_{\alpha=1}^{s}\frac{b_{\alpha}c_{\alpha}}{m} \varrho_{\alpha}\otimes\nu_{\alpha}.
\ee
Finally, 
\begingroup
\allowdisplaybreaks
\begin{align}
\sum_{\alpha=1}^{s}\frac{b_{\alpha}c_{\alpha}}{m}
&=\sum_{\alpha=1}^{s}\frac{c_{\alpha}}{m}\sum_{k=1}^{m}\frac{\<\mathbf{u}_{k}|\rho_{\alpha}|\mathbf{u}_{k}\>}{p_{k}} \nonumber\\
&=\sum_{k=1}^{m}\frac{1}{mp_{k}}\left\<\mathbf{u}_{k}\left|\left(\sum_{\alpha=1}^{s}c_{\alpha}\rho_{\alpha}\right)\right|\mathbf{u}_{k}\right\> \nonumber\\
&=\sum_{k=1}^{m}\frac{1}{mp_{k}}\<\mathbf{u}_{k}|\rho|\mathbf{u}_{k}\> \nonumber \\
&=\sum_{k=1}^{m}\frac{p_{k}}{mp_{k}}=1,
\end{align}
\endgroup
which proves that $\frac{1}{m}\Jamio[\mathcal{E}]$ is a separable state. 
\eprf

Proposition~\ref{prop:EBandQSOT} together with Theorem~\ref{thm:separableimpliesEB} beg the following two questions. 
\begin{enumerate}[(1)]
\item 
If $\mathcal{E}$ is entanglement breaking, is $\mathcal{E}\star\rho$ always a state, i.e., does $(\mathcal{E},\rho)$ exhibit causal confusion for all $\rho$? Answering this question would extend the results of Ref.~\cite{RAVJSR15} to beyond maximally-mixed input states. 
\end{enumerate}
In case the answer to the first question is no, we have a natural followup question. 
\begin{enumerate}[(1)]
\setcounter{enumi}{1}
\item 
If $\mathcal{E}$ is entanglement breaking \emph{and} if $\mathcal{E}\star\rho$ is a state, is $\mathcal{E}\star\rho$ necessarily separable (or even a PPT state)? 
\end{enumerate} 
The examples studied in this paper allow us to answer both of these questions in the negative, clearly indicating that entanglement-breaking channels \emph{do not} induce temporal correlations that are equivalent to spatial correlations for arbitrary input states (but they \emph{do} for the maximally-mixed state due to Proposition~\ref{prop:EBandQSOT}). 

\bx
\label{ex:EBdniState}
Let $\mathcal{E}(A):=(1-\lambda)A+\lambda\Tr[A]\frac{\mathds{1}_{2}}{2}$ be the depolarizing channel with parameter $\lambda\in[0,\frac{4}{3}]$. If $\rho=\frac{1}{2}(\mathds{1}_{2}+\mathbf{r}\cdot\boldsymbol{\sigma})$ with $\lVert\mathbf{r}\rVert\le 1$, then Proposition~\ref{prop:DepChanEstarRho} shows that $\mathcal{E}\star\rho$ is a state if and only if $\lVert\mathbf{r}\rVert\le R(\lambda)$, 
with $R(\lambda)$ defined by~\eqref{eqn:Rlambda} and $\lambda\in[\frac{2}{3},\frac{4}{3}]$. Therefore, Proposition~\ref{prop:EBandQSOT} tells us that $\mathcal{E}$ is entanglement breaking if and only if $\lambda \in[\frac{2}{3},\frac{4}{3}]$. We now answer the two questions posed. 
\begin{enumerate}[(1)]
\item 
Let $\mathbf{r}$ be a vector satisfying the condition 
\be
\label{eqn:DCstateradius}
R(\lambda)< \lVert \mathbf{r}\rVert < 1. 
\ee
Such a vector $\mathbf{r}$ always exists as long as $\lambda\in[\frac{2}{3},1)\cup(1,\frac{4}{3}]$. It follows that $\mathcal{E}\star\rho$ is \emph{not} positive. Therefore, even if $\mathcal{E}$ is entanglement breaking, it is not necessarily the case that $\mathcal{E}\star\rho$ is a state. As a concrete example, take 
\be
\lambda=\frac{2}{3}
\quad\text{ and }\quad 
\rho=\setlength{\arraycolsep}{6pt}\begin{bmatrix}\frac{3}{4}&0\\0&\frac{1}{4}\end{bmatrix}
\ee
so that $\mathbf{r}=(0,0,\frac{1}{2})$. The associated table of temporal joint expectation values is then given by 
\begin{center}
\noindent
\begin{tabular}{c|cccc}
 & $\sigma_0$ & $\sigma_1$ & $\sigma_2$ & $\sigma_3$ \\
\hline 
$\sigma_0$ & \,\;\;$1$\,\;\; & \,\;\;$0$\,\;\; & \,\;\;$0$\,\;\; & \,\;\;$\frac{1}{6}$\;\; \\ 
$\sigma_1$ & $0$ & $\frac{1}{3}$ & $0$ & $0$ \\
$\sigma_2$ & $0$ & $0$ & $\frac{1}{3}$ & $0$ \\
$\sigma_3$ & $\frac{1}{2}$ & $0$ & $0$ & $\frac{1}{3}$
\end{tabular}
\end{center}
This table cannot be obtained as the spatial joint expectation values of any bipartite state. This answers our first question. Thus, even if $\mathcal{E}$ is entanglement breaking, it need not be the case that $(\mathcal{E},\rho)$ leads to causal confusion for every input state $\rho$. 

\item
As for the second question, we first analyze the original channel $\mathcal{E}$ followed by the transpose map $\mathcal{T}$. This leads to the map
\be
\mathcal{E}^{\mathcal{T}}(A)=(1-\lambda)A^{T}+\lambda\Tr[A]\frac{\mathds{1}_{2}}{2}.
\ee
For this map to be completely positive, $\lambda$ must satisfy $\lambda\in[\frac{2}{3},2]$. A similar analysis to what was done in Section~\ref{sec:depolarizing} leads us to conclude that in order for $\mathcal{E}^{\mathcal{T}}\star\rho$ to be a state, $\mathbf{r}$ must satisfy the constraint 
\be
\label{eqn:TDCstateradius}
\lVert\mathbf{r}\rVert\le\frac{\sqrt{(4-3\lambda)\lambda}}{2-\lambda} =:g(\lambda)
\ee
and $\lambda$ must satisfy $\lambda\in[\frac{2}{3},\frac{4}{3}]$. 
The upper bounds on $\lVert\mathbf{r}\rVert$ from~\eqref{eqn:Rlambda} and~\eqref{eqn:TDCstateradius} are shown in Figure~\ref{fig:DCandTDCmaximumradii}.
\begin{figure}
\includegraphics{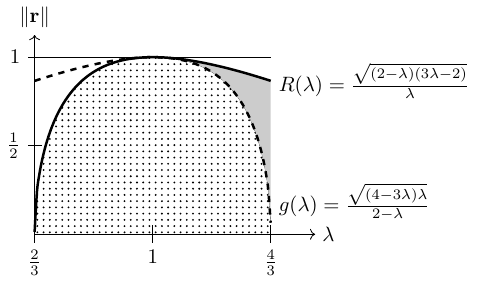}
    \caption{The values of $\lambda$ for which both $\mathcal{E}$ and $\mathcal{E}^{\mathcal{T}}$ are both channels must lie in the range $[\frac{2}{3},\frac{4}{3}]$. In this range of values for $\lambda$, the channel $\mathcal{E}$ (and also $\mathcal{E}^{\mathcal{T}}$) is entanglement breaking. The solid curve is the function 
    $R(\lambda)$ from~\eqref{eqn:Rlambda}. 
    For $\mathbf{r}$ satisfying $\lVert\mathbf{r}\rVert\le R(\lambda)$ and $\rho=\frac{1}{2}(\mathds{1}_{2}+\mathbf{r}\cdot\boldsymbol{\sigma})$, $\mathcal{E}\star\rho$ is a state. The dashed curve is the function 
    $g(\lambda)$ from~\eqref{eqn:TDCstateradius}. 
    For $\mathbf{r}$ satisfying $\lVert\mathbf{r}\rVert\le g(\lambda)$, $\mathcal{E}^{\mathcal{T}}\star\rho$ is a state. By the PPT criterion for two qubits~\cite{peres1996separability,horodecki1996necessary}, if the matrix $\mathcal{E}\star\rho$ is a state, then it is separable if and only if $\mathcal{E}^{\mathcal{T}}\star\rho$ is also a state. The dotted region underneath both curves is therefore the set of values $\lambda$ and $\lVert\mathbf{r}\rVert$ for which $\mathcal{E}\star\rho$ is a separable state. Meanwhile, the shaded region strictly between $R$ and $g$ with $\lambda\in(1,\frac{4}{3}]$ depicts the values of $\lambda$ and $\lVert\mathbf{r}\rVert$ for which $\mathcal{E}\star\rho$ is a state but is not separable, i.e., $\mathcal{E}\star\rho$ is entangled.}
    \label{fig:DCandTDCmaximumradii}
\end{figure}
Now suppose that $\mathbf{r}$ satisfies $\lVert\mathbf{r}\rVert\le R(\lambda)$ so that $\mathcal{E}\star\rho$ is a state. The PPT criterion for two qubits says that a state $\tau$ is separable if and only if its partial transpose is also a state~\cite{peres1996separability,horodecki1996necessary}. This is true for either the partial transpose operation $\mathcal{T}\otimes\id_{\Alg{B}}$ or of $\id_{\Alg{A}}\otimes\mathcal{T}$. But since the partial transpose of $\mathcal{E}\star\rho$ is 
\be
(\id_{\Alg{A}}\otimes\mathcal{T})\left(\mathcal{E}\star\rho\right)
=\mathcal{E}^{\mathcal{T}}\star\rho,
\ee
this means that $\mathcal{E}\star\rho$ is a separable state if and only if $\mathcal{E}^{\mathcal{T}}\star\rho$ is a separable state. Conversely, if $\mathcal{E}\star\rho$ is a state, then $\mathcal{E}\star\rho$ is entangled if and only if $\mathcal{E}^{\mathcal{T}}\star\rho$ is not a state. The set of possible values $(\lambda,\mathbf{r})$ for which $\mathcal{E}\star\rho$ is an entangled state is therefore given by 
\be
\label{eqn:DPentangledregion}
\left\{(\lambda,\mathbf{r})\,:\,
1\le\lambda\le\frac{4}{3} \text{ and } g(\lambda)<\lVert\mathbf{r}\rVert \le R(\lambda)\right\},
\ee
where $R(\lambda)$ and $g(\lambda)$ are defined in~\eqref{eqn:Rlambda} 
and~\eqref{eqn:TDCstateradius}, respectively. The region in~\eqref{eqn:DPentangledregion}
is the gray-shaded region in Figure~\ref{fig:DCandTDCmaximumradii}. Note that $\mathcal{E}$ is still entanglement breaking in this range because $\mathcal{E}$ is entanglement breaking whenever $\lambda\in[\frac{2}{3},\frac{4}{3}]$. This example shows that there is in fact a set of states $\rho$ of positive measure inside the Bloch ball for which $\mathcal{E}$ is entanglement breaking and yet $\mathcal{E}\star\rho$ is an entangled state. As a simple concrete example, take 
\be
\lambda=\frac{4}{3}
\quad\text{ and }\quad
\rho=\setlength{\arraycolsep}{6pt}\begin{bmatrix}\frac{3}{4}&0\\0&\frac{1}{4}\end{bmatrix}
\ee
so that $\mathbf{r}=(0,0,\frac{1}{2})$, which has an associated table of temporal joint expectation values given by 
\begin{center}
\noindent
\begin{tabular}{c|cccc}
 & $\sigma_0$ & $\sigma_1$ & $\sigma_2$ & $\sigma_3$ \\
\hline 
$\sigma_0$ & \,\;\;$1$\,\;\; & \,\;\;$0$\,\;\; & \,\;\;$0$\,\;\; & \,\;\;$-\frac{1}{6}$\;\; \\ 
$\sigma_1$ & $0$ & $-\frac{1}{3}$ & $0$ & $0$ \\
$\sigma_2$ & $0$ & $0$ & $-\frac{1}{3}$ & $0$ \\
$\sigma_3$ & $\frac{1}{2}$ & $0$ & $0$ & $-\frac{1}{3}$
\end{tabular}
\end{center}
This table can also be obtained as the spatial joint expectation values of a bipartite state, but it is necessarily the case that this bipartite state is entangled (and by the PPT criterion, it is not a PPT state either).
\end{enumerate}
\ex

\subsection{PPT Channels and PPT states over time}
\label{sec:PPT}

Let $\Alg{A}=\matr_{m}$ and $\Alg{B}=\matr_{n}$. A map $\mathcal{E}:\Alg{A}\to\Alg{B}$ is called \define{completely copositive} iff $\mathcal{E}\circ\mathcal{T}$ is completely positive, i.e., $\Choi[\mathcal{E}\circ\mathcal{T}]$ is positive semidefinite (this is equivalent to $\mathcal{E}^{\mathcal{T}}$ being completely positive). A map $\mathcal{E}$ is called a \define{PPT map} iff $\mathcal{E}$ is completely copositive. A quantum channel $\mathcal{E}$ that is also completely copositive is called a \define{PPT channel}~\cite{ChMuWo2019}. 
In this section, we will go beyond entanglement-breaking channels and relate PPT channels to spatiotemporal correlations. The main result is Theorem~\ref{thm:PPTEQSOTstate}, which says that if a channel $\mathcal{E}$ and an initial state $\rho$ exhibit causal confusion, i.e., $\mathcal{E}\star\rho$ is a state, then it is necessarily the case that $\mathcal{E}$ is a PPT channel. This implies Corollary~\ref{cor:causalconfusioniffPPT}, which says that a channel $\mathcal{E}$ admits causal confusion if and only if $\mathcal{E}$ is PPT. 

\blem
A map $\mathcal{E}:\Alg{A}\to\Alg{B}$ is a PPT channel if and only if $\frac{1}{m}\Choi[\mathcal{E}]=(\id_{\Alg{A}}\otimes\mathcal{E})\big(|\Omega\>\<\Omega|\big)$ is a PPT state, i.e., $(\id_{\Alg{A}}\otimes\mathcal{E})\big(|\Omega\>\<\Omega|\big)$ and 
$(\mathcal{T}\otimes\mathcal{E})\big(|\Omega\>\<\Omega|\big)$
are states. 
\elem

Although this result is well-known, we prove it because it directly relates to states over time in a way that is analogous to Proposition~\ref{prop:EBandQSOT} for entanglement-breaking channels. 

\bprf
We first note that 
\begin{align}
(\mathcal{T}\otimes\id_{\Alg{B}})\left(\frac{1}{m}\Choi[\mathcal{E}]\right)
&=(\mathcal{T}\otimes\mathcal{E})\big(|\Omega\>\<\Omega|\big)\nonumber \\
&=(\id_{\Alg{A}}\otimes\mathcal{E})(\mathcal{T}\otimes\id_{\Alg{B}})\big(|\Omega\>\<\Omega|\big) \nonumber \\
&=\frac{1}{m}\Jamio[\mathcal{E}]
\end{align}
due to~\eqref{eqn:TonChoi}. 
This shows that $\frac{1}{m}\Choi[\mathcal{E}]$ is a PPT state if and only if $\frac{1}{m}\Choi[\mathcal{E}]$ and $\frac{1}{m}\Jamio[\mathcal{E}]$ are both states. On the other hand, $\mathcal{E}$ is a PPT channel if and only if $\frac{1}{m}\Choi[\mathcal{E}]$ is a state and 
\be
\frac{1}{m}\Choi[\mathcal{E}\circ\mathcal{T}]=\big(\id_{\Alg{A}}\otimes (\mathcal{E} \circ\mathcal{T})\big)\big(|\Omega\>\<\Omega|\big) =\frac{1}{m}\Jamio[\mathcal{E}]
\ee
is a state. These two conditions are exactly the same, therefore proving the claim. 
\eprf

An immediate consequence of this standard fact is the following. 

\bn
\label{prop:PPTQSOT}
A map $\mathcal{E}:\Alg{A}\to\Alg{B}$ is a PPT channel if and only if $\mathcal{E}\star\frac{\mathds{1}_{m}}{m}=\frac{1}{m}\Jamio[\mathcal{E}]$ is a PPT state. Alternatively, a channel $\mathcal{E}:\Alg{A}\to\Alg{B}$ is PPT if and only if $\mathcal{E}\star\frac{\mathds{1}_{m}}{m}=\frac{1}{m}\Jamio[\mathcal{E}]$ is a state. 
\en

We now extend Theorem~\ref{thm:separableimpliesEB} from Ref.~\cite{SongParzygnat2025} regarding separable states over time and entanglement-breaking channels to \emph{positive} states over time and PPT channels. 

\bt
\label{thm:PPTEQSOTstate}
Let $\Alg{A}=\matr_{m}$,  $\Alg{B}=\matr_{n}$, and let $\mathcal{E}:\Alg{A}\to\Alg{B}$ be a quantum channel. If there exists a full rank state $\rho\in\Alg{A}$ such that $(\mathcal{E},\rho)$ exhibits causal confusion, then $\mathcal{E}$ is a PPT channel. 
\et

Notice that we only assume $\mathcal{E}\star\rho$ is positive in this statement, and not that $\mathcal{E}\star\rho$ is a PPT state. In this sense, this theorem differs a bit from Theorem~\ref{thm:separableimpliesEB}. 

\bprf[Proof of Theorem~\ref{thm:PPTEQSOTstate}]
By the assumption that $\mathcal{E}\star\rho$ is positive semidefinite, Lemma~\ref{lem:BhatiaRosenberg} implies
\be
\label{eqn:solvingforJE}
\Jamio[\mathcal{E}]=\int_{0}^{\infty}\left(e^{-\frac{t}{2}\rho\otimes\mathds{1}_{n}}\right)(\mathcal{E}\star\rho)\left(e^{-\frac{t}{2}\rho\otimes\mathds{1}_{n}}\right)\,dt
\ee
is positive semidefinite. Since the trace of this is $\Tr\big[\Jamio[\mathcal{E}]\big]=m$, it follows that $\frac{1}{m}\Jamio[\mathcal{E}]$ is a state. 
Thus, Proposition~\ref{prop:PPTQSOT} implies that $\mathcal{E}$ is a PPT channel. 
\eprf

The proof of Theorem~\ref{thm:PPTEQSOTstate} can also be used to prove the following, which is equivalent to the partial transpose theorem from Ref.~\cite{FuLi25}, and also implies Theorem~\ref{thm:PPTEQSOTstate}. 

\bt
\label{thm:Eisalwayscopositive}
Let $\Alg{A}=\matr_{m}$,  $\Alg{B}=\matr_{n}$, and let $\tau\in\Alg{A}\otimes\Alg{B}$ be a bipartite state. Suppose that $\Tr_{\Alg{B}}[\tau]$ is full rank. Then there exists a completely copositive trace-preserving map $\mathcal{E}:\Alg{A}\to\Alg{B}$ such that $\mathcal{E}\star\rho=\tau$.
\et

The map $\mathcal{E}$ in Theorem~\ref{thm:Eisalwayscopositive} is sometimes referred to as a \emph{causal map}~\cite{RAVJSR15}, \emph{temporal map}~\cite{SongParzygnat2025}, \emph{forward pseudo-channel}~\cite{SNREG23}, or a \emph{conditional}~\cite{PaQPL21,SongParzygnat2025} (the latter based on analogy with conditional distributions and conditioning from classical probability and their connection to Bayes' rule~\cite{FuPa22a,ChJa18,SongParzygnat2025}). Prior to this work, it was shown that $\mathcal{E}$ is positive and trace preserving in Ref.~\cite{SongParzygnat2025}, while Ref.~\cite{SNREG23} proved it was always Hermitian and trace preserving. 

\bprf
[Proof of Theorem~\ref{thm:Eisalwayscopositive}]
Let $\rho:=\Tr_{\Alg{B}}[\tau]$. First note that the specification of a map $\mathcal{E}:\Alg{A}\to\Alg{B}$ is equivalent to the specification of $\Jamio[\mathcal{E}]$ because 
\be
\Alg{A}\ni A\mapsto \mathcal{E}(A)=\Tr_{\Alg{A}}\big[\Jamio[\mathcal{E}] (A\otimes\mathds{1}_{n})\big]
\ee
establishes an isomorphism between linear maps $\mathcal{E}:\Alg{A}\to\Alg{B}$ and matrices $\Jamio[\mathcal{E}]\in\Alg{A}\otimes\Alg{B}$. Hence, it suffices to define $\mathcal{E}$ by specifying $\Jamio[\mathcal{E}]$, which is done using~\eqref{eqn:solvingforJE}. By Lemma~\ref{lem:BhatiaRosenberg}, this means $\Jamio[\mathcal{E}]$ is positive semidefinite. But this yields 
\be
\Jamio[\mathcal{E}]=\Choi[\mathcal{E}\circ\mathcal{T}]\ge0,
\ee
which exactly means that $\mathcal{E}$ is completely copositive. 
\eprf

\begin{figure*}[t]
\includegraphics{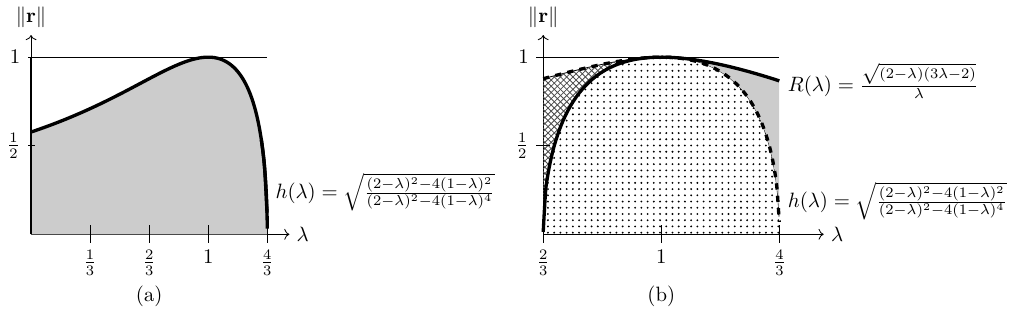}
\caption{(a) If $\mathcal{E}$ is the depolarizing channel parameterized by $\lambda\in[0,\frac{4}{3}]$, then for each $\lambda\in(0,\frac{4}{3}]$, the set of states $\rho=\frac{1}{2}(\mathds{1}_{2}+\mathbf{r}\cdot\boldsymbol{\sigma})$ for which $(\mathcal{E},\rho)$ has a Bayesian inverse corresponds to a ball whose radius $\lVert\mathbf{r}\rVert$ is given by 
$h(\lambda)$, as defined in~\eqref{eqn:DepChanrconstraint}. 
When $\lambda=0$, this is the entire Bloch ball. Thus, the shaded region, together with the vertical line $\lambda=0$ with $0\le\lVert\mathbf{r}\rVert\le1$, depicts the values of $\lambda$ and $\lVert\mathbf{r}\rVert$ such that $(\mathcal{E},\rho)$ is Bayesian invertible. This result was first obtained in Ref.~\cite{TiFuWu2026}. (b) The function $h(\lambda)$ (dashed) is plotted together with the curve $R(\lambda)$ (solid) from Figure~\ref{fig:DCandTDCmaximumradii}, which specifies the maximum $\lVert\mathbf{r}\rVert$ for which $\mathcal{E}\star\rho$ is a bipartite state. We note that despite the striking similarity, $h(\lambda)$ does \emph{not} equal the function $g(\lambda)$ obtained in Figure~\ref{fig:DCandTDCmaximumradii} (which specifies the maximum $\lVert\mathbf{r}\rVert$ such that $\mathcal{E}^{\mathcal{T}}\star\rho$ is a bipartite state), and we illustrate this in Figure~\ref{fig:hlambdaVglambda}. The dotted region depicts the values of $\lambda$ and $\lVert\mathbf{r}\rVert$ such that $(\mathcal{E},\rho)$ is Bayesian invertible \emph{and} $\mathcal{E}\star\rho$ is a state. The shaded region depicts the values of $\lambda$ and $\lVert\mathbf{r}\rVert$ such that $\mathcal{E}\star\rho$ is a state but $(\mathcal{E},\rho)$ is \emph{not} Bayesian invertible. And finally, the crosshatched region depicts the values of $\lambda$ and $\lVert\mathbf{r}\rVert$ such that $(\mathcal{E},\rho)$ is Bayesian invertible, but $\mathcal{E}\star\rho$ is \emph{not} a state. Note that in all of these examples, $\mathcal{E}$ is entanglement breaking.} 
\label{fig:DepChanBayesianInverse}
\end{figure*}

\bc
\label{cor:causalconfusioniffPPT}
A quantum channel $\mathcal{E}$ admits causal confusion if and only if $\mathcal{E}$ is a PPT channel. 
\ec

\bprf
Suppose a quantum channel $\mathcal{E}$ admits causal confusion. This means there exists a state $\rho$ such that $(\mathcal{E},\rho)$ exhibits causal confusion. Hence, Theorem~\ref{thm:PPTEQSOTstate} says $\mathcal{E}$ is a PPT channel. Conversely, $\mathcal{E}:\matr_{m}\to\matr_{n}$ is a PPT channel if and only if $\mathcal{E}\star\frac{\mathds{1}_{m}}{m}$ is a state by Proposition~\ref{prop:PPTQSOT}. Hence, $\mathcal{E}$ admits causal confusion. 
\eprf

These connections between quantum states over time and PPT channels allows us to formulate a new equivalent condition to the PPT${}^2$ conjecture~\cite{ChMuWo2019}. 

\bt
Set $\Alg{A}=\matr_{m}$. 
The following two statements are equivalent: 
\begin{enumerate}[(a)]
\item 
Given any PPT channel $\mathcal{E}:\Alg{A}\to\Alg{A}$, the composite $\mathcal{E}^{2}$ is entanglement breaking. 
\item Given any channel $\mathcal{E}:\Alg{A}\to\Alg{A}$ for which there exists a state $\rho$ such that $\mathcal{E}\star\rho$ is a state, $\mathcal{E}^{2}\star\frac{\mathds{1}_{m}}{m}$ is a separable state. 
\end{enumerate}
\et

Although this is an immediate consequence mathematically, it provides a new connection between PPT channels, entanglement-breaking channels, spatiotemporal correlations, pseudodensity matrices, and quantum states over time, which may be a useful perspective. 

\subsection{PPT channels and Bayesian inverses}
\label{sec:Bayesianinverses}

Given a channel $\mathcal{E}:\Alg{A}\to\Alg{B}$ and a state $\rho\in\Alg{A}$, a \define{Bayesian inverse} of $(\mathcal{E},\rho)$ is a channel $\mathcal{E}^{\star}_{\rho}:\Alg{B}\to\Alg{A}$ such that 
\be
\label{eqn:BayesianInverse}
\mathcal{E}\star\rho=\gamma\big(\mathcal{E}^{\star}_{\rho}\star\mathcal{E}(\rho)\big),
\ee
where $\gamma$ is the swap isomorphism that is uniquely determined by 
\be
\gamma(B\otimes A):=A\otimes B
\ee
for all $A\in\Alg{A}$ and $B\in\Alg{B}$. This definition was first proposed in Ref.~\cite{FuPa22a}, and the interpretation in terms of causal confusion between a forward and reverse temporal scenarios (with the reverse temporal scenario being that Bob measures first followed by Alice) was provided in Ref.~\cite{PaFu24TSC}, with followup work in Ref.~\cite{TiFuWu2026}. Bayesian inverses do not always exist~\cite{PaFu24TSC} (although they always exist as mathematical maps, called \emph{Bayes maps} in Ref.~\cite{FuPa22a} and \emph{reverse pseudo-channels} in Ref.~\cite{SNREG23}, which need not be quantum channels). 
Before stating several facts related to Bayesian inverses and PPT channels, we first recall a definition and a lemma for calculations. 

First, given a linear map $\mathcal{E}:\Alg{A}\to\Alg{B}$, its  \define{Hilbert--Schmidt adjoint} is the unique linear map $\mathcal{E}^*:\Alg{B}\to\Alg{A}$ satisfying the condition 
\be
\Tr\big[\mathcal{E}(A)^{\dag} B\big]=\Tr\big[A^{\dag}\mathcal{E}^*(B)\big]
\ee
for all $A\in\Alg{A}$ and $B\in\Alg{B}$. 
The Hilbert--Schmidt adjoint is not usually the Bayesian inverse of a channel, but they are closely related, as the following lemma describes. We let $\Alg{A}=\matr_{m}$ and $\Alg{B}=\matr_{n}$ throughout this section. 

\blem
\label{lem:whenBinverseexists}
Let $\mathcal{E}:\Alg{A}\to\Alg{B}$ be a quantum channel, and let $\rho\in\Alg{A}$ be a state such that $\mathcal{E}(\rho)$ is a full rank state. 
For each $t\in\R$, set  
\be
G(t):=e^{-t \mathcal{E}(\rho)^{T}} \otimes\mathds{1}_{n}.
\ee
A Bayesian inverse $\mathcal{E}^{\star}$ to $(\mathcal{E},\rho)$ exists if and only if 
\be
\label{eqn:ChoiOfBInverseDC}
\int_{0}^{\infty}G(t)\big\{\mathds{1}_{n}\otimes\rho,\Choi[\mathcal{E}^{*}]\big\}G(t)\,dt
\ee
is positive semidefinite. In fact, the expression in~\eqref{eqn:ChoiOfBInverseDC} is equal to the Choi matrix $\Choi[\mathcal{E}^{\star}_{\rho}]$ of $\mathcal{E}^{\star}_{\rho}$ (even if $\mathcal{E}^{\star}_{\rho}$ is only defined as a linear map). 
\elem

\bprf
We first note that~\eqref{eqn:BayesianInverse} yields 
\begin{align}
\big\{\mathcal{E}(\rho)\otimes\mathds{1}_{n},\Jamio[\mathcal{E}^{\star}_{\rho}]\big\}
&=\gamma\Big(\big\{\rho\otimes \mathds{1}_{n},\Jamio[\mathcal{E}]\big\}\Big) \nonumber \\
&=\big\{\mathds{1}_{n}\otimes\rho,\Jamio[\mathcal{E}^*]\big\}, 
\label{eqn:DepChanBIcondition}
\end{align}
(the fact that $\gamma\big(\Jamio[\mathcal{E}]\big)=\Jamio[\mathcal{E}^*]$ is proved in~\cite[Appendix A Lemma 2]{FuPa22a}). Applying the partial transpose $\mathcal{T}\otimes\id$ to~\eqref{eqn:DepChanBIcondition} and using~\eqref{eqn:PTofJamiol} gives
\be
\big\{\mathcal{E}(\rho)^{T}\otimes\mathds{1}_{n},\Choi[\mathcal{E}^{\star}_{\rho}]\big\}=\big\{\mathds{1}_{n}\otimes\rho,\Choi[\mathcal{E}^{*}]\big\}.
\ee
Using Lemma~\ref{lem:BhatiaRosenberg} then yields
\be
\label{eqn:ChoiOfBInverseDC2}
\Choi[\mathcal{E}^{\star}_{\rho}]=\int_{0}^{\infty}G(t)\big\{\mathds{1}_{n}\otimes\rho,\Choi[\mathcal{E}^{*}]\big\}G(t)\,dt.
\ee
By Choi's isomorphism theorem, $\mathcal{E}^{\star}_{\rho}$ exists as a linear map, and moreover $\Choi[\mathcal{E}^{\star}_{\rho}]$ is positive if and only if $\mathcal{E}^{\star}_{\rho}$ is a quantum channel (a Bayesian inverse is automatically trace-preserving). 
\eprf

\bn
\label{prop:Binvexistsmaxmix}
Using the same notation from Lemma~\ref{lem:whenBinverseexists}, set $\rho=\frac{\mathds{1}_{m}}{m}$, and suppose that $\mathcal{E}(\rho)$ is a full rank state. Then $(\mathcal{E},\rho)$ has a Bayesian inverse.  
\en

\bprf
Lemma~\ref{lem:whenBinverseexists} applied to this case yields
\be
\Choi[\mathcal{E}^{\star}_{\rho}]=2\int_{0}^{\infty}G(t)\Choi[\mathcal{E}^{*}]G(t)^{\dag}\,dt, 
\ee
which is positive because $\Choi[\mathcal{E}^{*}]$ is positive. 
\eprf

Operationally, Proposition~\ref{prop:Binvexistsmaxmix} says that the joint expectation values associated with the temporal scenario from Alice to Bob via $(\mathcal{E},\rho)$ are indistinguishable from those obtained by the temporal scenario from Bob to Alice via $(\mathcal{E}^{\star}_{\rho},\mathcal{E}(\rho))$ whenever $\rho$ is maximally mixed. More general situations for such operational time-reversal symmetry are discussed in Refs.~\cite{PaFu24TSC,TiFuWu2026}. Instead, in what follows, we focus on the connections to PPT channels. 

\bt
\label{thm:PPTBayesianinverse}
Let $\Alg{A}=\matr_{m}$ and $\Alg{B}=\matr_{n}$. 
Let $\mathcal{E}:\Alg{A}\to\Alg{B}$ be a PPT channel, set $\rho=\frac{\mathds{1}_{m}}{m}$, and suppose that $\mathcal{E}(\rho)$ is a full rank state. Then $(\mathcal{E},\rho)$ has a Bayesian inverse that is also a PPT channel. 
\et

\bprf
Proposition~\ref{prop:Binvexistsmaxmix} proves that $(\mathcal{E},\rho)$ has a Bayesian inverse, so that $\mathcal{E}\star\rho=\gamma(\mathcal{E}^{\star}_{\rho}\star\mathcal{E}(\rho))$. Since $\mathcal{E}$ is a PPT channel, $\tau:=\mathcal{E}\star\rho$ is a PPT state by Proposition~\ref{prop:PPTQSOT}. Therefore, $\gamma(\tau)=\mathcal{E}^{\star}_{\rho}\star\mathcal{E}(\rho)\in\Alg{B}\otimes\Alg{A}$ is also a PPT state because the full transpose map---not the partial transpose map---is a positive and trace-preserving map (even though it is not a quantum channel). 
By Theorem~\ref{thm:PPTEQSOTstate}, $\mathcal{E}^{\star}_{\rho}$ is a PPT channel. 
\eprf

We make a few remarks about the assumption that $\rho$ is maximally mixed in Theorem~\ref{thm:PPTBayesianinverse}. 

\begin{figure}
\includegraphics{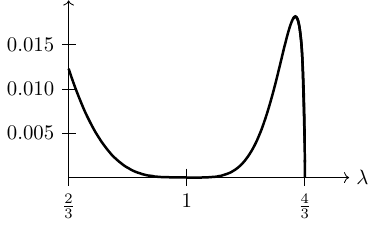}
\caption{Despite their striking similarity from Figure~\ref{fig:DCandTDCmaximumradii} and Figure~\ref{fig:DepChanBayesianInverse} (b), the function $h(\lambda)$ is different from $g(\lambda)$. This figure plots $h(\lambda)-g(\lambda)$. Note that $h(\lambda)\ge g(\lambda)$ for all $\lambda\in[\frac{2}{3},\frac{4}{3}]$. What this means for the depolarizing channel is that if $(\mathcal{E},\rho)$ is Bayesian invertible, then it is necessarily the case that $\mathcal{E}^{\mathcal{T}}\star\rho$ is a state, i.e., $(\mathcal{E}^{\mathcal{T}},\rho)$ exhibits causal confusion.}
\label{fig:hlambdaVglambda}
\end{figure}

\br
\label{rmk:BayesianInversesAndPPT}
Regarding Theorem~\ref{thm:PPTBayesianinverse}, if $\mathcal{E}:\Alg{A}\to\Alg{B}$ is a PPT channel (or even if $\mathcal{E}$ is an entanglement-breaking channel) and the state $\rho\in\Alg{A}$ is \emph{not} the maximally mixed state, then it is \emph{not} necessarily the case that $(\mathcal{E},\rho)$ has a Bayesian inverse. To see an explicit example, we will look at the depolarizing channel with parameter $\lambda$ as in Section~\ref{sec:depolarizing} and Example~\ref{ex:EBdniState}. In this case, a Bayesian inverse for $(\mathcal{E},\rho)$ exists if and only if 
\be
\label{eqn:DepChanrconstraint}
\lVert\mathbf{r}\rVert\le \sqrt{\frac{(2-\lambda)^2-4(1-\lambda)^2}{(2-\lambda)^2-4(1-\lambda)^4}} =:h(\lambda)
\ee
for all $\lambda\in(0,\frac{4}{3}]$, and for all $\mathbf{r}$ when $\lambda=0$, which is shown in Figure~\ref{fig:DepChanBayesianInverse}. 
\begin{figure}
\includegraphics{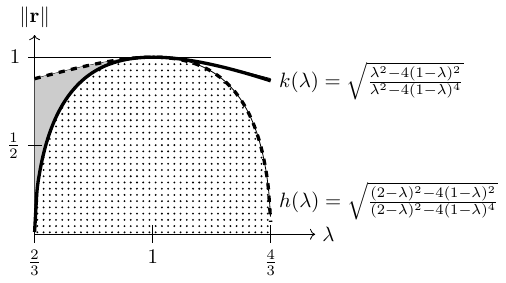}
\caption{The dotted region depicts values of $(\lambda,\mathbf{r})$ for which $(\mathcal{E},\rho)$ admits a Bayesian inverse $\mathcal{E}^{\star}_{\rho}$ that is also entanglement breaking. The original $\mathcal{E}$ is the depolarizing channel from Example~\ref{ex:EBdniState}, which is entanglement breaking for all $\lambda\in[\frac{2}{3},\frac{4}{3}]$. The shaded region depicts values of $(\lambda,\mathbf{r})$ for which $(\mathcal{E},\rho)$ admits a Bayesian inverse $\mathcal{E}^{\star}_{\rho}$ which is \emph{not} entanglement breaking. Despite the similarity, $k(\lambda)$ does not coincide with the function $R(\lambda)$ from Figure~\ref{fig:DCandTDCmaximumradii}. In fact, $k(\lambda)\ge R(\lambda)$ for all $\lambda\in[\frac{2}{3},\frac{4}{3}]$, which means that if $(\mathcal{E},\rho)$ admits an entanglement-breaking Bayesian inverse, then $\mathcal{E}\star\rho$ is guaranteed to be a state, i.e., $(\mathcal{E},\rho)$ exhibits causal confusion.}
\label{fig:BInotnecEB}
\end{figure}
This result, though not the explicit closed-form expression $h(\lambda)$, was also obtained in Ref.~\cite{TiFuWu2026}. 
This follows from Lemma~\ref{lem:whenBinverseexists}
and the fact that $\mathcal{E}^*=\mathcal{E}$. 
Requiring that $\Choi[\mathcal{E}^{\star}_{\rho}]$ be positive 
yields~\eqref{eqn:DepChanrconstraint}. Therefore, as long as $\lambda\in(0,\frac{4}{3}]$, if we pick $\mathbf{r}$ that violates the constraint~\eqref{eqn:DepChanrconstraint}, then $(\mathcal{E},\rho)$ will not admit a Bayesian inverse. Moreover, if $\lambda\in[\frac{2}{3},1)\cup(1,\frac{4}{3}]$, if we pick $\mathbf{r}$ that violates the constraint~\eqref{eqn:DepChanrconstraint}, then $(\mathcal{E},\rho)$ will not admit a Bayesian inverse despite the fact that $\mathcal{E}$ is entanglement breaking.
\er

\br
Regarding Theorem~\ref{thm:PPTBayesianinverse}, if $\mathcal{E}:\Alg{A}\to\Alg{B}$ is a PPT channel (or even if $\mathcal{E}$ is an entanglement-breaking channel) and $\rho\in\Alg{A}$ is \emph{not} the maximally mixed state but nevertheless $(\mathcal{E},\rho)$ \emph{has} a Bayesian inverse, then it is \emph{not} necessarily the case that the Bayesian inverse is a PPT channel. Again, an example can already be found with the depolarizing channel from Example~\ref{ex:EBdniState} and Remark~\ref{rmk:BayesianInversesAndPPT}. One can show using~\eqref{eqn:ChoiOfBInverseDC} and Proposition~\ref{prop:EBandQSOT} that in order for the Bayesian inverse of $(\mathcal{E},\rho)$ to be entanglement breaking, $\lambda$ and $\mathbf{r}$ must satisfy
\be
\lVert\mathbf{r}\rVert\le\sqrt{\frac{\lambda^2-4(1-\lambda)^2}{\lambda^2-4(1-\lambda)^4}}=:k(\lambda)
\ee
See Figure~\ref{fig:BInotnecEB} for details. 
Therefore, choosing $\lambda\in[\frac{2}{3},1)$ and $\mathbf{r}$ to satisfy 
\be
k(\lambda)<\lVert\mathbf{r}\rVert\le h(\lambda)
\ee
would provide an example of a process $(\mathcal{E},\rho)$ where $\mathcal{E}$ is entanglement breaking, a Bayesian inverse exists, but that Bayesian inverse is not entanglement breaking (in fact, not even PPT). 
\er

\section{Characterization of qubit causal confusion}
\label{sec:generalqubit}

In this section, we answer when causal confusion occurs for a general qubit quantum channel of the form~\cite{RSW02,WPVF25} 
\be
\label{eq:generalqubitcannel}
\mathcal{E}(A) = w_0\mathds{1} + (w_0\mathbf{t} + \mathbf{T}(\mathbf{w}))\cdot\boldsymbol{\sigma},
\ee 
where 
\be
A=\sum_{j=0}^{3}w_j\sigma_{j} 
\quad\text{ and }\quad
w_{j}=\frac{1}{2}\Tr[A\sigma_{j}] \;\; \forall j, 
\ee
\be
\textbf{T} = \begin{bmatrix}
    \lambda_1 & 0 & 0\\
    0 & \lambda_2 & 0\\
    0 & 0 & \lambda_3\\ 
\end{bmatrix} 
\quad\text{ and }\quad 
\mathbf{t} = \begin{bmatrix}
    t_1\\
    t_2\\
    t_3\\
\end{bmatrix},
\ee
noting that these parameters must lie in the range $\lambda_1,\lambda_2,\lambda_3,t_1,t_2,t_3\in[-1,1]$ (there are additional constraints---see Appendix~\ref{sec:Ruskai} for more details).

\bt
\label{thm:generalqubitchannelcausalconfusion}
Given a general qubit quantum channel $\mathcal{E}$ of the form~\eqref{eq:generalqubitcannel} described by $\boldsymbol{\lambda}=(\lambda_1,\lambda_2,\lambda_3)$ and $\mathbf{t}=(t_1,t_2,t_3)$, the channel $\mathcal{E}$ admits causal confusion if and only if $\mathcal{E}$ is an entanglement-breaking channel, i.e., 
\be
\label{eqn:generalchannelconstraint1}
(|\lambda_1|+|\lambda_2|)^2\leq(1-|\lambda_3|)^2 - t_3^2
\ee
and
\begin{align}
&4\left(\lambda_1^2(t_1^2+\lambda_2^2)+\lambda_2^2(t_2^2+\lambda_3^2)+\lambda_3^2(t_3^2+\lambda_1^2)+2|\lambda_1\lambda_2\lambda_3|\right) \nonumber \\
&\qquad\le
    (1-\lVert\boldsymbol{\lambda}\rVert^2-\lVert\mathbf{t}\rVert^2)^2.
    \label{eqn:generalchannelconstraint2}
\end{align}
\et

After the proof of this theorem, we illustrate some examples of the set of all possible $\boldsymbol{\lambda}$ for different fixed $\mathbf{t}$ where $\mathcal{E}$ admits causal confusion. 

\bprf
[Proof of Theorem~\ref{thm:generalqubitchannelcausalconfusion}]
The Choi matrix of $\mathcal{E}$ is 
\be
\frac{1}{2}\begin{bmatrix}
    1+t_3+\lambda_3 & t_1-i t_2 & 0 & \lambda_1+\lambda_2\\
    t_1+i t_2 & 1-t_3-\lambda_3 & \lambda_1-\lambda_2 & 0\\
    0 & \lambda_1-\lambda_2 & 1+t_3-\lambda_3 & t_1 -i t_2\\
    \lambda_1+\lambda_2 & 0 & t_1 +i t_2 & 1-t_3+\lambda_3\\
\end{bmatrix}
\ee
and the Jamio{\l}kowski matrix of $\mathcal{E}$ is 
\be
\frac{1}{2}\begin{bmatrix}
    1+t_3+\lambda_3 & t_1-i t_2 & 0 & \lambda_1-\lambda_2 \\
    t_1+i t_2 & 1-t_3-\lambda_3 & \lambda_1+\lambda_2 & 0 \\
    0 & \lambda_1+\lambda_2 & 1+t_3-\lambda_3 & t_1 -i t_2\\
    \lambda_1-\lambda_2 & 0 & t_1 +i t_2 & 1-t_3+\lambda_3\\
\end{bmatrix}.
\ee
Notice how the Jamio{\l}kowski matrix can be attained by simply multiplying all instances of $\lambda_2$ by $-1$.

By Corollary~\ref{cor:causalconfusioniffPPT}, a channel $\mathcal{E}$ admits causal confusion if and only if $\mathcal{E}$ is a PPT channel. But for qubits, a channel is PPT if and only if it is an entanglement-breaking channel by the PPT criterion~\cite{peres1996separability,horodecki1996necessary}. Hence, in the case of qubits, a channel $\mathcal{E}$ admits causal confusion if and only if it is entanglement breaking. This is equivalent to positivity of $\Choi[\mathcal{E}]$ and $\Jamio[\mathcal{E}]$. 

Therefore, we first prove that positivity of $\Choi[\mathcal{E}]$ and $\Jamio[\mathcal{E}]$ imply the constraints~\eqref{eqn:generalchannelconstraint1} and~\eqref{eqn:generalchannelconstraint2}. 
If $\Choi[\mathcal{E}]\ge0$, then its $1\times 1$ principal minors must be nonnegative, which implies $|t_3|+|\lambda_3|\leq1$. Thus, by \cite[Corollary 2]{RSW02}, we know that the constraints for $\mathcal{E}$ to be a channel are
\be
(\lambda_1+\lambda_2)^2 \leq (1+\lambda_3)^2-t_3^2,
\ee
\be
(\lambda_1-\lambda_2)^2 \leq (1-\lambda_3)^2 - t_3^2,
\ee 
and 
\begin{align}
&4\big(\lambda_1^2(t_1^2+\lambda_2^2)+\lambda_2^2(t_2^2+\lambda_3^2)+\lambda_3^2(t_3^2+\lambda_1^2)-2\lambda_1\lambda_2\lambda_3\big) \nonumber \\
&\qquad\le
    (1-\lVert\boldsymbol{\lambda}\rVert^2-\lVert\mathbf{t}\rVert^2)^2.
\end{align}
Keeping in mind that the Jamio{\l}kowski matrix of $\mathcal{E}$ can be obtained by reversing the signs in front of all the $\lambda_2$, a similar argument for positivity of $\Jamio[\mathcal{E}]=\Choi[\mathcal{E}^{\mathcal{T}}]$ leads to 
\be
(\lambda_1-\lambda_2)^2 \leq (1+\lambda_3)^2-t_3^2,
\ee
\be
(\lambda_1+\lambda_2)^2 \leq (1-\lambda_3)^2 - t_3^2,
\ee 
and 
\begin{align}
&4\big(\lambda_1^2(t_1^2+\lambda_2^2)+\lambda_2^2(t_2^2+\lambda_3^2)+\lambda_3^2(t_3^2+\lambda_1^2)+2\lambda_1\lambda_2\lambda_3\big) \nonumber \\
&\qquad\le
    (1-\lVert\boldsymbol{\lambda}\rVert^2-\lVert\mathbf{t}\rVert^2)^2.
\end{align}
Combining the above constraints inequalities yields exactly~\eqref{eqn:generalchannelconstraint1} and~\eqref{eqn:generalchannelconstraint2}, as needed. 
\eprf

\begin{figure*}
\begin{tabular}{cccc}
\;\;\includegraphics[width=3.5cm]{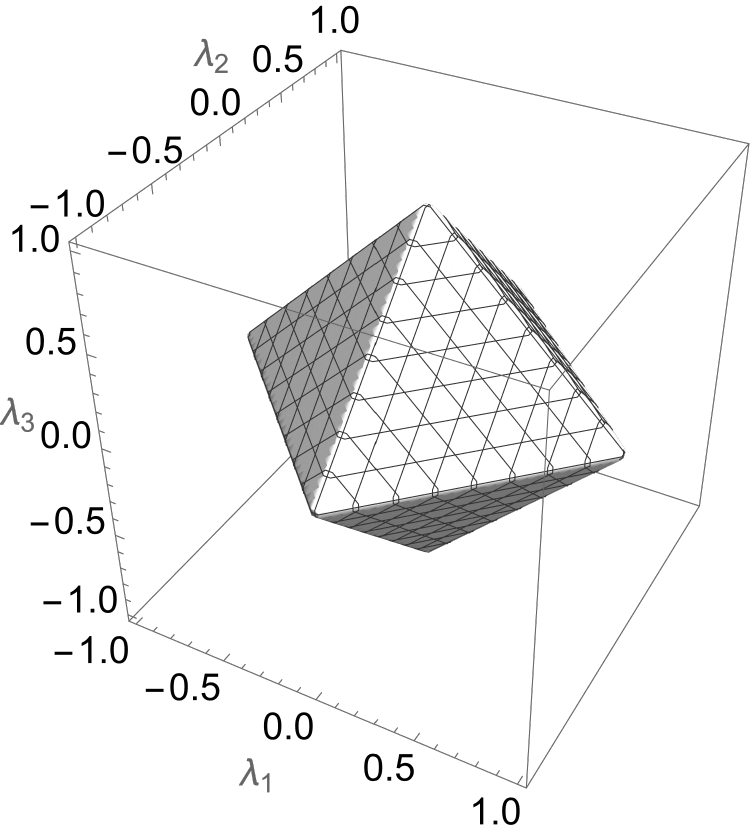}\;\;
& 
\;\;\includegraphics[width=3.5cm]{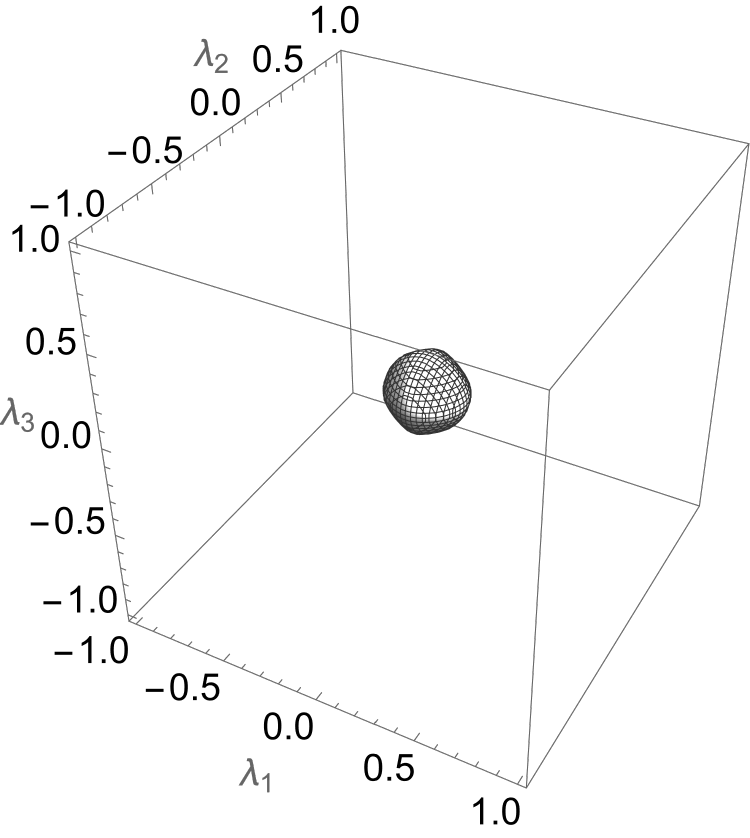} \;\;
& 
\;\;\includegraphics[width=3.5cm]{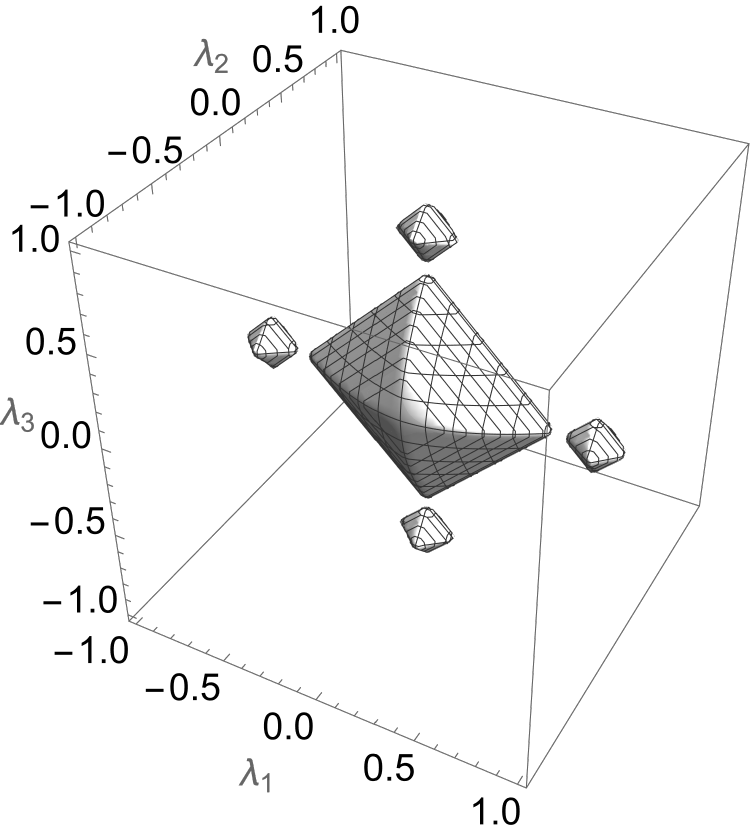} \;\;
& 
\;\;\includegraphics[width=3.5cm]{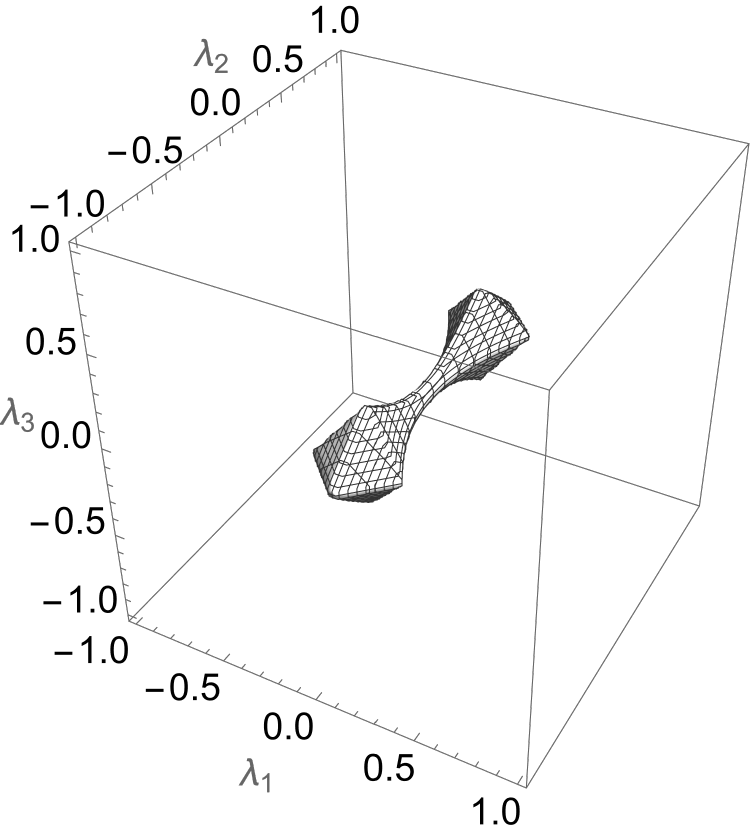} \;\;\\
$t_1=0,t_2=0,t_3=0$ 
&
$t_1=0.5,t_2=0.5,t_3=0.5$
& 
$t_1=0,t_2=0.75,t_3=0$ 
&
$t_1=0.9,t_2=0,t_3=0.5$ 
\end{tabular}
\caption{By setting $(t_1,t_2,t_3)$ to be certain fixed values, the regions in $(\lambda_1,\lambda_2,\lambda_3)$ space obtained from the inequalities~\eqref{eqn:generalchannelconstraint1} and~\eqref{eqn:generalchannelconstraint2} can be visualized. Such values of $t_1,t_2,t_3,\lambda_1,\lambda_2,\lambda_3$ are such that the channel $\mathcal{E}$ admits causal confusion, which is equivalent to $\mathcal{E}$ being an entanglement-breaking channel since the system is that of a qubit. These figures were made in Mathematica~\cite{Mathematica2024}.}
\label{fig:RuskaiImages}
\end{figure*}

\bx
We mention some special cases. 
If $\mathbf{t}=\mathbf{0}$, $\mathcal{E}$ admits causal confusion if and only if 
\be
|\lambda_1|+|\lambda_2|+|\lambda_3|\le 1. 
\ee
This follows from Theorem~\ref{thm:generalqubitchannelcausalconfusion}, but also~\cite[Theorem 1]{YLFW2019}. 
This region in $(\lambda_1,\lambda_2,\lambda_3)$ space corresponds to an octohedron with vertices at $(\pm1,0,0)$, $(0,\pm1,0)$, and $(0,0,\pm1)$. 
In general, the results of Ref.~\cite{YLFW2019} can be interpreted as providing sufficient conditions for $\mathcal{E}$ to be entanglement breaking and hence admitting causal confusion, which therefore extends the results of Ref.~\cite{KublerBraun2018}. A few sample regions in $(\lambda_1,\lambda_2,\lambda_3)$ space are shown in Figure~\ref{fig:RuskaiImages} for a variety of different $(t_1,t_2,t_3)$ values such that the conditions of Theorem~\ref{thm:generalqubitchannelcausalconfusion} are satisfied. 
\ex

\section{Discussion}

In summary, this paper investigated what properties of quantum channels and initial quantum states imply that the temporal correlations associated with a two-time sequential measurement scenario can be realized as the spatial correlations associated with local and independent measurements of a bipartite quantum state. When such temporal correlations can coincide with spatial correlations, we dubbed this \emph{causal confusion}. We looked at a variety of parameterized qubit quantum channels and isolated the subset of the Bloch ball that leads to causal confusion for each value of the parameter of that channel. Already in the examples of depolarizing and mixed two-qubit Pauli channels, we saw that determining that subset of states leading to causal confusion is a nontrivial task already in the case of qubits. 

Our investigations differ from earlier ones on spatiotemporal correlations in many respects. Many studies focus on repeated measurements or use general quantum instruments~\cite{HoSpGuBu2018,Ekblad2025,VieiraBudroni2022,ViKuBu2025,Hayashi2026,BuFaKl2019}, while others focus on correlation matrices associated with maximally-mixed marginals/input states and unital channels~\cite{RAVJSR15,KublerBraun2018}. 
In contrast, we concentrated on a single step in such a sequence, we analyzed a large class of observables and the L{\"u}ders instrument for the temporal scenario, and we identified which states lead to causal confusion. Although a single step in time and the utilization of the L{\"u}ders instrument sounds less general, the benefit is that we achieve mathematical tractability and richness in our tests for quantum causal inference and causal confusion. 
In close analogy, causal inference is already a challenging task classically, in the sense that it usually requires more assumptions beyond data. Such assumptions can come in the form of causal structures, which is why it becomes important to have computable methods to \emph{rule out} causal structures \emph{among} a set of plausible candidates~\cite{CaAnWoEv2025,PearlCausality2009,Evans2016,navascues2020inflation}. 

Although we worked in this more restricted setting consisting of two plausible causal structures associated with two parties, with each party making a single measurement of a light-touch observable, we have discovered intricate differences between spatial and temporal correlations even in this limited setting. In particular, we already see a surprising connection to entanglement-breaking channels and PPT channels. For example, one surprising result is that entanglement-breaking channels in a temporal two-time measurement scenario are \emph{not} necessarily compatible with spatial correlations, though the converse is always true if the spatial bipartite state is \emph{separable}. Another surprising result is that if a temporal scenario involving an initial state and a quantum channel is compatible with a spatial one given by a bipartite state, then the intermediary quantum channel between measurements is \emph{necessarily} a PPT channel (but not necessarily an entanglement-breaking channel), but again the converse need not be true.

Several open questions have surfaced due to the results of this paper. (1) Although we have necessary conditions specifying that causal confusion implies that the intermediary channel must be PPT, one wonders if there are simple necessary and sufficient conditions for general input states. It seems plausible that one could utilize the results from Ref.~\cite{Hayashi2026} to answer this question. Alternatively, is there a suitable notion of distance among states and/or channels that captures causal confusion~\cite{BorovykVellambi2025}? (2) Can the distinction between spatial and temporal correlations of Pauli observables, or more generally light-touch observables, be utilized for some practically useful quantum protocol or algorithm, say akin to Ekert's utilization of Bell inequalities for quantum security~\cite{Ekert1991}? What genuine quantum advantage can be obtained from distinguishing quantum causation from mere correlation, eg., is there a resource theory associated with causal distinguishability~\cite{CFS16}? (3) Do the connections between quantum states over time, entanglement-breaking channels, and PPT channels offer a new and productive perspective for the PPT${}^{2}$ conjecture~\cite{ChMuWo2019,Park2026PPT}? (4) Is there an insightful explanation behind Figure~\ref{fig:hlambdaVglambda} that more closely relates causal confusion to time-symmetric correlations from Bayesian inversion? (5) Finally, motivated by the additional connections between PPT states and other states over time in Ref.~\cite{LorenzBrunnerBenedetti2026}, it is natural to compare and contrast these works.

\vspace{3mm}
\noindent
\textbf{Acknowledgements.} RBM and KT acknowledge support from the Massachusetts Institute of Technology (MIT) Undergraduate Research Opportunities Program (UROP). AJP thanks Owen Ekblad, James Fullwood, Felix Leditzky, Marina Maciel Ansanelli, Minjeong Song, and Yil\`e Y\={\i}ng for discussions. AJP also thanks the organizers (James Fullwood, Aming Liu, and Yulai Wu) of the Horizons of Quantum Information II conference in Hainan University during July 28th to August 1st, 2025, which is where some of the questions answered here were brought up during discussions. Mathematica~\cite{Mathematica2024} and Desmos~\cite{Desmos2026} were used for working through some examples (algebraic calculations and visualizations). Mathematica was used to generate coordinates which were then used in images created with TikZ. 

\vspace{3mm}
\noindent
\textbf{AI usage declaration.}
In this manuscript, Generative AI (GenAI) was used primarily as a tool to learn some concepts, locate references, debug/modify code, and to simplify calculations. 
AJP used Copilot to assist with the decompositions in footnote%~\ref{note:wernerseparable}. %https://tex.stackexchange.com/questions/187863/how-to-refer-to-same-footnote-twice-in-revtex (this is such an awful solution, but I don't know what else to do at the moment) 
~\cite{Note1}, suggest the usage of $\ell(\theta)$ in the parametric equation~\eqref{eqn:MPCparametric}, polish the TikZ code for some figures, and polish a few sentences once the manuscript was written. 
KT used Claude Opus 4.8 to debug Mathematica code (used in the process of analyzing examples), locate references, and assist in the learning process. 
RBM used ChatGPT-5.6 to help simplify complicated expressions, specifically when trying to find a closed-form expression for $\ell^2$ in Proposition~\ref{prop:mixedPaulichannel}. 
All of these results were checked. AI was not used in any other substantial way. In particular, all technical results have been worked out by the authors, and the entire text of the paper was written by the authors.

\appendix

\section{Background in linear algebra and quantum theory}
\label{sec:background}

Here, we establish some standard definitions from quantum information theory, including density matrices, completely positive maps, and the Choi matrix. Importantly, we establish the usual correspondence between qubit density matrices and elements of the Bloch ball. We also justify the usage of expectation values and connect them to probabilities of outcomes. 

\subsection{Observables and density matrices}

The conjugate transpose of a matrix $A$ is denoted by $A^{\dag}$. An \define{observable} is a Hermitian matrix, i.e., a matrix $A$ such that $A^{\dag}=A$. The eigenvalues of an observable $A$ are called the \define{outcomes} of $A$. A \define{positive matrix} (also called \define{positive semidefinite matrix}) is an observable whose outcomes are nonnegative. Equivalently, a matrix $A$ is positive if and only if all of its principal minors are nonnegative~\cite[X.\S4 Theorem 4]{Gantmacher1959} (recall, a \define{principal minor} of a matrix $A$ is obtained by removing any rows and the corresponding columns from $A$). 
If $A\in\matr_{m}$ is an observable, then there exists a unitary matrix $U$, i.e., $U^{\dag}U=\mathds{1}_{m}=UU^{\dag}$, and a diagonal matrix $\Lambda$ such that $A=U\Lambda U^{\dag}=\lambda_1 \mathbf{u}_1 \mathbf{u}_1^{\dag}+\lambda_2 \mathbf{u}_2 \mathbf{u}_2^{\dag}+\cdots+\lambda_m \mathbf{u}_m \mathbf{u}_m^{\dag}$, where $\mathbf{u}_{1},\dots,\mathbf{u}_{m}$ are the columns of $U$, the eigenvectors of $A$, and $\lambda_{1},\dots,\lambda_{m}$ are the diagonal entries of $\Lambda$, the corresponding eigenvalues of $A$. Note that the $\mathbf{u}_{k}\mathbf{u}_{k}^{\dag}$ are rank-1 projections. In case the eigenvalues repeat, it is often useful to group together the corresponding rank-1 projections into a projection onto the corresponding eigenspace. When this is done, the spectral theorem says that every observable $A$ can be uniquely expressed as $A=\lambda_{1}P_{1}+\cdots+\lambda_r P_r$, where $\lambda_{1},\dots,\lambda_{r}$ are the distinct eigenvalues of $A$ and $P_{1},\dots,P_{r}$ are the projections onto the corresponding eigenspaces. This is called the spectral decomposition of $A$. 

Vectors $\mathbf{u}$ may also be written as $|\mathbf{u}\>:=\mathbf{u}$ using Dirac notation, in which case $\<\mathbf{u}|:=\mathbf{u}^{\dag}$. In this way, the decomposition $A=U\Lambda U^{\dag}$ can also be written as $A=\lambda_{1}|\mathbf{u}_{1}\>\<\mathbf{u}_{1}|+\cdots+\lambda_{m} |\mathbf{u}_{m}\>\<\mathbf{u}_{m}|$. Meanwhile, the dot product of $\mathbf{u}$ with $\mathbf{v}$ is expressed in any of the following equivalent ways $\mathbf{u}^{\dag}\mathbf{v}\equiv\<\mathbf{u},\mathbf{v}\>\equiv\<\mathbf{u}|\mathbf{v}\>$. 

A \define{density matrix}, also called a \define{state}, is a matrix that is Hermitian, positive, and has trace $1$. By decomposing a density matrix $\rho\in\matr_{m}$ as $\rho=\sum_{j}p_{j}\mathbf{u}_{j}\mathbf{u}^{\dag}$, the collection of numbers $(p_1,\dots,p_m)$ defines a probability distribution. 

\subsection{The expectation value}

The \define{(quantum) expectation value} of an observable $A$ in the state $\rho$ is the number 
\be
\mathbb{E}_{\rho}[A]:=\Tr[\rho A].
\ee
The justification for this terminology is important for this paper, so we review it. Write the spectral decomposition of $A$ as $A=\sum_{k=1}^{r}\lambda_{k}P_{k}$ so that 
\be
\mathbb{E}_{\rho}[A]=\sum_{k}\lambda_{k} \Tr[\rho P_{k}].
\ee
Notice that 
\be
\Tr[\rho P_{k}]\ge0
\ee
for all $k$ and that 
\be
\sum_{k}\Tr[\rho P_{k}]
=\Tr\left[\rho \sum_{k}P_{k}\right]
=\Tr[\rho\mathds{1}_{m}]
=\Tr[\rho]
=1.
\ee
This proves that 
\be
p:=\big(\Tr[\rho P_{1}],\dots,\Tr[\rho P_{r}]\big)
\ee
defines a probability distribution. The interpretation that outcome $\lambda_{k}$ occurs with probability $p_{k}:=\Tr[\rho P_{k}]$ is called the \define{Born rule} of quantum mechanics. Therefore, if we let $\lambda$ denote the random variable whose values are given by $\lambda=(\lambda_{1},\dots,\lambda_{r})$, then we have that 
\be
\mathbb{E}_{\rho}[A]=\mathbb{E}_{p}[\lambda],
\ee
where the \define{(classical) expectation value} on the right-hand-side is the usual expectation value of the random variable $\lambda$ with respect to the probability distribution $p=(p_1,\dots,p_r)$~\cite{Ja03}, namely, 
\be
\mathbb{E}_{p}[\lambda]:=\sum_{k}p_{k}\lambda_{k}. 
\ee
Now, if an experimenter measures the observable $A$ in the state $\rho$, each such measurement will induce only a single outcome. If the experimenter can repeat the experiment, with the capability of measuring the observable $A$ in the same state $\rho$ during each run of the experiment, then the experimenter will end up with a frequency chart of the outcomes. Let $f_{k}$ denote the number of times the outcome $\lambda_{k}$ occurs among a total number of $F=\sum_{k}f_{k}$ runs of the experiment. Then, in the limit of infinitely many runs of the experiment, one expects 
\be
\lim_{F\to\infty}\frac{f_k}{F}= p_{k}=\Tr[\rho P_{k}]. 
\ee
This connects the idealized probabilities $p_{k}$ with the sample-based probabilities obtained from a finite number of runs of the experiment. In this paper, we mostly work with the idealized probabilities when proving theorems, but we often discuss the experimental setups as if they are done in a finite number of runs. 

\subsection{Pauli observables and qubit density matrices}

The four \define{Pauli observables} are given by 
\be
\sigma_0\!:=\!\begin{bmatrix}1&0\\0&1\end{bmatrix}\!,
\;
\sigma_{1}\!:=\!\begin{bmatrix}0&1\\1&0\end{bmatrix}\!,
\;
\sigma_{2}\!:=\!\begin{bmatrix}0&-i\\i&0\end{bmatrix}\!,
\;
\sigma_{3}\!:=\!\begin{bmatrix}1&0\\0&-1\end{bmatrix}\!.
\ee
Additionally, let
\begin{align}
|0\>&:=\begin{bmatrix}1\\0\end{bmatrix}, & 
|+\>&:=\frac{1}{\sqrt{2}}\begin{bmatrix}1\\1\end{bmatrix}, & 
|+i\>&:=\frac{1}{\sqrt{2}}\begin{bmatrix}1\\i\end{bmatrix}, \nonumber\\
|1\>&:=\begin{bmatrix}0\\1\end{bmatrix}, & 
|-\>&:=\frac{1}{\sqrt{2}}\begin{bmatrix}1\\-1\end{bmatrix}, & 
|-i\>&:=\frac{1}{\sqrt{2}}\begin{bmatrix}1\\-i\end{bmatrix},
\end{align}
so that $|0\>$ and $|1\>$ are eigenvectors of $\sigma_{3}$, $|+\>$ and $|-\>$ are eigenvectors of $\sigma_{1}$, and $|+i\>$ and $|-i\>$ are eigenvectors of $\sigma_{2}$ (with corresponding eigenvalues $+1$ and $-1$, respectively). An important fact about Pauli matrices is that they are orthogonal with respect to the \define{Hilbert--Schmidt} inner product on square matrices, which is defined by 
\be
(A,B)\mapsto \Tr[A^{\dag}B]
\ee
for all $A,B\in\matr_{m}$. The statement about the orthogonality of the Pauli matrices reads
\be
\Tr[\sigma_{j}\sigma_{k}]=2\delta_{jk}
\ee
for all $j,k\in\{0,1,2,3\}$ (the equation was simplified by the fact that $\sigma_{j}^{\dag}=\sigma_{j}$), where $\delta_{jk}=1$ if $j=k$ and $0$ otherwise. This orthogonality relation implies that every $2\times 2$ matrix $A$ can be written as the linear combination 
\be
A=\frac{1}{2}\sum_{k=0}^{3}\Tr[\sigma_{k}A]\sigma_{k}.
\ee

\begin{figure}
\includegraphics{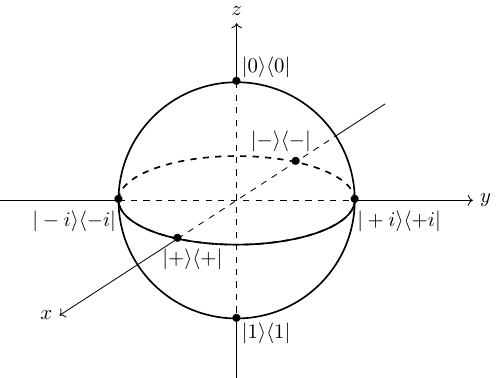}
\caption{The Bloch ball (the set of points $(x,y,z)$ satisfying $x^2+y^2+z^2\le 1$) parametrizing single qubit density matrices is shown. The six density matrices corresponding to the projectors onto the eigenspaces of the Pauli matrices $\sigma_1,\sigma_2,\sigma_3$ are also shown.}
\label{fig:Blochball}
\end{figure}

In particular, every qubit (i.e., $2\times 2$) density matrix $\rho\in\matr_{2}$ can be written as 
\be
\label{eqn:qubitrho}
\rho=\frac{1}{2}(\mathds{1}_{2}+\mathbf{r}\cdot\boldsymbol{\sigma})
=\frac{1}{2}\begin{bmatrix}1+z&x-iy\\x+iy&1-z\end{bmatrix}
\ee
with $\mathbf{r}=(x,y,z)\in\R^{3}$ satisfying $\lVert \mathbf{r}\rVert\le1$ (i.e., $\mathbf{r}$ is inside the standard unit 3-ball in $\R^{3}$). In this notation, $\mathbf{r}\cdot\boldsymbol{\sigma}:=x\sigma_1+y\sigma_2+z\sigma_3$, where 
$\lVert\mathbf{r}\rVert$ is the length of $\mathbf{r}$, which is equivalently $\lVert\mathbf{r}\rVert=\sqrt{\mathbf{r}\cdot\mathbf{r}}$. In this way, the set of qubit density matrices is in one-to-one correspondence with the unit ball in $\R^3$, and which is often called the \define{Bloch ball} (cf.\ Figure~\ref{fig:Blochball}). One can show that the components of $\mathbf{r}$ are the expectation values of the Pauli observables $\sigma_1,\sigma_2,\sigma_3$ in the state $\rho$, i.e.,  
\be
\mathbf{r}=\big(\Tr[\rho\sigma_1],\Tr[\rho\sigma_2],\Tr[\rho\sigma_3]\big).
\ee
In other words, every $2\times 2$ density matrix is uniquely determined by its expectation values for the Pauli matrices $\sigma_1,\sigma_2,\sigma_3$ (the expectation value of $\sigma_0$ is automatically $1$ by the definition of a density matrix). 

\subsection{Quantum channels}

Let $\Alg{A}$ and $\Alg{B}$ be full matrix algebras. A \define{positive linear map} from $\Alg{A}$ to $\Alg{B}$ is a function $\mathcal{E}:\Alg{A}\to\Alg{B}$ such that 
\begin{enumerate}[i.]
\item
$\mathcal{E}(A_1+A_2)=\mathcal{E}(A_1)+\mathcal{E}(A_2)$ for all $A_1,A_2\in\Alg{A}$
\item
$\mathcal{E}(c A)=c\mathcal{E}(A)$ for all $A\in\Alg{A}$ and $c\in\C$
\item
$\mathcal{E}(A)$ is always positive whenever $A$ is positive.
\end{enumerate}
The first two conditions say that $\mathcal{E}$ is \define{linear}, and the third that $\mathcal{E}$ is \define{positive}. A \define{quantum channel} requires two additional conditions. The first is that $\mathcal{E}$ is \define{trace-preserving}, i.e., $\Tr\big[\mathcal{E}(A)\big]=\Tr[A]$ for all $A\in\Alg{A}$. To state the second, first define the \define{Choi matrix} of $\mathcal{E}$, which is 
\be
\label{eqn:Choimatrix}
\Choi[\mathcal{E}]=\sum_{i,j}|i\>\<j|\otimes\mathcal{E}\big(|i\>\<j|\big)
=(\id_{\Alg{A}}\otimes\mathcal{E})\big(|\Upsilon\>\<\Upsilon|\big), 
\ee
where $|\Upsilon\>:=\sum_{i}|i\>\otimes|i\>$ and $\{|i\>\}$ is the standard basis~\cite{Ch75}. The additional condition for $\mathcal{E}$ to be a quantum channel is that $\Choi[\mathcal{E}]$ is positive semidefinite. 

An example of a quantum channel is the partial trace. Given full matrix algebras $\Alg{A}$ and $\Alg{B}$, the assignment $\Tr_{\Alg{B}}:\Alg{A}\otimes\Alg{B}\to\Alg{A}$ uniquely determined by linear extension by sending tensors of the form $A\otimes B$ to $\Tr[B]A$ is called a \define{partial trace}. There is also a partial trace of the form $\Tr_{\Alg{A}}:\Alg{A}\otimes\Alg{B}\to\Alg{B}$ uniquely determined by sending tensors of the form $A\otimes B$ to $\Tr[A]B$. The reason these are uniquely defined by this is because, although not every element of $\Alg{A}\otimes\Alg{B}$ is of the form $A\otimes B$, it is the case that every element of $\Alg{A}\otimes\Alg{B}$ is of the form $\sum_{k} A_{k}\otimes B_{k}$ for some finite collections $A_{k}\in\Alg{A}$ and $B_{k}\in\Alg{B}$. So, for example, $\Tr_{\Alg{B}}[\sum_{k} A_{k}\otimes B_{k}]=\sum_{k} \Tr[ B_{k}] A_{k}$.

\section{Examples of joint expectation value tables}
\label{sec:exampletables}

In this appendix, we provide several examples of tables of joint expectation values that are obtained either from a common-cause scenario or a direct-cause scenario.

\begin{example}
A bipartite state $\tau$ is a \define{factorized density matrix} iff $\tau=\rho\otimes\nu$, where $\rho$ and $\nu$ are each $2\times 2$ density matrices. In this case, 
\be
\mathbb{E}_{\rho\otimes\nu}[A\otimes B]=\mathbb{E}_{\rho}[A]\,\mathbb{E}_{\nu}[B]
\ee
for all observables $A\in\Alg{A}$ and $B\in\Alg{B}$. This means that Alice's and Bob's measurement outcomes are independent of each other. If we write out each density matrix as 
\begin{align}
\rho&=\frac{1}{2}(\mathds{1}_{2}+x_1\sigma_1+y_1\sigma_2+z_1\sigma_3) \quad\text{ and } \nonumber \\
\nu&=\frac{1}{2}(\mathds{1}_{2}+x_2\sigma_1+y_2\sigma_2+z_2\sigma_3),
\end{align}
then the table of joint expectation values for all Pauli observables is given by 
\begin{center}
\noindent
\begin{tabular}{c|cccc}
 & $\sigma_0$ & $\sigma_1$ & $\sigma_2$ & $\sigma_3$ \\
\hline 
$\sigma_0$ & \,\;\;$1$\,\;\; & \,\;\;$x_2$\,\;\; & \,\;\;$y_2$\,\;\; & \,\;\;$z_2$\;\; \\ 
$\sigma_1$ & $x_1$ & $x_1 x_2$ & $x_1 y_2$ & $x_1 z_2$ \\
$\sigma_2$ & $y_1$ & $y_1 x_2$ & $y_1 y_2$ & $y_1 z_2$ \\
$\sigma_3$ & $z_1$ & $z_1 x_2$ & $z_1 y_2$ & $z_1 z_3$
\end{tabular}
\end{center}
\end{example}

\begin{example}
\label{ex:WernerSeparable}
A bipartite state $\tau$ is \define{separable} iff there exists a probability distribution $(t_1,\dots,t_r)$ and density matrices $\rho_1,\dots,\rho_r\in\Alg{A}$ and $\nu_1,\dots,\nu_r\in\Alg{B}$ such that $\tau=\sum_{l=1}^{r}t_{l}\rho_{l}\otimes\nu_{l}$. A simple example of a separable density matrix can be obtained from the one-parameter family of 2-qubit \define{Werner states}~\cite{We89} given by 
\begin{equation}
\label{eqn:wernerstates}
\tau=
s\left(\frac{\mathds{1}_{4}+\SWAP}{6}\right)+(1-s)\left(\frac{\mathds{1}_{4}-\SWAP}{2}\right),
\end{equation}
where 
\be
\SWAP=\frac{1}{2}\sum_{k=0}^{3}\sigma_k\otimes\sigma_k=\setlength{\arraycolsep}{5pt}\begin{bmatrix}1&0&0&0\\0&0&1&0\\0&1&0&0\\0&0&0&1\end{bmatrix}
\ee
is the 2-qubit $\SWAP$ operator and $s$ is in the range $s\in[\frac{1}{2},1]$~%
%footnote
\footnote{%\label{note:wernerseparable}
The reason $s\in[\frac{1}{2},1]$ in order for the Werner state to be separable can be seen in the following way. One decomposition of $\tau$ into an affine combination of tensor products of density matrices is given by 
\[
\tau=(4s-2)\frac{\mathds{1}_{4}}{4}+\frac{3-4s}{6}\sum_{k=1}^{3}\Big(\rho_{k}^{+}\otimes\rho_{k}^{-}+\rho_{k}^{-}\otimes\rho_{k}^{+}\Big),
\]
where 
$
\rho_{k}^{\pm}:=\frac{1}{2}(\mathds{1}_{2}\pm\sigma_k).
$
Notice that 
$
\frac{\mathds{1}_{4}}{4}=
\frac{\mathds{1}_{2}}{2}\otimes \frac{\mathds{1}_{2}}{2}.
$
Therefore, this decomposition explicitly shows that $\tau$ is separable for the values $s\in[\frac{1}{2},\frac{3}{4}]$ since the coefficients are nonnegative for those values. Another decomposition of $\tau$ is given by 
\[
\tau=(4-4s)\frac{\mathds{1}_{4}}{4}+\frac{4s-3}{6}\sum_{k=1}^{3}\Big(\rho_{k}^{+}\otimes\rho_{k}^{+}+\rho_{k}^{-}\otimes\rho_{k}^{-}\Big),
\]
which explicitly shows that $\tau$ is separable for values $s\in[\frac{3}{4},1]$.
}.
%end footnote
As a concrete example, the case when $s=\frac{1}{2}$ is 
\be
\tau=\frac{1}{6}\begin{bmatrix}1&0&0&0\\0&2&-1&0\\0&-1&2&0\\0&0&0&1\end{bmatrix}.
\ee
The corresponding table of joint expectation values is given by 
\begin{center}
\noindent
\begin{tabular}{c|cccc}
 & $\sigma_0$ & $\sigma_1$ & $\sigma_2$ & $\sigma_3$ \\
\hline 
$\sigma_0$ & \,\;\;$1$\,\;\; & \,\;\;$0$\,\;\; & \,\;\;$0$\,\;\; & \,\;\;$0$\;\; \\ 
$\sigma_1$ & $0$ & $-\frac{1}{3}$ & $0$ & $0$ \\
$\sigma_2$ & $0$ & $0$ & $-\frac{1}{3}$ & $0$ \\
$\sigma_3$ & $0$ & $0$ & $0$ & $-\frac{1}{3}$
\end{tabular}
\end{center}
This is clearly not of the form provided in Example 1 thus showing that the bipartite state in this example is not factorized. 
\end{example}

\begin{example}
\label{ex:Wernerentangled}
A bipartite state $\tau$ is \define{entangled} iff $\tau$ is not separable. The 2-qubit Werner state~\eqref{eqn:wernerstates} is entangled whenever $s\in[0,\frac{1}{2})$~%
%footnote
\footnote{This can be seen by Peres' positive partial transpose (PPT) criterion~\cite{peres1996separability,horodecki1996necessary}. Namely, if $\mathcal{T}:\matr_{m}\to\matr_{m}$ denotes the transpose map that sends an $m\times m$ matrix $A$ to its transpose $A^T$, then the PPT criterion states that if $\tau\in\matr_{m}\otimes\matr_{m}$ is a density matrix such that $(\id\otimes\mathcal{T})(\tau)$ is \emph{not} positive, then $\tau$ is entangled (this is only a sufficient condition for entanglement, not a necessary condition). The map $\id\otimes\mathcal{T}$ is called the \define{partial transpose}. The partial transpose of the Werner state is 
\[
(\id\otimes\mathcal{T})(\tau)=s\left(\frac{\mathds{1}_{4}+2|\Omega\>\<\Omega|}{6}\right)+(1-s)\left(\frac{\mathds{1}_{4}-2|\Omega\>\<\Omega|}{2}\right),
\]
where 
$
|\Omega\>=\frac{1}{\sqrt{2}}
\big(|0\>\otimes|0\>+|1\>\otimes|1\>\big).
$
The eigenvalues of $|\Omega\>\<\Omega|$ are just $0$ (with multiplicity 3) and $1$ (with multiplicity 1). Hence, since $\mathds{1}_{4}$ commutes with $|\Omega\>\<\Omega|$, the eigenvalues of $(\id\otimes\mathcal{T})(\tau)$ are 
\[
\frac{s}{6}+\frac{1-s}{2}=\frac{3-2s}{6} 
\;\;\text{ and }\;\;
\frac{3s}{6}+\frac{(1-s)(-1)}{2}=\frac{2s-1}{2}.
\]
Among the values of $s\in[0,1]$ for which at least one of these eigenvalues is negative leads to $s\in[0,\frac{1}{2})$. This proves that the Werner states are entangled states when $s\in[0,\frac{1}{2})$. 
}.
%end footnote
As a concrete example, the case when $s=\frac{1}{4}$ is 
\be
\tau=\frac{1}{12}\begin{bmatrix}1&0&0&0\\0&5&-4&0\\0&-4&5&0\\0&0&0&1\end{bmatrix}.
\ee
The corresponding table of joint expectation values is given by 
\begin{center}
\noindent
\begin{tabular}{c|cccc}
 & $\sigma_0$ & $\sigma_1$ & $\sigma_2$ & $\sigma_3$ \\
\hline 
$\sigma_0$ & \,\;\;$1$\,\;\; & \,\;\;$0$\,\;\; & \,\;\;$0$\,\;\; & \,\;\;$0$\;\; \\ 
$\sigma_1$ & $0$ & $-\frac{2}{3}$ & $0$ & $0$ \\
$\sigma_2$ & $0$ & $0$ & $-\frac{2}{3}$ & $0$ \\
$\sigma_3$ & $0$ & $0$ & $0$ & $-\frac{2}{3}$
\end{tabular}
\end{center}
\end{example}

\begin{example}
\label{ex:temporaldepolarizing}
Let $\rho=\frac{\mathds{1}_{2}}{2}$ be the maximally mixed qubit state and let 
\be
\label{eqn:PDchannel}
\mathcal{E}(A):=(1-\lambda)A+\lambda \Tr[A] \frac{\mathds{1}_{2}}{2}
\ee
be the partially depolarizing channel parameterized by $\lambda\in[0,\frac{4}{3}]$. As a particular case, take $\lambda=\frac{1}{2}$. In this case, the corresponding table of joint expectation values is 
\begin{center}
\noindent
\begin{tabular}{c|cccc}
 & $\sigma_0$ & $\sigma_1$ & $\sigma_2$ & $\sigma_3$ \\
\hline 
$\sigma_0$ & \,\;\;$1$\,\;\; & \,\;\;$0$\,\;\; & \,\;\;$0$\,\;\; & \,\;\;$0$\;\; \\ 
$\sigma_1$ & $0$ & $\frac{1}{2}$ & $0$ & $0$ \\
$\sigma_2$ & $0$ & $0$ & $\frac{1}{2}$ & $0$ \\
$\sigma_3$ & $0$ & $0$ & $0$ & $\frac{1}{2}$
\end{tabular}
\end{center}
The entries $\mathbb{E}_{(\rho,\mathcal{E})}[\sigma_{j},\sigma_{k}]$ of this table seems to involve a lot of calculation steps due to the fact that one uses the spectral decomposition of each observable. We will later review a significantly simpler way to compute the entries of this table that is analogous to that for bipartite states.
\end{example}

\begin{example}
Let $\rho=\frac{\mathds{1}_{2}}{2}$ and let $\mathcal{E}$ be the partially depolarizing channel~\eqref{eqn:PDchannel}. If $\lambda=\frac{4}{3}$, the corresponding table of temporal joint expectation values is given by 
\begin{center}
\noindent
\begin{tabular}{c|cccc}
 & $\sigma_0$ & $\sigma_1$ & $\sigma_2$ & $\sigma_3$ \\
\hline 
$\sigma_0$ & \,\;\;$1$\,\;\; & \,\;\;$0$\,\;\; & \,\;\;$0$\,\;\; & \,\;\;$0$\;\; \\ 
$\sigma_1$ & $0$ & $-\frac{1}{3}$ & $0$ & $0$ \\
$\sigma_2$ & $0$ & $0$ & $-\frac{1}{3}$ & $0$ \\
$\sigma_3$ & $0$ & $0$ & $0$ & $-\frac{1}{3}$
\end{tabular}
\end{center}
This should look familiar, because it happens to coincide with the \emph{spatial} joint expectation values we found in Example~\ref{ex:WernerSeparable} for a particular bipartite Werner state. 
\end{example}

\section{Proofs of scattered results in the literature}
\label{sec:proofs}

In this appendix, we compile some proofs that are scattered throughout the literature but which we utilize here. 

\subsection{Proof that identity channel never admits causal confusion}
\label{sec:idcausalconf}

Here, we prove Proposition~\ref{prop:identitychannel}, which says that $\id_{\Alg{A}}\star\rho$ is never positive for any state $\rho$. Before doing so, we state some additional properties of the swap operator. These include
\be
\SWAP_{\Alg{A}}^{\dag}=\SWAP_{\Alg{A}},
\qquad\SWAP_{\Alg{A}}^2=\mathds{1}_{\Alg{A}},
\ee 
\be
\SWAP_{\Alg{A}} \big(|\mathbf{u}\>\otimes|\mathbf{v}\>\big)=|\mathbf{v}\>\otimes|\mathbf{u}\>
\ee
for all vectors $|\mathbf{u}\>,|\mathbf{v}\>\in\C^{m}$, 
\be
\SWAP_{\Alg{A}}(A_{1}\otimes A_{2})\SWAP_{\Alg{A}}
=A_{2}\otimes A_{1} 
\ee
for all $A_1,A_2\in\Alg{A}$, and 
\be
(U\otimes V)\SWAP_{\Alg{A}}(U^{\dag}\otimes V^{\dag})=\SWAP_{\Alg{A}}
\ee 
for all unitaries $U,V\in\Alg{A}$. This last property says that $\SWAP_{\Alg{A}}$ can be expressed via the same formula~\eqref{eqn:SWAP} using any orthonormal basis of $\C^{m}$. 

\bprf[Proof of Proposition~\ref{prop:identitychannel}]
First notice that $\id_{\Alg{A}}\star\rho$ can be re-expressed as
\begin{align}
\id_{\Alg{A}}\star\rho
&=\frac{1}{2}\Big((\rho\otimes\mathds{1}_{\Alg{A}})\SWAP_{\Alg{A}}+\SWAP_{\Alg{A}}(\rho\otimes\mathds{1}_{\Alg{A}})\Big) \nonumber \\
&=\frac{1}{2}\big(\rho\otimes\mathds{1}_{\Alg{A}}+\mathds{1}_{\Alg{A}}\otimes\rho\big)\SWAP_{\Alg{A}}. 
\end{align}
Similarly, 
\be
\id_{\Alg{A}}\star\rho=\frac{1}{2}\SWAP_{\Alg{A}}\big(\rho\otimes\mathds{1}_{\Alg{A}}+\mathds{1}_{\Alg{A}}\otimes\rho\big). 
\ee
This shows that the operator $\rho\otimes\mathds{1}_{\Alg{A}}+\mathds{1}_{\Alg{A}}\otimes\rho$ commutes with $\SWAP_{\Alg{A}}$, which implies that there exists a common orthonormal basis of $\C^{m}\otimes\C^{m}$ consisting of eigenvectors for both operators. In fact, if we write write an eigenvector decomposition of $\rho$ as 
\be
\rho=\sum_{i=1}^{m}p_{i}|\mathbf{u}_{i}\>\<\mathbf{u}_{i}|, 
\ee 
then set 
\be
\label{eqn:viivectors}
|\mathbf{v}_{ii}\>:=|\mathbf{u}_{i}\>\otimes|\mathbf{u}_{i}\>
\ee
for $i\in\{1,\dots,m\}$ and 
\be
\label{eqn:vijvectors}
|\mathbf{v}_{ij}^{\pm}\>:=\frac{1}{\sqrt{2}}\Big(|\mathbf{u}_{i}\>\otimes|\mathbf{u}_{j}\>\pm|\mathbf{u}_{j}\>\otimes|\mathbf{u}_{i}\>\Big)
\ee
for all integers $i,j$ satisfying $1\le i<j\le m$. The set of vectors from~\eqref{eqn:viivectors} and~\eqref{eqn:vijvectors} together form an orthonormal basis of $\C^{m}\otimes\C^{m}$ consisting of \emph{common} eigenvectors of $\rho\otimes\mathds{1}_{\Alg{A}}+\mathds{1}_{\Alg{A}}\otimes\rho$ and $\SWAP_{\Alg{A}}$. This is because 
\be
(\rho\otimes\mathds{1}_{\Alg{A}}+\mathds{1}_{\Alg{A}}\otimes\rho)|\mathbf{v}_{ii}\>=2p_{i}|\mathbf{v}_{ii}\>
\ee
and
\be
\SWAP_{\Alg{A}}|\mathbf{v}_{ii}\>=|\mathbf{v}_{ii}\>
\ee
for all $i\in\{1,\dots,m\}$, and because
\be
(\rho\otimes\mathds{1}_{\Alg{A}}+\mathds{1}_{\Alg{A}}\otimes\rho)|\mathbf{v}_{ij}^{\pm}\>
=(p_{i}+p_{j})|\mathbf{v}_{ij}^{+}\>
\ee
and
\be
\SWAP_{\Alg{A}}|\mathbf{v}^{\pm}_{ij}\>=\pm |\mathbf{v}^{\pm}_{ij}\>
\ee
for all integers $i,j$ satisfying $1\le i<j\le m$. Therefore, 
\be
(\id_{\Alg{A}}\star\rho)|\mathbf{v}_{ij}^{-}\>=-\left(\frac{p_i+p_j}{2}\right)|\mathbf{v}_{ij}^{-}\>
\ee
for all integers $i,j$ satisfying $1\le i<j\le m$, which shows that $\id_{\Alg{A}}\star\rho$ always has negative eigenvalues. 
\eprf

\subsection{Solution to the Jordan product matrix equation}
\label{sec:Sylvester}

Here we prove Lemma~\ref{lem:BhatiaRosenberg}, which says that the solution to $\{A,X\}=B$ is $X=\int_{0}^{\infty}e^{-tA} B e^{-tA}dt$ whenever $A$ is positive definite. 

\bprf
[Proof of Lemma~\ref{lem:BhatiaRosenberg}]
First notice that if $A$ is any $m\times m$ matrix, then the function 
\be
\label{eqn:bracketfunction}
\matr_{m}\ni X\mapsto \{A,X\}\in\matr_{m}
\ee
on matrices is linear. Second, if $A$ is positive definite the kernel (nullspace~\cite{Strang2022}) of~\eqref{eqn:bracketfunction} is trivial, i.e., if $\{A,X\}=0$, then $X=0$. To see this, first notice that 
\be
\{A,X\}=0 
\iff
AX=-XA.
\ee
Next, let $A=\sum_{k=1}^{m}\lambda_{k}|\mathbf{u}_{k}\>\<\mathbf{u}_{k}|$ be an eigendecomposition of $A$ so that $\{\lambda_{k}\}$ are all strictly positive and so that $\{\mathbf{u}_{k}\}$ defines an orthonormal basis of $\C^{m}$. Then this last equation is equivalent to 
\be
\<\mathbf{u}_{j}|AX|\mathbf{u}_{k}\>
=-\<\mathbf{u}_{j}|XA|\mathbf{u}_{k}\>,
\ee
which is itself equivalent to 
\be
\lambda_{j}\<\mathbf{u}_{j}|X|\mathbf{u}_{k}\>
=-\lambda_{k}\<\mathbf{u}_{j}|X|\mathbf{u}_{k}\>
\ee
for all $j,k\in\{1,\dots,m\}$. This implies $\<\mathbf{u}_{j}|X|\mathbf{u}_{k}\>=0$ for all $j,k\in\{1,\dots,m\}$. But since $\{\mathbf{u}_{k}\}$ is a basis, this implies $X=0$. Thus,~\eqref{eqn:bracketfunction} is a linear transformation with trivial kernel. Therefore, if a solution to the inhomogeneuous system $\{A,X\}=B$ exists, it is necessarily unique~\cite{Strang2022}. 

We next show that~\eqref{eqn:SylvesterSolution} solves~\eqref{eqn:SylvesterEquation}. Since
\be
\<\mathbf{u}_{j}|B|\mathbf{u}_{k}\>
=\<\mathbf{u}_{j}|\{A,X\}|\mathbf{u}_{k}\>
=(\lambda_{j}+\lambda_{k})\<\mathbf{u}_{j}|X|\mathbf{u}_{k}\>, 
\ee
we arrive at the solution
\be
\<\mathbf{u}_{j}|X|\mathbf{u}_{k}\>=\frac{\<\mathbf{u}_{j}|B|\mathbf{u}_{k}\>}{\lambda_{j}+\lambda_{k}}.
\ee
To see that this agrees with~\eqref{eqn:SylvesterSolution}, we compute the $jk$ entry 
\begin{align}
\bigg\<\mathbf{u}_{j}\bigg| \int_{0}^{\infty} e^{-tA}B&e^{-tA}\,dt \bigg| \mathbf{u}_{k}\bigg\> \nonumber \\
&=\int_{0}^{\infty} \<\mathbf{u}_{j}| e^{-tA}Be^{-tA} |\mathbf{u}_{k}\> \, dt \nonumber \\
&=\left(\int_{0}^{\infty} e^{-t (\lambda_{j}+\lambda_{k})} \,dt \right) \<\mathbf{u}_{j}| B |\mathbf{u}_{k}\> \nonumber \\
&=\frac{\<\mathbf{u}_{j}|B|\mathbf{u}_{k}\>}{\lambda_{j}+\lambda_{k}}.
\label{eqn:uSylvestoru}
\end{align}
This proves the first claim. 

For the second claim, if $B$ is positive semidefinite, then we'll give two arguments to show $X$ is positive semidefinite. First, notice that the integral expression for $X$ in~\eqref{eqn:SylvesterSolution} is a limit of Riemann sums of the form
\be
\lim_{\Delta t\to0}\sum_{k=0}^{\infty} (\Delta t) \left(B^{\frac{1}{2}}e^{-k (\Delta t)A}\right)^{\dag} \left(B^{\frac{1}{2}}e^{-k (\Delta t) A}\right),
\ee
with each term being positive semidefinite by the Cholesky test~\cite{Strang2022}. For a different proof using eigenvectors, using~\eqref{eqn:uSylvestoru} gives
\be
X=\sum_{j,k=1}^{m}\left(\frac{1}{\lambda_{j}+\lambda_{k}}\right) \<\mathbf{u}_{j}|B|\mathbf{u}_{k}\> |\mathbf{u}_{j}\> \<\mathbf{u}_{k}|
\ee
is the Hadamard--Schur product~\cite{Bh07} of the matrix 
\be
\Gamma:=\sum_{k,l=1}^{m} \left(\frac{1}{\lambda_{j}+\lambda_{k}}\right)|\mathbf{u}_{j}\> \<\mathbf{u}_{k}|
\ee 
with the matrix $B$ with respect to the $\{\mathbf{u}_{k}\}$ orthonormal basis. Since the Hadamard--Schur product of two positive semidefinite matrices is positive semidefinite (this is called the Schur Product Theorem---see Exercise 1.2.5 in Ref.~\cite{Bh07}), it suffices to show $\Gamma$ is positive semidefinite. This is the case (see Exercise 1.1.2 in Ref.~\cite{Bh07} for example), and for completeness we include an argument here. It follows from the integral expression for its entries, namely, if $\mathbf{x}=\sum_{j=1}^{m}x_{j}\mathbf{u}_{j}$ is an arbitrary vector, then 
\begin{align}
\<\mathbf{x}|\Gamma|\mathbf{x}\>
&=\int_{0}^{\infty}\sum_{i,j,k,l}\overline{x_{i}}x_{j}e^{-\lambda_{k} t}e^{-\lambda_{l} t} \<\mathbf{u}_{i}|\mathbf{u}_{k}\>\<\mathbf{u}_{l}|\mathbf{u}_{j}\>\,dt\nonumber \\
&=\int_{0}^{\infty}\sum_{k,l}\overline{x_{k}}x_{j}e^{-\lambda_{k}t}e^{-\lambda_{l}t}\,dt \nonumber\\
&=\int_{0}^{\infty}\left|\sum_{k}x_{k}e^{-\lambda_{k}t}\right|^2\,dt
\ge0.
\end{align}
Moreover, if $A$ is positive definite and $\mathbf{x}\ne0$, then this reads $\<\mathbf{x}|\Gamma|\mathbf{x}\>>0$. Thus, by the energy test for positivity~\cite{Strang2022}, this shows that $\Gamma$ is positive definite. 
\eprf

\section{The most general qubit quantum channel}
\label{sec:Ruskai}

In this section, we recall the characterization of qubit quantum channels from Ref.~\cite{RSW02} (see also Ref.~\cite{WPVF25}). We first note that every qubit quantum channel $\mathcal{F}:\matr_2\to\matr_2$ is of the form 
\be
\mathcal{F}=\Ad_{V}\circ\mathcal{E}\circ\Ad_{U}, 
\ee
for some unitary matrices $U,V\in\matr_2$ and for $\mathcal{E}$ of the form~\eqref{eq:generalqubitcannel}. In the above expression, $\mathrm{Ad}_{U}(A):=UAU^{\dag}$ for all $A\in\matr_{2}$ and similarly for $\mathrm{Ad}_{V}$. Since Section~\ref{sec:generalqubit} focused only on $\mathcal{E}$, we prove here why that can be done without loss of generality. 

\bn
A quantum channel $\mathcal{E}$ admits causal confusion if and only if $\mathcal{F}=\Ad_{V}\circ\mathcal{E}\circ\Ad_{U}$ admits causal confusion for all unitary matrices $U$ and $V$. 
\en

\bprf
The Choi matrix of $\mathcal{F}$ is 
\begin{align}
\Choi[\mathcal{F}]&= \sum_{i,j} |\mathbf{e}_{i}\>\<\mathbf{e}_{j}|\otimes V\mathcal{E}\big(U|\mathbf{e}_{i}\> \<\mathbf{e}_{j}|U^{\dag}\big)V^{\dag} \nonumber \\
&=(U^{\dag}\otimes V)\left(\sum_{i,j} |\mathbf{u}_i\> \<\mathbf{u}_{j}| \otimes \mathcal{E}(|\mathbf{u}_i\> \<\mathbf{u}_{j}|) \right) (U\otimes V^{\dag}),
\end{align}
where $|\mathbf{u}_{i}\>:=U|\mathbf{e}_{i}\>$ for all $i$. This is positive if and only if the term inside the large parentheses is positive. Although the term inside the large parentheses is not equal to the standard Choi matrix of $\mathcal{E}$, it is the Choi matrix of $\mathcal{E}$ expressed in terms of a different orthonormal basis, and is therefore positive if and only if $\Choi[\mathcal{E}]$ is positive. Similarly, and more simply, the Jamio{\l}kowski matrix becomes
\be
\Jamio[\mathcal{F}]=(U^{\dag}\otimes V)\Jamio[\mathcal{E}](U\otimes V^{\dag})
\ee
due to the fact that the Jamio{\l}kowski matrix is independent of the choice of basis. Hence, $\Jamio[\mathcal{F}]$ is positive if and only if $\Jamio[\mathcal{E}]$ is positive. Hence, by Proposition~\ref{prop:PPTQSOT}, $\mathcal{F}$ is a PPT channel if and only if $\mathcal{E}$ is a PPT channel. Thus, by Corollary~\ref{cor:causalconfusioniffPPT}, $\mathcal{F}$ admits causal confusion if and only if $\mathcal{E}$ admits causal confusion. 
\eprf
        
%%%%%%%%BIBLIOGRAPHY%%%%%%%%%%%%
\addcontentsline{toc}{section}{\numberline{}Bibliography}
\bibliography{references}
\end{document}